\documentclass[prd,preprintnumbers,twocolumn,amsmath,nofootinbib,amssymb]{revtex4}
\usepackage{graphicx,color,dcolumn,booktabs,bm}
\usepackage{longtable,lscape}
\usepackage{txfonts}
\usepackage{overpic}
\usepackage{amssymb}
\usepackage{epstopdf}
\usepackage{indentfirst}
\usepackage{feynmf}   
\usepackage{slashed}  
\usepackage{cases}
\usepackage{color}
\usepackage{float}
\usepackage{multirow}
\usepackage{makecell}%
\usepackage{ulem}
\usepackage{graphicx,color,dcolumn,booktabs,bm}
\usepackage{epsfig,dsfont,amssymb,amsmath,amsfonts,amsbsy,mathrsfs}
\usepackage{ulem}

\graphicspath{{Figures/}} %

\usepackage{hyperref}
\hypersetup{colorlinks,citecolor=blue,anchorcolor=red,menucolor=red, linkcolor=red,filecolor=red,runcolor=red,urlcolor=blue,frenchlinks=red}

\makeatletter
\@addtoreset{equation}{section}
\makeatother

\allowdisplaybreaks

\begin{document}

\title{Correlation of the hidden-charm molecular tetraquarks and the charmoniumlike structures existing in the $B\to XYZ+K$}
\author{Fu-Lai Wang$^{1,2}$}
\email{wangfl2016@lzu.edu.cn}
\author{Xin-Dian Yang$^{1,2}$}
\email{yangxd20@lzu.edu.cn}
\author{Rui Chen$^{4,5}$}
\email{chen$_$rui@pku.edu.cn}
\author{Xiang Liu$^{1,2,3}$\footnote{Corresponding author}}
\email{xiangliu@lzu.edu.cn}
\affiliation{$^1$School of Physical Science and Technology, Lanzhou University, Lanzhou 730000, China\\
$^2$Research Center for Hadron and CSR Physics, Lanzhou University and Institute of Modern Physics of CAS, Lanzhou 730000, China\\
$^3$Lanzhou Center for Theoretical Physics, Key Laboratory of Theoretical Physics of Gansu Province, and Frontiers Science Center for Rare Isotopes, Lanzhou University, Lanzhou 730000, China\\
$^4$Key Laboratory of Low-Dimensional Quantum Structures and Quantum Control of Ministry of Education, Department of Physics and Synergetic Innovation Center for Quantum Effects and Applications, Hunan Normal University, Changsha 410081, China\\
$^5$Center of High Energy Physics, Peking University, Beijing 100871, China}

\begin{abstract}
The molecular assignments to the three $P_c$ states and the similar production mechanism between the $\Lambda_b\to P_c+K$ and $B\to XYZ+K$ convince us the $B$ decaying to a charmonium state plus light mesons could be the appropriate production process to search for the charmoniumlike molecular tetraquarks. In this work, we systematically study the interactions between a charmed (charmed-strange) meson and an anti-charmed (anti-charm-strange) meson, which include the $D^{(*)}\bar{D}^{(*)}$, $\bar{D}^{(*)}\bar{D}_1$, $D^{(*)}\bar{D}_2^*$, $D_s^{(*)}\bar{D}_s^{(*)}$, ${D}_s^{(*)}\bar{D}_{s0}^*$, $D_s^{(*)}\bar{D}_{s1}^{\prime}$, ${D}_s^{(*)}\bar{D}_{s1}$, $D_s^{(*)}\bar{D}_{s2}^*$ systems. After adopting the one-boson-exchange effective potentials, our numerical results indicate that, on one hand, there can exist a serial of isoscalar charmoniumlike $\mathcal{D}\bar{\mathcal{D}}$ and $\mathcal{D}_s\bar{\mathcal{D}}_s$ molecular states, on the other hand, we can fully exclude the charged charmoniumlike states as the isovector charmoniumlike molecules. Meanwhile, we discuss the two-body hidden-charm decay channels for the obtained $\mathcal{D}\bar{\mathcal{D}}$ and $\mathcal{D}_s\bar{\mathcal{D}}_s$ molecules, especially the $D^{*}\bar{D}^{*}$ molecular tetraquarks. By analyzing the experimental data collected from the $B\to XYZ+K$ and the mass spectrum and two-body hidden-charm decay channels for the obtained $\mathcal{D}\bar{\mathcal{D}}$ and $\mathcal{D}_s\bar{\mathcal{D}}_s$ molecules, we find several possible hints of the existence of the charmoniumlike molecular tetraquarks, i.e., a peculiar characteristic mass spectrum of the isoscalar $D^*\bar{D}^*$ molecular systems can be applied to identify the charmoniumlike molecule. We look forward to the future experiments like the LHCb, Belle II, and BESIII Collaborations can test our results with more precise experimental data.
\end{abstract}

\maketitle

\section{Introduction}\label{sec1}

In 2015, the LHCb Collaboration analyzed the $\Lambda_b\to J/\psi p K$ decay and reported two $P_c$ structures ($P_c(4380)$ and $P_c(4450)$) existing in the $J/\psi p$ invariant mass spectrum \cite{Aaij:2015tga}. After four years, the LHCb revised $\Lambda_b\to J/\psi p K$ process with higher precision data and found that former $P_c(4450)$ contains two substructures $P_c(4440)$ and $P_c(4457)$ \cite{Aaij:2019vzc}. Besides, a new enhancement structure $P_c(4312)$ was announced \cite{Aaij:2019vzc}. In fact, this updated result of $P_c$ states provides a direct evidence to support the existence of the hidden-charm molecular pentaquark states \cite{Li:2014gra, Wu:2010jy, Karliner:2015ina, Wang:2011rga, Yang:2011wz, Wu:2012md, Chen:2015loa}.
\begin{figure}[!htbp]
\centering
\begin{tabular}{cc}
\includegraphics[scale=0.50]{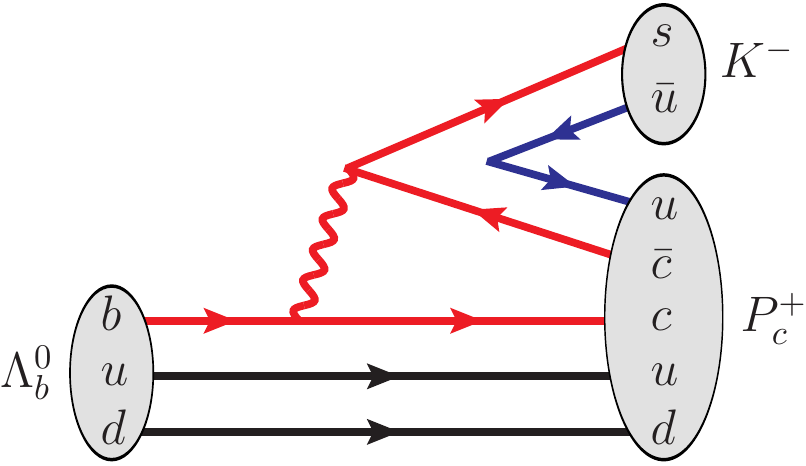}&\includegraphics[scale=0.50]{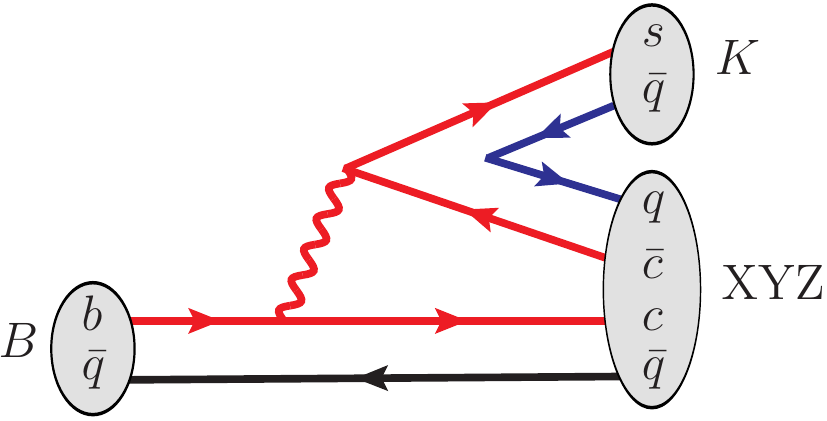}
\end{tabular}
\caption{The production mechanisms of the $P_c$ states from the $\Lambda_b$ baryon decays and the $XYZ$ states from the $B$ meson decays.}
\label{Productionmechanisms}
\end{figure}

In Fig. \ref{Productionmechanisms}, we present the quark level description of the $\Lambda_b\to P_c K$, which is a typical hadronic weak decay. If replacing $ud$ quarks of the $\Lambda_b$ by an antiquark $\bar{q}$, we may get the $B\to XYZ+K$ process, where $XYZ$ denote the charmoniumlike structures. Obviously, due to the similar production mechanism between the $\Lambda_b\to P_c K$ and $B\to XYZ+K$ (see Fig. \ref{Productionmechanisms}), we naturally conjecture that the $B\to XYZ+K$ should be the ideal processes to produce the hidden-charm molecular tetraquark states, especially with establishing the hidden-charm molecular tetraquark states. {As is well known, the $X(3872)$ is a typical example of hidden-charm molecular tetraquark state, which was found in the $B^{\pm}\to J/\psi\pi^+\pi^- K^{\pm}$ process by the Belle collaboration in 2003 \cite{Choi:2003ue}. The $X(3872)$ was suggested to be   an isoscalar $D\bar{D}^*$ molecular state with $J^{PC}=1^{++}$ \cite{Wong:2003xk,Swanson:2003tb,Suzuki:2005ha,Liu:2008fh,Thomas:2008ja,Liu:2008tn,Lee:2009hy}.}

In fact, the road of identifying the hidden-charm molecular tetraquark states is getting confusing \cite{Chen:2016qju,Liu:2013waa,Hosaka:2016pey,Liu:2019zoy,Brambilla:2019esw,Olsen:2017bmm,Guo:2017jvc}. In 2004, the Belle Collaboration reported the observation of the charmoniumlike state $Y(3940)$ in the $J/\psi\omega$ invariant mass spectrum of the $B\to J/\psi \omega K$ \cite{Abe:2004zs}. In 2009, the CDF Collaboration observed another charmoniumlike state $Y(4140)$ in the $J/\psi\phi$ invariant mass spectrum of the $B\to J/\psi \phi K$ \cite{Aaltonen:2009tz}. In Ref. \cite{Liu:2009ei}, Liu and Zhu noticed the similarity between the $Y(3940)$ and $Y(4140)$, and proposed $S-$wave $D^*\bar{D}^*$ and $D_s^*\bar{D}_s^*$ molecular states assignment to the $Y(3940)$ and $Y(4140)$, respectively, where the $J^{PC}$ quantum numbers for the $Y(3940)$ and $Y(4140)$ should be either $0^{++}$ or $2^{++}$ due to a simple selection rules from the parity and angular momentum conservation \cite{Liu:2009ei}. In the
subsequent experiments, the LHCb analyzed the $B\to J/\psi \phi K$ process, where the $Y(4140)$ was confirmed and other three new structures $X(4274)$, $X(4500)$ and $X(4700)$ were discovered in the $J/\psi\phi$ invariant mass spectrum \cite{Aaij:2016iza}. However, the LHCb announced that the preferred $J^{PC}$ quantum number of the $Y(4140)$ is $1^{++}$ \cite{Aaij:2016nsc}, which obviously cannot support the molecular assignment to the $Y(4140)$ suggested in Ref. \cite{Liu:2009ei}. If checking the experimental and theoretical research status of the $Z(4430)$, which was reported by the Belle in the $B\to \psi(3686)\pi K$ \cite{Choi:2007wga}, the same situation again happens. Since the $Z(4430)$ is near the threshold of $D^*\bar{D}_1^{(\prime)}$ channel, the authors of Refs. \cite{Liu:2007bf,Liu:2008xz} suggested that the $Z(4430)$ can be the good candidate of the $S-$wave $D_1\bar{D}^*$ molecular state\footnote{To be convenient, we use the shorthand notation $\mathcal{A}\overline{\mathcal{B}}$ to denote the $\mathcal{A}\overline{\mathcal{B}}+c.c.$ system in the following parts, where the notations $\mathcal{A}$ and $\mathcal{B}$ represent two different charmed (charmed-strange) mesons, respectively.} with $J^{P}=0^-,1^-,2^-$ by the one-boson-exchange (OBE) model calculation. However, the Belle \cite{Chilikin:2013tch} and LHCb \cite{Aaij:2014jqa} reanalyzed the $B\to \psi(3686)\pi K$ and indicated that the $Z(4430)$ has quantum number $J^P=1^+$, which is contradict with the $S-$wave $D_1\bar{D}^*$ molecular state assignment \cite{Liu:2007bf,Liu:2008xz}.

Although there were a dozen of charmoniumlike structures reported by experiments in the $B$ meson decays \cite{Brambilla:2019esw}, unfortunately we have not definitely identify one hidden-charm molecular tetraquark state until now \cite{Chen:2016qju,Liu:2013waa,Hosaka:2016pey,Liu:2019zoy,Brambilla:2019esw,Olsen:2017bmm,Guo:2017jvc}. It is big challenge to assign the molecular state explanations to the charmoniumlike structures from the $B\to XYZ+K$ processes.

If we still believe the similar production mechanism between the $\Lambda_b\to P_c K$ and $B\to XYZ+K$ (see Fig. \ref{Productionmechanisms}), we have reason to believe the existence of the hidden-charm molecular tetraquark state in the $B\to XYZ+K$ processes. Facing this situation mentioned above, we should try to find possible solutions existing in the reported experimental data of the $B\to XYZ+ K$. In fact, the lesson of observation of the $P_c$ states in 2015 \cite{Aaij:2015tga} and 2019 \cite{Aaij:2019vzc} may inspire us. In 2015, the LHCb measured the $J^P$ quantum numbers of the $P_c(4380)$ and $P_c(4450)$, gave that their preferred $J^P$ quantum numbers are of opposite parity \cite{Aaij:2015tga}, which is a challenge to the hadronic molecular assignment \cite{Chen:2015loa}. This situation was dramatically changed with further LHCb experiment in 2019, where the $P_c(4450)$ is composed of two substructures $P_c(4440)$ and $P_c(4457)$ \cite{Aaij:2019vzc}, which means that the measurement of spin-parity quantum number of the observed $P_c(4450)$ can be ignored \cite{Aaij:2015tga}. To some extent, this fact reflects the importance of higher precision to the study of hadron spectroscopy.

With the running of the Belle II \cite{Kou:2018nap} and the accumulation of Run II and Run III data at the LHCb \cite{Bediaga:2018lhg}, obviously investigation of the charmoniumlike $XYZ$ states must enter a new era. Thus, we should systematically reexamine the correlation of the hidden-charm molecular states and the charmoniumlike structures existing in the $B\to XYZ+K$ processes \cite{Chen:2016qju,Liu:2013waa,Hosaka:2016pey,Liu:2019zoy,Brambilla:2019esw,Olsen:2017bmm,Guo:2017jvc}, where the constraint from $J^{PC}$ quantum numbers and so-called resonance parameters of depicting these observed charmoniumlike $XYZ$ structures should be more cautious in the interpretation of these resonances as the molecular states.

Along this line, we systematically restudy the $S-$wave interactions between a charmed (charmed-strange) meson and an anticharmed (anticharmed-strange) meson in the framework of the one-boson-exchange (OBE) model in this work \cite{Chen:2016qju,Liu:2019zoy}. In concrete calculations, both the $S-$$D$ wave mixing effect and the coupled channel effect are taken into account. Additionally, we discuss the two-body hidden-charm decay channels for the obtained $\mathcal{D}\bar{\mathcal{D}}$ and $\mathcal{D}_s\bar{\mathcal{D}}_s$ molecules, especially the $D^{*}\bar{D}^{*}$ molecular tetraquarks. Based on above discussion, we find a series of possible hints of the hidden-charm molecular tetraquarks existing in the released experimental data of the $B$ meson decays, which will be a main task of the present work.

The remainder of this paper is organized as follows. In Sec. \ref{sec2}, a comparison of the relevant experimental data and the corresponding thresholds will be given. In Sec. \ref{sec3}, the mass spectrum and the two-body hidden-charm decay channels of these discussed hidden-charm molecular tetraquark systems will be given, and a series of possible hints of the hidden-charm molecular tetraquarks existing in the reported experimental data of the $B$ meson decays will be presented. Finally, a brief summary will be given in Sec. \ref{sec4}.

\section{A comparison of the experimental data and the corresponding thresholds}\label{sec2}

In this work, the main task is to find possible hints of the hidden-charm molecular tetraquarks existing in the reported experimental data of the $B\to XYZ+ K$ \cite{Brambilla:2019esw}, and the hidden-charm molecular tetraquarks are composed of a charmed (charmed-strange) meson and an anticharmed (anticharmed-strange) meson. Thus, we need to make comparison of the involved experimental data and the corresponding thresholds of charmed meson pairs or charmed-strange meson pairs.
\renewcommand\tabcolsep{0.09cm}
\renewcommand{\arraystretch}{1.50}
\begin{table}[!htpb]
\centering
\caption{The thresholds of charmed (charmed-strange) meson pairs (in unit of MeV).}\label{massthresholds}
\begin{tabular}{ccccccc}\toprule[1pt]\toprule[1pt]
\multicolumn{7}{c}{Without hidden-strange quantum number}\\ \hline
$DD$&$DD^*$&$D^*D^*$&$DD_0^*$&$DD_1$&$DD_1^{\prime}$&$DD_2^*$\\
3734.48&3875.80&4017.12&4191.74&4289.24&4294.24&4330.29\\
$D^*D_0^*$&$D^*D_1$&$D^*D_1^{\prime}$&$D^*D_2^*$&$D_0^*D_0^*$&$D_0^*D_1$&$D_0^*D_1^{\prime}$\\
4333.06&4430.56&4435.56&4471.61&4649.00&4746.50&4751.50\\
$D_0^*D_2^*$&$D_1D_1$&$D_1^{\prime}D_1$&$D_1^{\prime}D_1^{\prime}$&$D_1D_2^*$&$D_1^{\prime}D_2^*$&$D_2^*D_2^*$\\
4787.55&4844.00&4849.00&4854.00&4885.05&4890.05&4926.10\\\midrule[1.0pt]
\multicolumn{7}{c}{With hidden-strange quantum number}\\ \hline
$D_sD_s$&$D_sD_s^*$&$D_s^*D_s^*$&$D_sD_{s0}^*$&$D_sD_{s1}^{\prime}$&$D_s^*D_{s0}^*$&$D_sD_{s1}$\\
3936.68&4080.54&4224.40&4286.14&4427.84&4430&4503.45\\
$D_sD_{s2}^*$&$D_s^*D_{s1}^{\prime}$&$D_{s0}^*D_{s0}^*$&$D_s^*D_{s1}$&$D_s^*D_{s2}^*$&$D_{s0}^*D_{s1}^{\prime}$&$D_{s0}^*D_{s1}$\\
4537.44&4571.70&4635.60&4647.31&4681.30&4777.30&4852.91\\
$D_{s0}^*D_{s2}^*$&$D_{s1}^{\prime}D_{s1}^{\prime}$&$D_{s1}^{\prime}D_{s1}$&$D_{s1}D_{s1}$&$D_{s1}^{\prime}D_{s2}^*$&$D_{s1}D_{s2}^*$
&$D_{s2}^*D_{s2}^*$\\
4886.90&4919.00&4994.61&5028.60&5070.22&5104.20&5138.20\\
\bottomrule[1pt]\bottomrule[1pt]
\end{tabular}
\end{table}

Usually, the $S$-wave and $P$-wave charmed mesons can be grouped into three doublets $H =[D(0^-),D^{\ast}(1^-)]$, $S =[D_0^{\ast}(0^+),D^{\prime}_1(1^+)]$, and $T =[D_1(1^+),D^{\ast}_2(2^+)]$ according to the heavy quark spin symmetry \cite{Wise:1992hn}. Similarly, there also exist three doublets for the $S-$wave and $P$-wave charmed-strange mesons, i.e., $H =[D_s,D_s^{\ast}]$, $S =[D_{s0}^{\ast},D_{s1}^{\prime}]$, and $T =[D_{s1},D_{s2}^{\ast}]$ \cite{Wise:1992hn}. In Table \ref{massthresholds}, we list these thresholds of charmed meson pairs or charmed-strange meson pairs \cite{Zyla:2020zbs}, which are distributed over a very wide energy range $3.7\sim 5.2$ GeV. However, if only considering the kinetics of the $B\to XYZ+ K$ decays, the maximum mass of the involved $XYZ$ structures should be $M(XYZ)_{max}=M(B)-M(K)=4783~{\rm MeV}$, which makes us select these thresholds with mass lower than 4783 MeV when making further analysis. Additionally, the charmed mesons in $S-$doublet have broad widths around several hundred MeV \cite{Zyla:2020zbs}, which may be a obstacle for the formation of the hadronic molecular states \cite{Hanhart:2010wh,Filin:2010se}. Different from the charmed mesons in $S-$doublet, these charmed mesons in the $H$-doublet and $T$-doublet have narrow widths \cite{Zyla:2020zbs}, which can be regarded as the suitable components to form the hadronic molecular states \cite{Liu:2008xz,Hanhart:2010wh,Filin:2010se}. Just considering the above fact, we consider the cases of $H\bar{H}$ and $H\bar{T}$ for charmed meson pairs and the cases of $H\bar{H}$, $H\bar{S}$, and $H\bar{T}$ for charmed-strange meson pairs when performing a comparison of these thresholds with the released experimental data.

At present, the experimental information of the $B\to XYZ+ K$ are very abundant \cite{Chen:2016qju,Liu:2013waa,Hosaka:2016pey,Liu:2019zoy,Brambilla:2019esw,Olsen:2017bmm,Guo:2017jvc}. Here, the $XYZ$ data are from the $J/\psi\pi\pi$ invariant mass spectrum of the $B\to J/\psi\pi\pi K$ \cite{Choi:2003ue}, the $J/\psi\omega$ invariant mass spectrum of the $B\to J/\psi\omega K$ \cite{Abe:2004zs,Aubert:2007vj,delAmoSanchez:2010jr}, the $J/\psi\eta$ invariant mass spectrum of the $B\to J/\psi\eta K$ \cite{Aubert:2004fc}, the $J/\psi \phi$ invariant mass spectrum of  the $B\to J/\psi \phi K$ \cite{Aaltonen:2009tz,Chatrchyan:2013dma,Aaij:2016iza,Aaij:2012pz,Abazov:2013xda,Lees:2014lra,Aaltonen:2011at,Aaij:2016nsc,Aaij:2021ivw}, the $\eta_c \pi$ invariant mass spectrum of the $B\to \eta_c\pi K$ \cite{Aaij:2018bla}, the $J/\psi \pi$ invariant mass spectrum of the $B\to J/\psi\pi K$ \cite{Chilikin:2014bkk,Abazov:2018cyu,Aaij:2019ipm}, the $\psi(3686)\pi$ invariant mass spectrum of the $B\to \psi(3686)\pi K$ \cite{Choi:2007wga,Mizuk:2009da,Chilikin:2013tch,Aaij:2014jqa}, the $\chi_{c1}\pi$ invariant mass spectrum of the $B\to \chi_{c1}\pi K$ \cite{Mizuk:2008me,Lees:2011ik,Bhardwaj:2015rju,Bhardwaj:2019spn}, and the $\chi_{c2}\pi$ invariant mass spectrum of the $B\to \chi_{c2}\pi K$ \cite{Bhardwaj:2015rju}. For clearly discussing the present issue, we categorized these $XYZ$ data into three groups:
\begin{itemize}
\item Isoscalar $XYZ$ data without hidden-strange quantum number are involved in the $B\to J/\psi\omega K$ and $B\to J/\psi\eta K$;
\item Isoscalar $XYZ$ data with hidden-strange quantum number are relevant to the $B\to J/\psi\phi K$;
\item Isovector $XYZ$ data without hidden-strange quantum number have relation to five decay processes, i.e., the $B\to \eta_c\pi K$, $B\to J/\psi\pi K$, $B\to \psi(3686)\pi K$, $B\to \chi_{c1}\pi K$, and $B\to \chi_{c2}\pi K$.
\end{itemize}

In the following, we adopt this line to make comparison of the released experimental data and the corresponding thresholds.

\subsection{Isoscalar $XYZ$ data without hidden-strange quantum number}\label{subsec21}

\begin{figure}[!htbp]
\centering
\includegraphics[width=8.4cm]{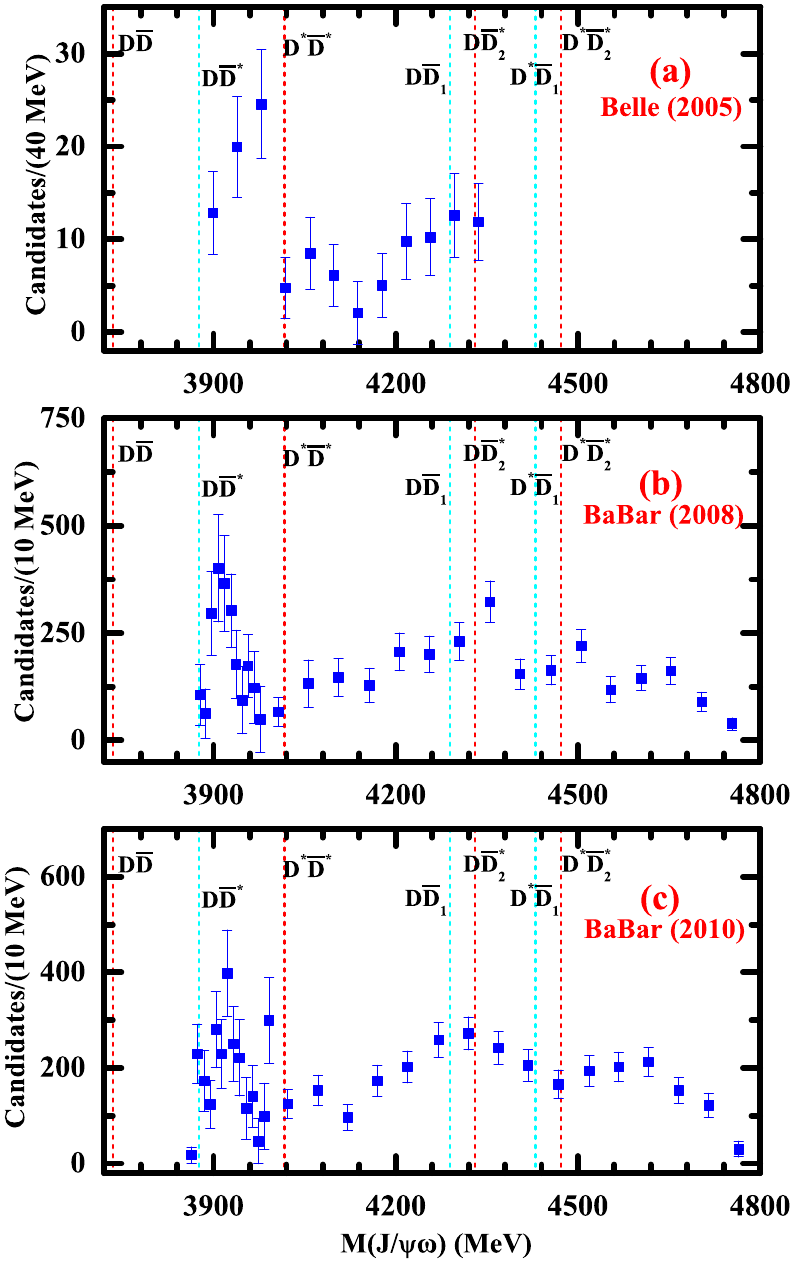}\\
\includegraphics[width=7.7cm]{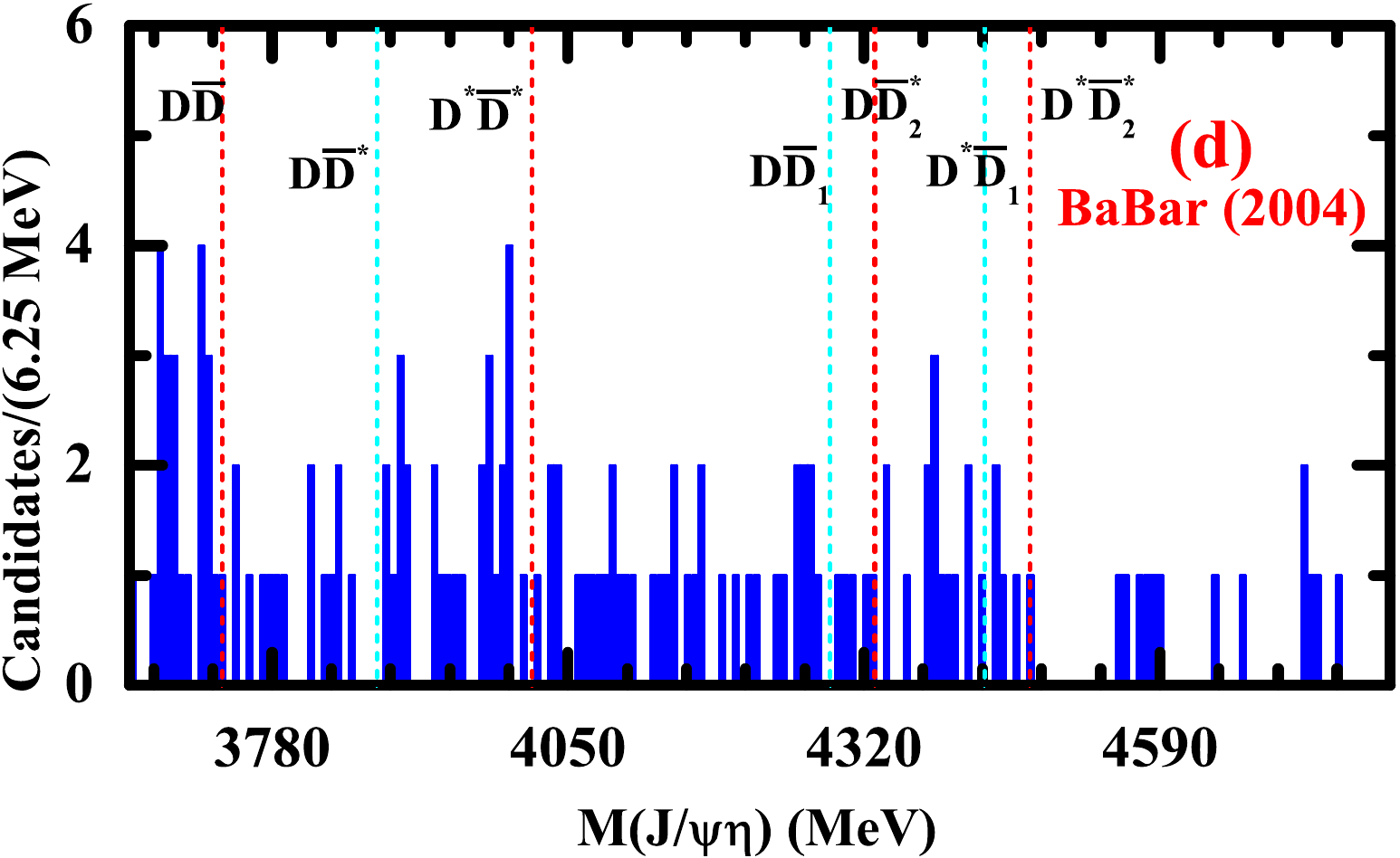}\\
\caption{(color online) The $J/\psi \omega$ and $J/\psi\eta$ invariant mass distributions in the $B \to  J/\psi \omega K$ and $B \to  J/\psi \eta K$, respectively, and the comparison with the thresholds of charmed meson pairs. Here, the experimental data of the $J/\psi \omega$ invariant mass spectrum are taken from the Belle \cite{Abe:2004zs}, BaBar \cite{Aubert:2007vj}, and BABAR \cite{delAmoSanchez:2010jr}, which correspond to diagrams (a)-(c), and the experimental data of the $J/\psi \eta$ invariant mass spectrum is from the BABAR measurement \cite{Aubert:2004fc} (see diagram (d)).}
\label{Jpsiomegaeta}
\end{figure}

In 2005, the Belle Collaboration analyzed the $J/\psi\omega$ invariant mass spectrum of the $B\to J/\psi \omega K$ decay, where the charmoniumlike structure $Y(3940)$ was reported, which can be depicted by resonance parameters $M = (3943 \pm 11 \pm 13)~{\rm MeV}$ and $\Gamma = (87 \pm 22 \pm 26)~{\rm MeV}$ \cite{Abe:2004zs}.  In 2008, the BABAR Collaboration confirmed this observation in the same process with a lower mass, named as the $Y(3915)$, and its mass and width were measured to be $(M, \Gamma) = (3914.6^{+3.8}_{-3.4} \pm 2.0~{\rm MeV}, 34^{+12}_{-8} \pm 5~{\rm MeV})$ \cite{Aubert:2007vj}. In the subsequent BABAR experiment in 2010 \cite{delAmoSanchez:2010jr}, the mass and width of the $Y(3915)$ were measured to be $M = (3919.1^{+3.8}_{-3.5} \pm 2.0)~{\rm MeV}$ and $\Gamma = (31^{+10}_{-8} \pm 5)~{\rm MeV}$, respectively.

If checking these data of the $J/\psi\omega$ invariant mass spectrum of the $B\to J/\psi \omega K$ from the Belle and BABAR \cite{Abe:2004zs,Aubert:2007vj,delAmoSanchez:2010jr}, we may find the impact of experimental precision on reflecting the details. In the Belle data \cite{Abe:2004zs}, there are only four experimental points sandwiched by the $D\bar{D}^*$ and $D^*\bar{D}^*$ thresholds. After several years, the number of experimental points in this energy range reach up to 12 and 14, which correspond to the BABAR data measured in 2008 \cite{Aubert:2007vj} and 2010 \cite{delAmoSanchez:2010jr}, respectively. In Fig. \ref{Jpsiomegaeta}, we collect all reported data of the $J/\psi\omega$ invariant mass spectrum from the $B\to J/\psi \omega K$ \cite{Abe:2004zs,Aubert:2007vj,delAmoSanchez:2010jr}. In fact, the BABAR data released in 2008 \cite{Aubert:2007vj} show the possibility of existing two structures around 3.9 GeV below the $D^*\bar{D}^*$ threshold in the $J/\psi\omega$ invariant mass spectrum. Especially, the BABAR measurement in 2010 further enforce this possibility \cite{delAmoSanchez:2010jr}.

If $D^*$ and $\bar D^*$ can be bound together to form the hidden-charm molecular tetraquarks, the $J^{PC}$ quantum numbers of the $S-$wave isoscalar $D^*\bar D^*$ molecular system must be $0^{++}$, $1^{+-}$, and $2^{++}$  \cite{Liu:2009ei}. Under this assumption, the behavior of mass spectrum of the $S-$wave isoscalar $D^*\bar D^*$ molecular states can explain why two substructures around 3.9 GeV exist in the $J/\psi\omega$ invariant mass spectrum of the $B\to J/\psi \omega K$ \cite{Abe:2004zs,Aubert:2007vj,delAmoSanchez:2010jr}. Thus, we strongly encourage our experimental colleagues to focus on the detail of the structures around 3.9 GeV with more precise data. If these substructures can be confirmed in future experiments discussed above, it will provide strong evidence of existing the hidden-charm molecular tetraquark. Later, we will revisit a dynamics study of the $S-$wave isoscalar $D^*\bar D^*$ system, and come back to address this point.

Besides the enhancement structures around 3.9 GeV exist in the $J/\psi\omega$ invariant mass spectrum, we may find a very broad structure around 4.3 GeV exists in the $J/\psi\omega$ invariant mass spectrum of the $B\to J/\psi \omega K$ \cite{Abe:2004zs,Aubert:2007vj,delAmoSanchez:2010jr}, where there exist four thresholds ($D\bar{D}_1$, $D\bar{D}^*_2$, $D^*\bar{D}_1$, and $D^*\bar{D}_2^*$) in this energy range, which inspire our interest in exploring whether the isoscalar $D\bar{D}_1$, $D\bar{D}^*_2$, $D^*\bar{D}_1$, and $D^*\bar{D}_2^*$ molecular tetraquarks exist in nature, which will be one of tasks in this work. Obviously, the details of such broad structure around 4.3 GeV should be given in future experiments with more precise data accumulation.

For the isoscalar $XYZ$ data without hidden-strange quantum number, we should mention the measurement of the $J/\psi\eta$ invariant mass spectrum in the $B \to  J/\psi \eta K$ decay. In 2004, the BABAR released the result of the $J/\psi\eta$ invariant mass spectrum in the $B \to  J/\psi \eta K$ \cite{Aubert:2004fc}. In Fig. \ref{Jpsiomegaeta}, we compare the BABAR data with the several thresholds of charmed meson pairs, and may find the evidence of one structure below $D^*\bar{D}^*$ threshold and possible enhancement structure around 4.3 GeV which overlaps with the $D\bar{D}_1$, $D\bar{D}^*_2$, $D^*\bar{D}_1$, and $D^*\bar{D}_2^*$ thresholds. In fact, this phenomenon again shows studying the isoscalar $D^*\bar D^*$, $D\bar{D}_1$, $D\bar{D}^*_2$, $D^*\bar{D}_1$, and $D^*\bar{D}_2^*$ hadronic molecular states is an urgent research issue, which may provide crucial information to find the evidence of the existence of the hidden-charm molecular tetraquark.

\subsection{Isoscalar $XYZ$ data with hidden-strange quantum number}\label{subsec22}

\begin{figure}[!htbp]
\centering
\includegraphics[width=8.4cm,keepaspectratio]{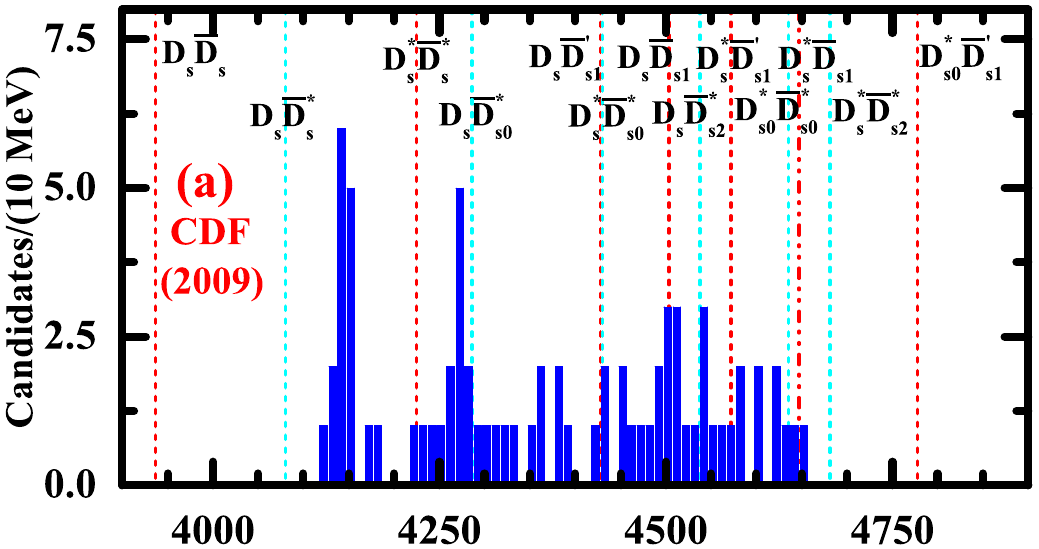}\\
\includegraphics[width=8.4cm,keepaspectratio]{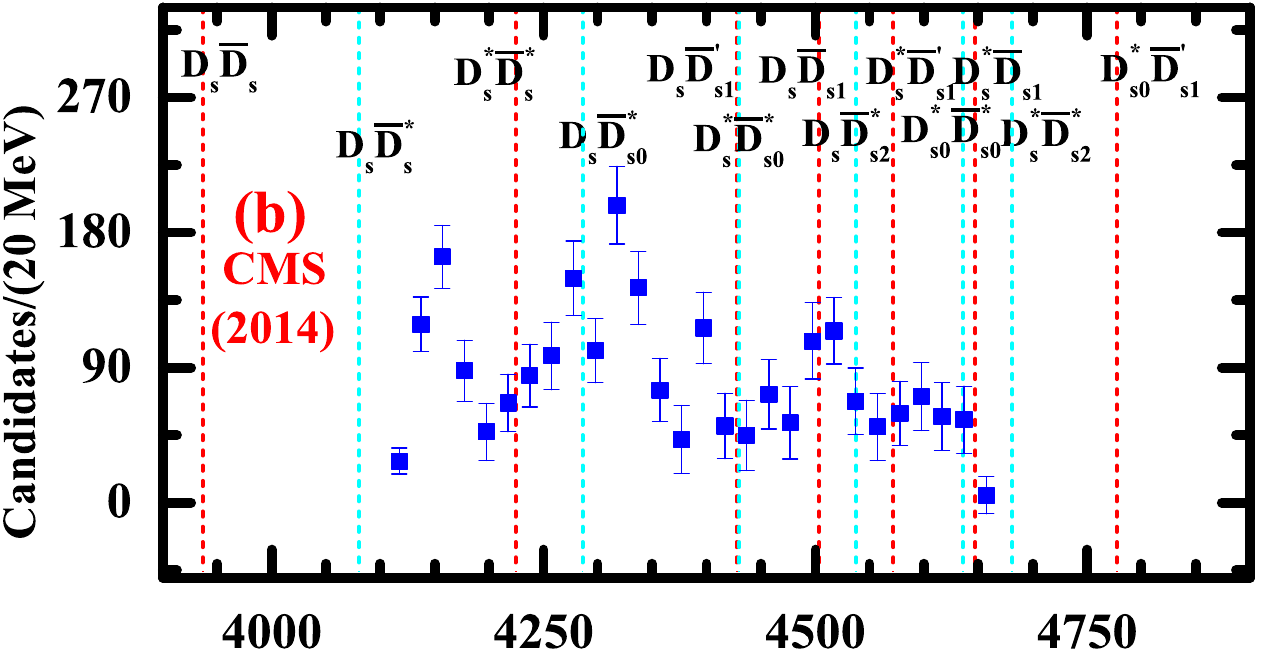}\\
\includegraphics[width=8.3cm,keepaspectratio]{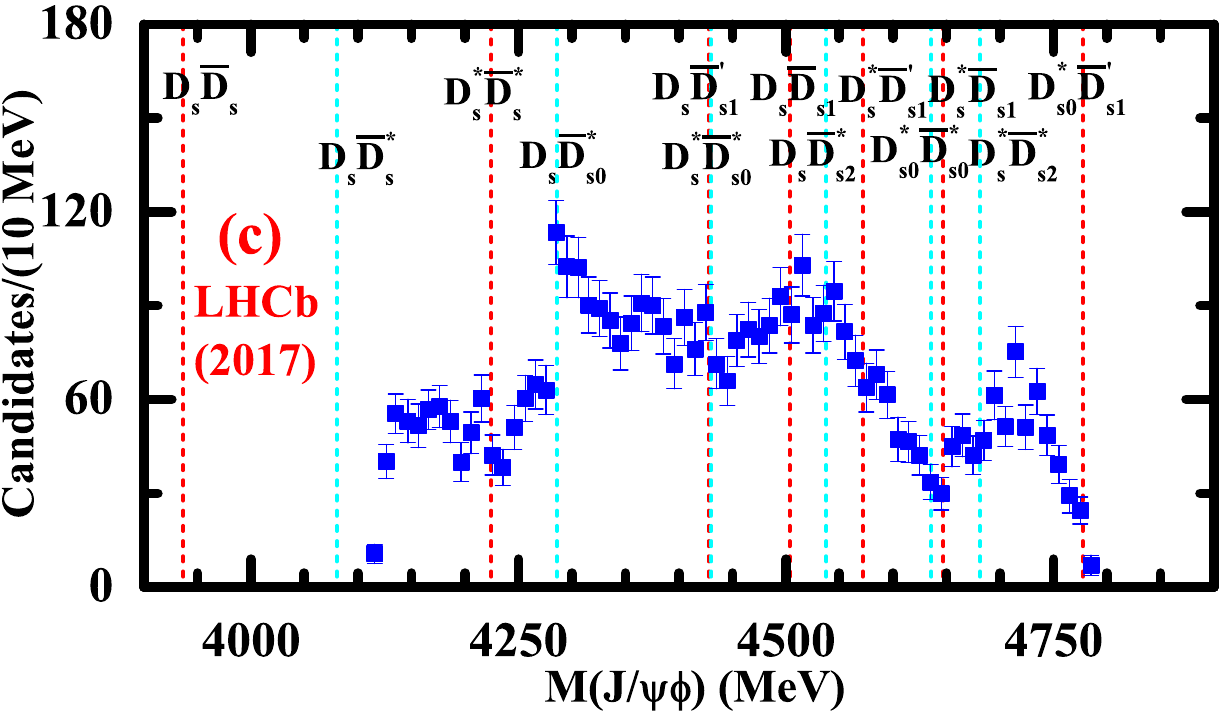}
\caption{(color online) The measured $ J/\psi \phi$ invariant mass distribution of the $B \to  J/\psi \phi K$ and the thresholds of charmed-strange meson pairs. Here, the experimental data are taken from the CDF \cite{Aaltonen:2009tz}, CMS \cite{Chatrchyan:2013dma}, and LHCb \cite{Aaij:2016iza}, which correspond to diagrams (a)-(c).}
\label{Jpsiphi}
\end{figure}

Up to now, the $B \to  J/\psi \phi K$ decay have been analyzed by the different experiment collaborations \cite{Aaltonen:2009tz,Chatrchyan:2013dma,Aaij:2016iza,Aaij:2012pz,Abazov:2013xda,Lees:2014lra,Aaltonen:2011at,Aaij:2016nsc,Aaij:2021ivw}, where we are mainly interested in the experimental data of the $J/\psi \phi$ invariant mass spectrum from the CDF \cite{Aaltonen:2009tz}, CMS \cite{Chatrchyan:2013dma}, and LHCb \cite{Aaij:2016iza} in the following discussion. Since the quark components of the $\phi$ and $J/\psi$ are the $s\bar s$ and $c\bar c$, respectively, we present the thresholds of charmed-strange meson pairs in Fig.~\ref{Jpsiphi} when making comparison with these released experimental data.

First, we should brief introduce the experimental status of the obtained $J/\psi \phi$ invariant mass spectrum in the $B \to  J/\psi \phi K$ decay \cite{Aaltonen:2009tz,Chatrchyan:2013dma,Aaij:2016iza}. As shown in Fig.~\ref{Jpsiphi} (a), the CDF Collaboration reported the charmoniumlike structure $Y(4140)$ in the $J/\psi \phi$ mass spectrum of the $B \to  J/\psi \phi K$ decay in 2009 \cite{Aaltonen:2009tz}. When depicting this enhancement structure, the resonance parameters can be obtained, i.e., the mass and width of the $Y(4140)$ were measured to be $M = (4143.0 \pm 2.9 \pm 1.2)~{\rm MeV}$ and $\Gamma = (11.7 ^{+8.3}_{-5.0} \pm 3.7)~{\rm MeV}$ \cite{Aaltonen:2009tz}, respectively. Besides the $Y(4140)$ structure, the CDF also reported another enhancement structure, named as the $Y(4274)$, in the $J/\psi \phi$ invariant mass spectrum, where its mass and width are $(M, \Gamma) = (4274.4 ^{+8.4}_{-6.7}\pm 1.9~{\rm MeV}, 32.3^{+21.9}_{-15.3} \pm 7.6~{\rm MeV})$ \cite{Aaltonen:2009tz}, respectively. Later, the CMS confirmed these two structures $Y(4140)$ and $Y(4274)$ in the same decay process in 2014, where the resonance parameters of the $Y(4140)$ and $Y(4274)$ were measured to be $(M, \Gamma)_{Y(4140)} = (4148.0 \pm 2.4\pm 6.3~{\rm MeV}, 28 ^{+15}_{-11} \pm 19~{\rm MeV})$ and $(M, \Gamma)_{Y(4274)} = (4313.8 \pm 5.3\pm 7.3~{\rm MeV}, 38 ^{+30}_{-15} \pm 16~{\rm MeV})$ \cite{Chatrchyan:2013dma}, respectively. Surprisingly, the LHCb Collaboration announced four charmoniumlike resonances in the $J/\psi \phi$ mass spectrum of the $B \to  J/\psi \phi K$ decay in 2017 \cite{Aaij:2016iza}. Here, we collect their resonance parameters, i.e., $(M, \Gamma)_{Y(4140)} = (4146.5 \pm 4.5^{+4.6}_{-2.8}~{\rm MeV}, 83 \pm 21 ^{+21}_{-14}~{\rm MeV})$, $(M, \Gamma)_{Y(4274)} = (4273.3 \pm 8.3^{+17.2}_{-3.6}~{\rm MeV}, 56.2 \pm 10.9 ^{+8.4}_{-11.1}~{\rm MeV})$, $(M, \Gamma)_{X(4500)} = (4506 \pm 11^{+12}_{-15}~{\rm MeV}, 92 \pm 21 ^{+21}_{-20}~{\rm MeV})$, and $(M, \Gamma)_{X(4700)} = (4704 \pm 10^{+14}_{-24}~{\rm MeV}, 120 \pm 31 ^{+42}_{-33}~{\rm MeV})$ \cite{Aaij:2016iza}.

If comparing three experimental data listed in Fig. \ref{Jpsiphi}, we find that the precision of the LHCb data \cite{Aaij:2016iza} is higher than the CDF and CMS data \cite{Aaltonen:2009tz,Chatrchyan:2013dma}, where there are 12 experimental points in the energy range sandwiched by the $D_s\bar{D}_s^*$ and $D_s^*\bar{D}_s^*$ thresholds for the LHCb data \cite{Aaij:2016iza}, which reflect some abundant details difference from former CDF and CMS \cite{Aaltonen:2009tz,Chatrchyan:2013dma}. When only adopting a Breit-Wigner formula to describe this resonance structure around 4140 MeV, the width given by the LHCb \cite{Aaij:2016iza} becomes wider than that from the CDF and CMS \cite{Aaltonen:2009tz,Chatrchyan:2013dma}. This phenomenon is puzzling for us since the LHCb released this result \cite{Aaij:2016iza}, especially the recent LHCb result of the $Y(4140)$ \cite{Aaij:2021ivw}.

We notice an interesting fact that this structure around 4140 MeV reported by the CDF is below the $D_s^*\bar{D}_s^*$ threshold \cite{Aaltonen:2009tz}, and Liu and Zhu proposed that the $Y(4140)$ observed by the CDF can be regarded as partner of the $Y(3940)$ due to the similarity between the $Y(3940)$ and $Y(4140)$ in 2009 \cite{Liu:2009ei}. Thus, the $D_s^*\bar{D}_s^*$ hadronic molecular state explanation to the $Y(4140)$ was given \cite{Liu:2009ei}. Along this line,  there may exist $S-$wave $D_s^*\bar{D}_s^*$ molecular states, which is similar to the situation of the $S-$wave isoscalar $D^*\bar{D}^*$ molecular system discussed in Sec. \ref{subsec21}. Facing such abundant details given by the LHCb in 2017 \cite{Aaij:2016iza}, we may conjecture that the $Y(4140)$ structure may be contain at least two substructures, which can be tested by future experiments based on more precise data. If introducing this proposal, the conclusion of the spin-parity quantum number $J^{PC}=1^{++}$ for the $Y(4140)$ given by the LHCb seems unreliable\footnote{For the $S-$wave $D_s^*\bar{D}_s^*$ molecular system, its $J^{PC}$ quantum numbers are either $0^{++}$ or $2^{++}$, which is due to a selection rule for the quantum numbers \cite{Liu:2009ei}. However, the LHCb measurement suggested $J^{PC}=1^{++}$ for the $Y(4140)$ \cite{Aaij:2016iza}, which results in the difficulty to understand the $Y(4140)$ under the hadronic molecular state assignment.} \cite{Aaij:2016iza}. Here, we need to mention that the recent LHCb Collaboration updated the amplitude analysis of the $B^+\to J/\psi\phi K^+$ decay with higher precision data and reported the $X(4140)$ with $J^{PC}=1^{++}$ and $X(4150)$ with $J^{PC}=2^{-+}$ \cite{Aaij:2021ivw}, which imply the very complicated structures around the $D_s^*\bar{D}_s^*$ threshold. In the following, the investigation of the $S-$wave $D_s^*\bar{D}_s^*$ hadronic molecular system will become a main point of this work, which will be further discussed in the next section.

Besides the structure around 4140 MeV existing in the $J/\psi\phi$ invariant mass spectrum, we should focus on the $Y(4274)$ \cite{Aaltonen:2009tz,Chatrchyan:2013dma,Aaij:2016iza}, which is just near the $D_s\bar{D}_{s0}^*$ threshold. Thus, the study of the $D_s\bar{D}_{s0}^*$ molecular system will be paid attention in this work. Although the LHCb reported $X(4500)$ and $X(4700)$ in 2017 \cite{Aaij:2016iza}, we can find more abundant structures above 4250 MeV existing in the $J/\psi\phi$ invariant mass spectrum if carefully checking the LHCb data \cite{Aaij:2016iza}. Indeed, the recent LHCb Collaboration announced the observation of other new resonance structures existing in this energy range with more precise data, i.e., the $X(4630)$ with $J^{PC}=1^{-+}$ and $X(4685)$ with $J^{PC}=1^{++}$ \cite{Aaij:2021ivw}. We also notice that there are abundant thresholds of charmed-strange meson pairs in this interesting energy range. Thus, we will investigate the hidden-charm and hidden-strange molecular tetraquarks involved these thresholds in this work.

\subsection{Isovector $XYZ$ data without hidden-strange quantum number}\label{sec23}

\begin{figure*}[!htbp]
\centering
\begin{tabular}{cl}
\includegraphics[width=8.4cm,keepaspectratio]{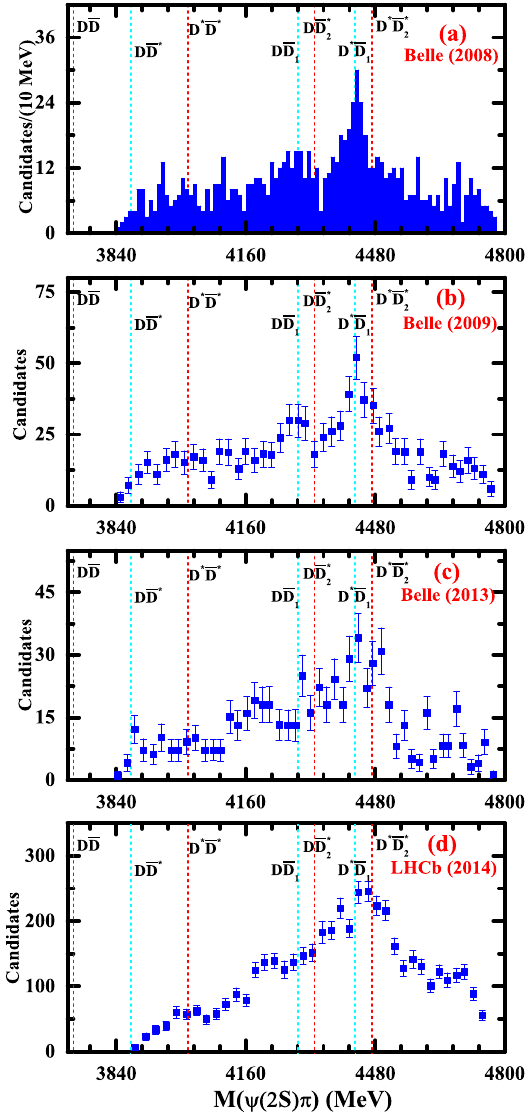}&\includegraphics[width=8.0cm,keepaspectratio]{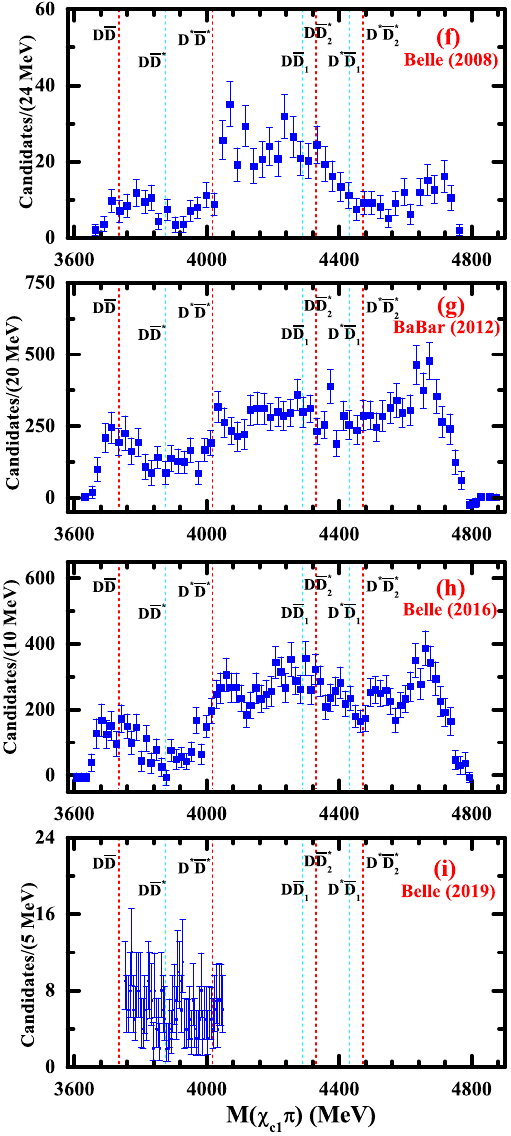}\\
\includegraphics[width=8.0cm,keepaspectratio]{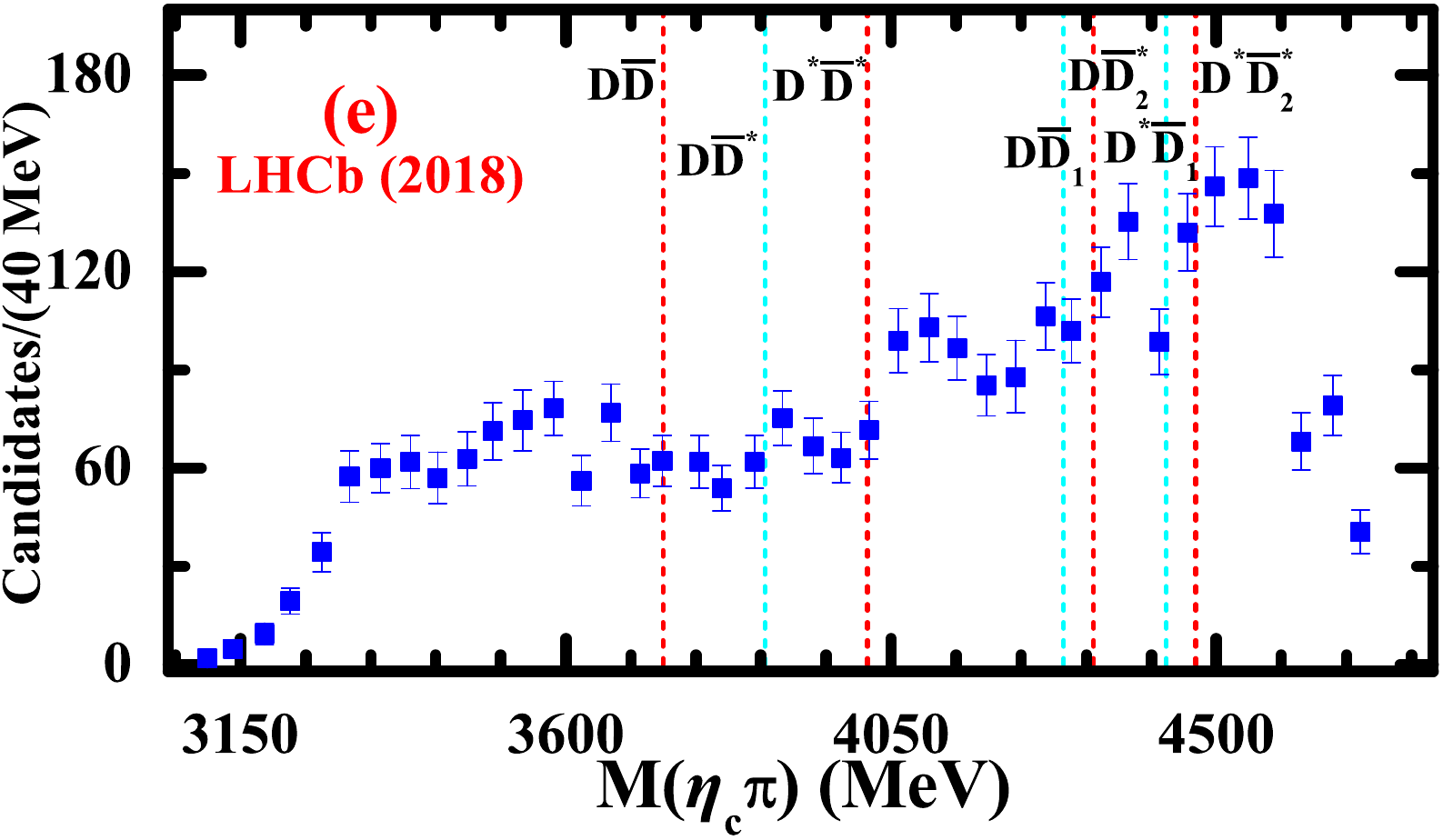}   &\includegraphics[width=8.2cm,keepaspectratio]{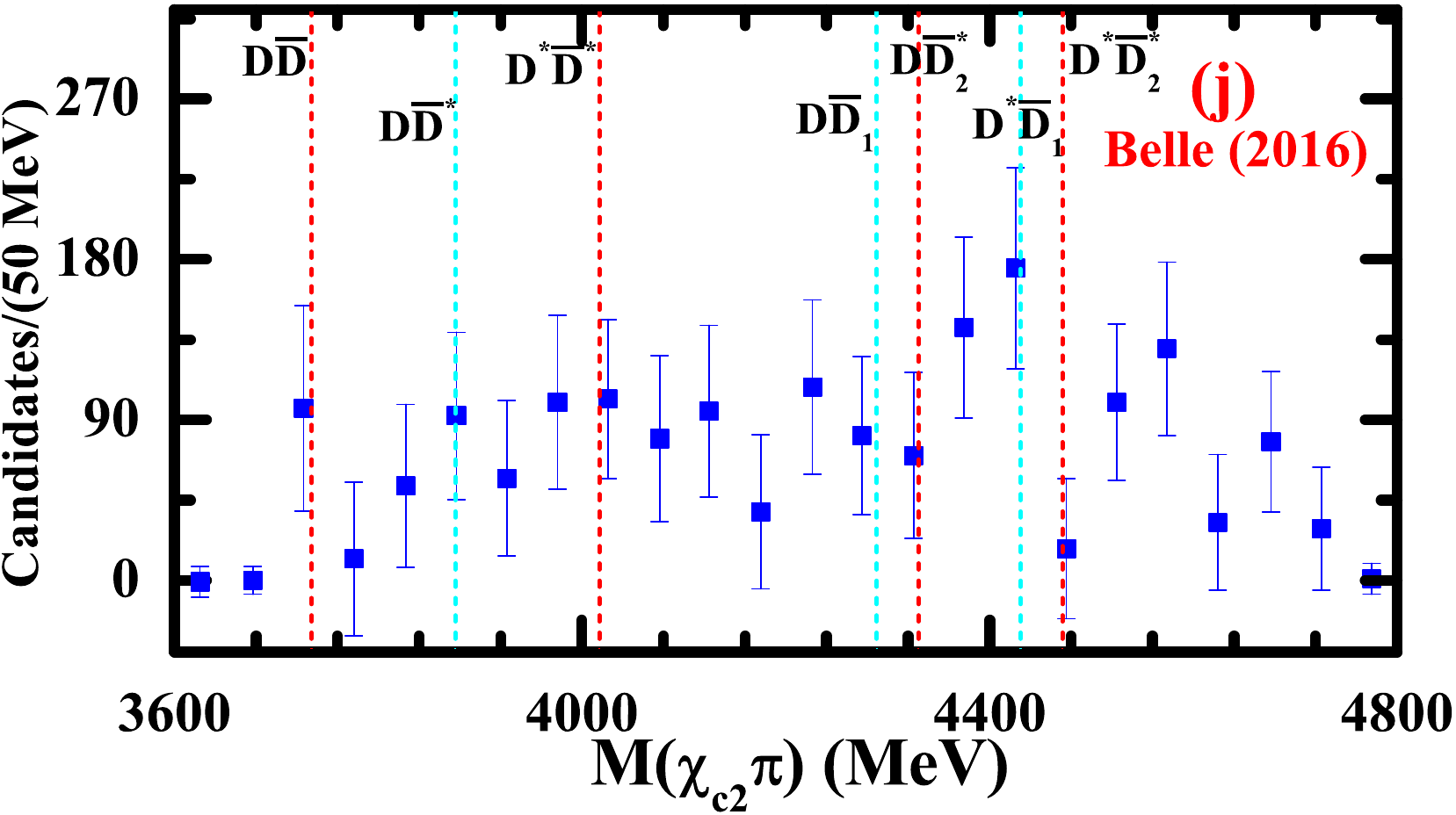}
\end{tabular}
\caption{(color online) The measured $\psi(2S)\pi$, $\eta_c\pi$, $\chi_{c1}\pi$ and $\chi_{c2}\pi$ invariant mass distributions in the $B \to  XYZ+ K$ decays, and the comparison with the thresholds of charmed meson pairs. Here, the experimental data of the $\psi(2S) \pi$ are taken from (a) the Belle \cite{Choi:2007wga}, (b) the Belle \cite{Mizuk:2009da}, (c) the Belle \cite{Chilikin:2013tch}, and (d) the LHCb \cite{Aaij:2014jqa}, while the experimental data of $\eta_c \pi$ is taken from the LHCb \cite{Aaij:2018bla} (see diagram (e)). In addition, the experimental data of $\chi_{c1} \pi$ are taken from (f) the Belle \cite{Mizuk:2008me}, (g) the BABAR \cite{Lees:2011ik}, (h) the Belle \cite{Bhardwaj:2015rju}, and (i) the Belle \cite{Bhardwaj:2019spn}, while the experimental result of the $\chi_{c2} \pi$ invariant mass spectrum is given by the Belle \cite{Bhardwaj:2015rju} (see diagram (j)).}
\label{ccpion}
\end{figure*}

\begin{figure}[!htbp]\centering
\includegraphics[width=7.9cm,keepaspectratio]{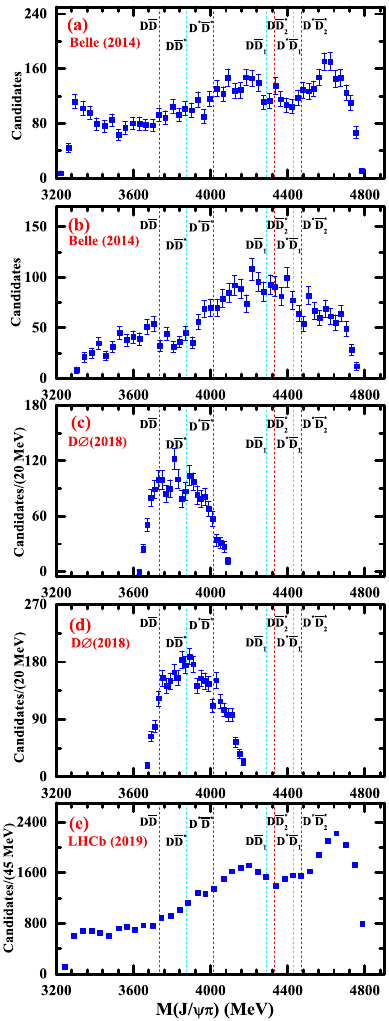}
\caption{(color online) The $J/\psi \pi$ invariant mass distribution in the $B \to  J/\psi \pi K$ and the relevant thresholds of charmed meson pairs. Here, the experimental datas are taken from the Belle \cite{Chilikin:2014bkk} ((a) for $1.20~{\rm GeV}^2 < m^2(K\pi) < 2.05~{\rm GeV}^2$ and (b) for $2.05~{\rm GeV}^2 < m^2(K\pi) < 3.20~{\rm GeV}^2$), the D\O\ \cite{Abazov:2018cyu} ((c) for $4.25~{\rm GeV} < m(J/\psi \pi^+ \pi^-) < 4.30~{\rm GeV}$ and (d) for $4.30~{\rm GeV} < m(J/\psi \pi^+ \pi^-) < 4.40~{\rm GeV}$), and (e) the LHCb \cite{Aaij:2019ipm}, respectively.}
\label{Jpsipion}
\end{figure}

In the past 18 years, the isovector $XYZ$ data from the $B\to XYZ +K$ were accumulated \cite{Chen:2016qju,Liu:2013waa,Hosaka:2016pey,Liu:2019zoy,Brambilla:2019esw,Olsen:2017bmm,Guo:2017jvc}. In Figs.~\ref{ccpion}-\ref{Jpsipion}, we collected all relevant experimental data and make a comparison with several typical thresholds of charmed meson pairs. For different decay processes listed in Figs.~\ref{ccpion}-\ref{Jpsipion}, we briefly review the experimental information:
\begin{enumerate}
\item The $B \to  \psi(2S) \pi K$ decay:
  As a super star among these reported charged charmoniumlike structures, the $Z^+(4430)$ structure was first observed by the Belle Collaboration in the $\psi(2S) \pi$ invariant mass distribution of the $B \to  \psi(2S) \pi K$ decay in 2008 \cite{Choi:2007wga}. Here, its mass and width were measured to be $M = (4433 \pm 4 \pm 2)~{\rm MeV}$ and $\Gamma = (45^{+18}_{-13}{}^{+30}_{-13})~{\rm MeV}$, respectively \cite{Choi:2007wga}. In the following years, the Belle Collaboration continued their studies on the $Z^+(4430)$, and the measured mass of the $Z^+(4430)$ structure are $M=(4443^{+15}_{-12}{}^{+19}_{-13})~{\rm MeV}$ \cite{Mizuk:2009da} and $M=(4485^{+22}_{-22}{}^{+28}_{-11})~{\rm MeV}$ \cite{Chilikin:2013tch}, respectively. After ten years, the LHCb Collaboration confirmed the existence of the $Z^+(4430)$ structure \cite{Aaij:2014jqa}, and the resonance parameters are $M= (4475\pm 7^{+15}_{-25})~{\rm MeV}$ and $\Gamma= (172\pm 13^{+37}_{-34})~{\rm MeV}$, respectively. But, the width from the LHCb \cite{Aaij:2014jqa} is more broad than that from the Belle \cite{Choi:2007wga,Mizuk:2009da,Chilikin:2013tch}. Additionally, the LHCb found a new structure $Z^+(4240)$ in the $\psi(2S) \pi$ invariant mass distribution of the $B \to  \psi(2S) \pi K$, which can be depicted by resonance parameters $M =(4239 \pm 18 ^{+45}_{-10})~{\rm MeV}$ and $\Gamma =(220\pm 47^{+108}_{-74})~{\rm MeV}$, respectively \cite{Aaij:2014jqa}. Comparing with the Belle data \cite{Choi:2007wga,Mizuk:2009da,Chilikin:2013tch}, the LHCb data have smaller error bars \cite{Aaij:2014jqa}. Since the $Z^+(4430)$ structure is near several thresholds of charmed meson pair, there were extensive discussion of the hidden-charm molecular tetraquark explanation to the $Z^+(4430)$ \cite{Liu:2007bf,Liu:2008xz,Close:2009ag}. In this work, we will still dedicate this topic, which is the study of the isovector hidden-charm molecular tetraquark involved in this energy range.

\item The $B \to  \eta_c \pi K$ decay:
  In 2018, the LHCb found the evidence of a broad charmonium-like
   structure in the $\eta_c \pi$ invariant mass spectrum of the $B \to  \eta_c \pi K$ decay, which was named as the $X(4100)$ with the mass $M = (4096\pm 20^{+18}_{-22})~{\rm MeV}$ and the width $\Gamma = (152 \pm 58^{+60}_{-35})~{\rm MeV}$  \cite{Aaij:2018bla}. In fact, there exists event accumulation around 4.5 GeV, where several thresholds of charmed meson pairs can be found (see Fig. \ref{ccpion} (e)).

\item The $B \to  \chi_{c1} \pi K$ decay:
  For the $\chi_{c1} \pi$ invariant mass distribution in the $B \to  \chi_{c1} \pi K$ decay, two charged charmoniumlike structures $Z^+(4051)$ and $Z^+(4248)$ were announced by the Belle Collaboration in the exclusive $B \to  \chi_{c1} \pi K$ decay \cite{Mizuk:2008me}, where their resonance parameters were measured to be $(M, \Gamma)_{Z^+(4051)} = (4051\pm 14^{+20}_{-41}~{\rm MeV}, 82^{+21}_{-17}{}^{+47}_{-22}~{\rm MeV})$ and $(M, \Gamma)_{Z^+(4248)} = (4248^{+44}_{-29}{}^{+180}_{-35}~{\rm MeV}, 177^{+54}_{-39}{}^{+316}_{-61}~{\rm MeV})$ \cite{Mizuk:2008me}, respectively. However, these two charged charmoniumlike structures were not seen in the following BABAR experiment \cite{Lees:2011ik}. In 2016, the Belle Collaboration carried out a new measurement of the $B \to  \chi_{c1} \pi K$, where the distribution of the $\chi_{c1} \pi$ invariant mass spectrum was given \cite{Bhardwaj:2015rju}. In fact, the line shape of the $\chi_{c1} \pi$ invariant mass spectrum is complicated. Additionally, for searching for the $X(3872)$ and $X(3915)$ decaying to the final state $\chi_{c1} \pi$ in the $B$ meson decay, the Belle measured the $\chi_{c1}\pi$ invariant mass spectrum from the $B$ meson decay, by which they did not find the significant signals of the $X(3872)$ and $X(3915)$ \cite{Bhardwaj:2019spn}.

\item The $B \to  \chi_{c2} \pi K$ decay:
 In 2016, the Belle Collaboration provided the measured $\chi_{c2}\pi$ invariant mass spectrum with low precision \cite{Bhardwaj:2015rju}. Here, the comparison of data and thresholds is shown.

\item The $B \to  J/\psi \pi K$ decay:
  As shown in Fig.~\ref{Jpsipion} (a)-(b), the Belle\footnote{Besides announcing the $Z_c(4200)^+$ structure, the Belle also gave the evidence for the $Z(4430)^+$ structure \cite{Chilikin:2014bkk}.} Collaboration presented the results of an amplitude analysis of the $B \to  J/\psi \pi K$ decay in 2014, and reported a very broad charged charmoniumlike structure $Z_c(4200)^+$, where its mass and width were measured to be $(M, \Gamma) = (4196 ^{+31}_{-29}{}^{+17}_{-13}~{\rm MeV}, 370 ^{+70}_{-70} {}^{+70}_{-132}~{\rm MeV})$ \cite{Chilikin:2014bkk}, respectively. We may find such broad structure overlaps with many thresholds (see Fig.~\ref{Jpsipion} (a)-(b) for more details). The correlation of this broad structure with the corresponding isovector hidden-charm molecular tetraquarks should be investigated, which will become an important issue in this work. In 2018, the D\O\, Collaboration analyzed the data of the $J/\psi \pi$ invariant mass spectrum from the $B \to  J/\psi \pi K$ \cite{Abazov:2018cyu}, where they only focused on $3600<m_{J/\psi\pi}<4200$ MeV range inspired by the observed $Z_c(3900)^{\pm}$ from the BESIIII \cite{Ablikim:2013mio} and Belle \cite{Liu:2013dau}. And the D\O\,  data show the evidence of charged charmoniumlike structure similar to the $Z_c(3900)^{\pm}$, where the measured mass of the $Z_c(3900)^{\pm}$ is $3895.0 \pm 5.2 ^{+4.0}_{-2.7}~{\rm MeV}$. Since the width information of such structure is still absent in the D\O\, measurement, it is hard to conclude that the D\O\, evidence truly corresponds to the $Z_c(3900)^{\pm}$ \cite{Ablikim:2013mio,Liu:2013dau}. Additionally, the $J/\psi \pi$ invariant mass spectrum of the $B \to  J/\psi \pi K$ decay has been studied by the LHCb Collaboration in 2019 \cite{Aaij:2019ipm}, and two enhancement structures are visible at 4200 MeV and 4600 MeV.
\end{enumerate}

As introduced above, the isovector $XYZ$ data of the $B\to XYZ+K$ decays stimulate our interest in exploring the isovector hidden-charm molecular systems, which will be illustrated in Sec. \ref{sec3}. By this study, we want to answer whether these isovector $XYZ$ structures have close relation to the isovector hidden-charm molecular tetraquarks.

\section{Mass spectrum of the charmoniumlike molecular tetraquark systems}\label{sec3}

In this section, we restudy the mass spectrum behaviors from a pair of charmed (charm-strange) meson and anticharmed (anti-charm-strange) meson interactions, these charmed or anticharmed mesons are in the $H$, $S$, and $T$ doublets. Here, we still adopt the OBE model and consider the $S-D$ wave mixing effect and the coupled channel effect. As is well known, since Yukawa firstly proposed the nucleon-nucleon interaction is mediated through the pion-exchange in 1935 \cite{Yukawa:1935xg}, the OBE model obtains great promotion. On one hand, theorists consider the scalar meson ($\sigma$) and vector meson $(\rho/\omega)$ exchanges interactions to depict the intermediate and short range interactions, respectively. On the other hand, several various corrections are introduced to discuss the fine properties of the hadron-hadron interactions, like the isospin breaking effects, the coupled-channel effects, the spin-orbit force, and the recoil corrections. Up to now, the OBE model have been frequently adopted to study the hadron-hadron interactions in the heavy flavor sector \cite{Chen:2016qju,Liu:2019zoy}.

\subsection{OBE effective potentials}\label{sec31}

For deducing the OBE effective potentials for the hadron-hadron interactions at the hadronic level quantitatively, we usually adopt the effective Lagrangian approach. The general procedures include three typical steps \cite{Wang:2020dya,Wang:2019nwt,Wang:2019aoc,Wang:2020bjt,Wang:2021hql,Chen:2018pzd,Yang:2021sue}. First, we can write out the scattering amplitude $\mathcal{M}(h_1h_2\to h_3h_4)$ for the relevant scattering process $h_1h_2\to h_3h_4$ according to the effective Lagrangians. And then, the effective potentials in the momentum space $\mathcal{V}^{h_1h_2\to h_3h_4}_E(\bm{q})$ can be related to the corresponding scattering amplitudes $\mathcal{M}(h_1h_2\to h_3h_4)$ by using the Breit approximation \cite{Breit:1929zz,Breit:1930zza}, i.e.,
\begin{eqnarray}
\mathcal{V}^{h_1h_2\to h_3h_4}_E(\bm{q})=-\frac{\mathcal{M}(h_1h_2\to h_3h_4)}{\sqrt{\prod_i 2 m_i\prod_f 2 m_f}},
\end{eqnarray}
where $m_{i}\,(i=h_1,\,h_2)$ and $m_{f}\,(f=h_3,\,h_4)$ denote the masses of the initial and final states, respectively. Finally, the effective potentials in the coordinate space $\mathcal{V}^{h_1h_2\to h_3h_4}_E(\bm{r})$ can be obtained by performing the Fourier transformation, i.e.,
\begin{eqnarray}
\mathcal{V}^{h_1h_2\to h_3h_4}_E(\bm{r}) =\int \frac{d^3\bm{q}}{(2\pi)^3}e^{i\bm{q}\cdot\bm{r}}\mathcal{V}^{h_1h_2\to h_3h_4}_E(\bm{q})\mathcal{F}^2(q^2,m_E^2),
\end{eqnarray}
which will be applied to search for the bound state solutions by solving the coupled channel Schr$\rm \ddot{o}$dinger equation, and we can further extract the bound state properties from the obtained bound state solutions. Because the discussed hadrons are not pointlike particles, we introduce the monopole type form factor in each interaction vertex \cite{Tornqvist:1993ng, Tornqvist:1993vu}, i.e.,
\begin{eqnarray}
\mathcal{F}(q^2,m_E^2) = \frac{\Lambda^2-m_E^2}{\Lambda^2-q^2},
\end{eqnarray}
which reflects the finite size effect of the discussed hadrons and compensate the off-shell effect of the exchanged light mesons \cite{Wang:2020dya}. Here, $\Lambda$, $m_E$, and $q$ are the cutoff parameter, the mass, and the four momentum of the exchanged light mesons, respectively.

Subsequently, let us construct the relevant effective Lagrangians. According to the heavy quark limit~\cite{Wise:1992hn}, the relevant super-fields $H^{(Q)}_a$, $H^{(\overline{Q})}_a$, $S^{(Q)}_a$, $S^{(\overline{Q})}_a$, $T^{(Q)\mu}_a$, and $T^{(\overline{Q})\mu}_a$ can be defined as as~\cite{Ding:2008gr}
\begin{eqnarray}
H^{(Q)}_a&=&{\mathcal P}_{+}\left(D^{*(Q)\mu}_a\gamma_{\mu}-D^{(Q)}_a\gamma_5\right),\nonumber\\
H^{(\overline{Q})}_a&=&\left(\bar{D}^{*(\overline{Q})\mu}_a\gamma_{\mu}-\bar{D}^{(\overline{Q})}_a\gamma_5\right){\mathcal P}_{-},\nonumber\\
S^{(Q)}_a&=&{\mathcal P}_{+}\left(D^{\prime(Q)\mu}_{1a}\gamma_{\mu}\gamma_5-D^{*(Q)}_{0a}\right),\nonumber\\
S^{(\overline{Q})}_a&=&\left(D^{\prime(\overline{Q})\mu}_{1a}\gamma_{\mu}\gamma_5-D^{*(\overline{Q})}_{0a}\right){\mathcal P}_{-},\nonumber\\
T^{(Q)\mu}_a&=&{\mathcal P}_{+}\left[D^{*(Q)\mu\nu}_{2a}\gamma_{\nu}-\sqrt{\frac{3}{2}}D^{(Q)}_{1a\nu}\gamma_5\left(g^{\mu\nu}-\frac{1}{3}\gamma^{\nu}
\left(\gamma^{\mu}-v^{\mu}\right)\right)\right],\nonumber\\
T^{(\overline{Q})\mu}_a&=&\left[\bar{D}^{*(\overline{Q})\mu\nu}_{2a}\gamma_{\nu}-\sqrt{\frac{3}{2}}\bar{D}^{(\overline{Q})}_{1a\nu}\gamma_5
\left(g^{\mu\nu}-\frac{1}{3}\gamma^{\nu}\left(\gamma^{\mu}-v^{\mu}\right)\right)\right]{\mathcal P}_{-},\nonumber\\
\end{eqnarray}
respectively. Here, ${\mathcal P}_{\pm}=(1\pm{v}\!\!\!\slash)/2$ are the projection operators, and $v^{\mu}=(1, \bf{0})$ denotes the four velocity in the nonrelativistic approximation. Their conjugate fields read as $\overline{X}=\gamma_0X^{\dagger}\gamma_0$ with $X=H^{(Q)}_a$, $H^{(\overline{Q})}_a$, $S^{(Q)}_a$, $S^{(\overline{Q})}_a$, $T^{(Q)\mu}_a$, and $T^{(\overline{Q})\mu}_a$.

According to the heavy quark symmetry, the chiral symmetry, and the hidden local symmetry \cite{Casalbuoni:1992gi,Casalbuoni:1996pg,Yan:1992gz,Harada:2003jx,Bando:1987br}, one can construct the effective Lagrangians describing the interactions between the (anti)charmed mesons in the $H/S/T$-doublet and the light scalar, pseudoscalar, and vector mesons \cite{Ding:2008gr},
\begin{eqnarray}\label{eq:compactlag}
{\mathcal L}&=&g_{\sigma}\left\langle H^{(Q)}_a\sigma\overline{H}^{(Q)}_a\right\rangle+g_{\sigma}\left\langle \overline{H}^{(\overline{Q})}_a\sigma H^{(\overline{Q})}_a\right\rangle\nonumber\\
&&+g^{\prime}_{\sigma}\left\langle S^{(Q)}_a\sigma\overline{S}^{\,(Q)}_a\right\rangle+g^{\prime}_{\sigma}\left\langle\overline{S}^{\,(\overline{Q})}_a\sigma S^{(\overline{Q})}_a\right\rangle\nonumber\\
&&+g^{\prime\prime}_{\sigma}\left\langle T^{(Q)\mu}_a\sigma\overline{T}^{(Q)}_{a\mu}\right\rangle+g^{\prime\prime}_{\sigma}\left\langle\overline{T}^{(\overline{Q})\mu}_a\sigma T^{(\overline{Q})}_{a\mu}\right\rangle\nonumber\\
&&+\frac{h_{\sigma}}{f_{\pi}}\left[\left\langle S^{(Q)}_a\gamma^{\mu}\partial_{\mu}\sigma\overline{H}^{\,({Q})}_a\right\rangle-\left\langle\overline{H}^{\,(\overline{Q})}_a\gamma^{\mu}\partial_{\mu}\sigma S^{(\overline{Q})}_a\right\rangle+H.c.\right]\nonumber\\
&&+\frac{h^{\prime}_{\sigma}}{f_{\pi}}\left[\left\langle T^{(Q)\mu}_a\partial_{\mu}\sigma\overline{H}^{(Q)}_b\right\rangle+\left\langle\overline{H}^{(\overline{Q})}_a\partial_{\mu}\sigma T^{(\overline{Q})\mu}_b\right\rangle+H.c.\right]\nonumber\\
&&+ig\left\langle H^{(Q)}_b{\mathcal A}\!\!\!\slash_{ba}\gamma_5\overline{H}^{\,({Q})}_a\right\rangle+ig\left\langle \overline{H}^{(\overline{Q})}_a{\mathcal A}\!\!\!\slash_{ab}\gamma_5 H^{\,(\overline{Q})}_b\right\rangle\nonumber\\
&&+i\tilde{k}\left\langle S^{(Q)}_b{\cal A}\!\!\!\slash_{ba}\gamma_5\overline{S}^{\,(Q)}_a\right\rangle+i\tilde{k}\left\langle\overline{S}^{\,(\overline{Q})}_a{\cal A}\!\!\!\slash_{ab}\gamma_5S^{(\overline{Q})}_b\right\rangle\nonumber\\
&&+ik\left\langle T^{\,(Q)\mu}_b{\mathcal A}\!\!\!\slash_{ba}\gamma_5\overline{T}^{(Q)}_{a\mu}\right\rangle+ik\left\langle\overline{T}^{\,(\overline{Q})\mu}_a{\mathcal A}\!\!\!\slash_{ab}\gamma_5T^{(\overline{Q})}_{b\mu}\right\rangle\nonumber\\
&&+\left[ih\left\langle S^{(Q)}_b{\cal A}\!\!\!\slash_{ba}\gamma_5\overline{H}^{\,(Q)}_a\right\rangle+ih\left\langle\overline{H}^{\,(\overline{Q})}_a{\cal
A}\!\!\!\slash_{ab}\gamma_5S^{(\overline{Q})}_b\right\rangle+H.c.\right]\nonumber\\
&&+\left[i\left\langle T^{(Q)\mu}_b\left(\frac{h_1}{\Lambda_{\chi}}D_{\mu}{\mathcal A}\!\!\!\slash+\frac{h_2}{\Lambda_{\chi}}D\!\!\!\!/ {\mathcal A}_{\mu}\right)_{ba}\gamma_5\overline{H}^{\,(Q)}_a\right\rangle+H.c.\right]\nonumber\\
&&+\left[i\left\langle\overline{H}^{\,(\overline{Q})}_a\left(\frac{h_1}{\Lambda_{\chi}}{\mathcal A}\!\!\!\slash\stackrel{\leftarrow}{D_{\mu}'}+\frac{h_2}{\Lambda_{\chi}}{\mathcal A}_{\mu}\stackrel{\leftarrow}{D\!\!\!\slash'}\right)_{ab}\gamma_5T^{(\overline{Q})\mu}_b\right\rangle+H.c.\right]\nonumber\\
&&+\left\langle iH^{(Q)}_b\left(\beta v^{\mu}({\mathcal V}_{\mu}-\rho_{\mu})+\lambda \sigma^{\mu\nu}F_{\mu\nu}(\rho)\right)_{ba}\overline{H}^{\,(Q)}_a\right\rangle\nonumber\\
&&-\left\langle i\overline{H}^{(\overline{Q})}_a\left(\beta v^{\mu}({\mathcal V}_{\mu}-\rho_{\mu})-\lambda \sigma^{\mu\nu}F_{\mu\nu}(\rho)\right)_{ab}H^{\,(\overline{Q})}_b\right\rangle\nonumber\\
&&+\left\langle iS^{(Q)}_b\left(\beta^{\prime} v^{\mu}({\mathcal V}_{\mu}-\rho_{\mu})+\lambda^{\prime} \sigma^{\mu\nu}F_{\mu\nu}(\rho)\right)_{ba}\overline{S}^{\,(Q)}_a\right\rangle\nonumber\\
&&-\left\langle i\overline{S}^{(\overline{Q})}_a\left(\beta^{\prime} v^{\mu}({\mathcal V}_{\mu}-\rho_{\mu})-\lambda^{\prime} \sigma^{\mu\nu}F_{\mu\nu}(\rho)\right)_{ab}S^{\,(\overline{Q})}_b\right\rangle\nonumber\\
&&+\left\langle iT^{\,(Q)}_{b\lambda}\left(\beta^{\prime\prime} v^{\mu}({\mathcal V}_{\mu}-\rho_{\mu})+\lambda^{\prime\prime}\sigma^{\mu\nu}F_{\mu\nu}(\rho)\right)_{ba}\overline{T}^{(Q)\lambda}_{a}\right\rangle\nonumber\\
&&-\left\langle i\overline{T}^{\,(\overline{Q})}_{a\lambda}\left(\beta^{\prime\prime} v^{\mu}({\mathcal V}_{\mu}-\rho_{\mu})-\lambda^{\prime\prime}\sigma^{\mu\nu}F_{\mu\nu}(\rho)\right)_{ab}T^{(\overline{Q})\lambda}_{b}\right\rangle\nonumber\\
&&+\left[\left\langle H^{(Q)}_b (i\zeta\gamma^{\mu}({\cal V}_{\mu}-\rho_{\mu})+i\mu \sigma^{\lambda\nu}F_{\lambda\nu}(\rho))_{ba}\overline{S}^{\,(Q)}_a\right\rangle+H.c.\right]\nonumber\\
&&+\left[\left\langle\overline{S}^{\,(\overline{Q})}_a (i\zeta\gamma^{\mu}({\cal V}_{\mu}-\rho_{\mu})+i\mu\sigma^{\lambda\nu}F_{\lambda\nu}(\rho))_{ab}H^{(\overline{Q})}_b\right\rangle+H.c.\right]\nonumber\\
&&+\left[\left\langle T^{(Q)\mu}_b\left(i\zeta_1({\mathcal V}_{\mu}-\rho_{\mu})+\mu_{1}\gamma^{\nu}F_{\mu\nu}(\rho)\right)_{ba}\overline{H}^{\,(Q)}_a\right\rangle+H.c.\right]\nonumber\\
&&-\left[\left\langle\overline{H}^{\,(\overline{Q})}_a\left(i\zeta_1({\mathcal V}_{\mu}-\rho_{\mu})-\mu_1\gamma^{\nu}F_{\mu\nu}(\rho)\right)_{ab}T^{(\overline{Q})\mu}_b\right\rangle+H.c.\right],\nonumber\\
\end{eqnarray}
Here, the covariant derivatives are written as $D_{\mu}=\partial_{\mu}+\mathcal{V}_{\mu}$ and $D^{\prime}_{\mu}=\partial_{\mu}-\mathcal{V}_{\mu}$. And the axial current $\mathcal{A}_\mu$, the vector current ${\cal V}_{\mu}$, the vector meson field $\rho_{\mu}$, and the vector meson strength tensor $F_{\mu\nu}(\rho)$ are defined as
\begin{eqnarray}
&&{\mathcal A}_{\mu} =\frac{1}{2}\left(\xi^{\dagger}\partial_{\mu}\xi-\xi\partial_{\mu}\xi^{\dagger}\right)_{\mu}, \nonumber\\
&&{\mathcal V}_{\mu}=\frac{1}{2}\left(\xi^{\dagger}\partial_{\mu}\xi+\xi\partial_{\mu}\xi^{\dagger}\right)_{\mu},\nonumber\\
&&\xi = \exp(i\mathbb{P}/f_\pi), \quad     \rho_{\mu}=\frac{i{g_V}}{{\sqrt{2}}}\mathbb{V}_{\mu},\nonumber\\
&&F_{\mu\nu}(\rho)=\partial_{\mu}\rho_{\nu}-\partial_{\nu}\rho_{\mu}+\left[\rho_{\mu},\rho_{\nu}\right].
\end{eqnarray}
respectively. The light pseudoscalar meson matrix $\mathbb{P}$ and the light vector meson matrix $\mathbb{V}_{\mu}$ have the conventional form \cite{Wang:2020bjt,Wang:2021hql,Chen:2018pzd}, which can be expressed as
\begin{eqnarray}
\left.\begin{array}{l}
{\mathbb{P}} = {\left(\begin{array}{ccc}
       \frac{\pi^0}{\sqrt{2}}+\frac{\eta}{\sqrt{6}} &\pi^+ &K^+\\
       \pi^-       &-\frac{\pi^0}{\sqrt{2}}+\frac{\eta}{\sqrt{6}} &K^0\\
       K^-         &\bar K^0   &-\sqrt{\frac{2}{3}} \eta     \end{array}\right)},\\
{\mathbb{V}}_{\mu} = {\left(\begin{array}{ccc}
       \frac{\rho^0}{\sqrt{2}}+\frac{\omega}{\sqrt{2}} &\rho^+ &K^{*+}\\
       \rho^-       &-\frac{\rho^0}{\sqrt{2}}+\frac{\omega}{\sqrt{2}} &K^{*0}\\
       K^{*-}         &\bar K^{*0}   & \phi     \end{array}\right)}_{\mu},
\end{array}\right.
\end{eqnarray}
respectively. After expanding the compact effective Lagrangians in Eq. (\ref{eq:compactlag}) to the leading order of the pseudo-Goldstone field $\xi$, we can further obtain the concrete effective Lagrangians (see Refs. \cite{Wang:2019nwt,Wang:2019aoc,Wang:2020dya,Shen:2010ky,Hu:2010fg} for more information). The normalized relations for these discussed charmed mesons are written as
\begin{eqnarray}
\left.\begin{array}{ll}
\langle 0|D|c\bar{q}(0^-)\rangle=\sqrt{m_{D}},\quad&\langle 0|D^{*\mu}|c\bar{q}(1^-)\rangle=\epsilon^\mu\sqrt{m_{D^*}},\\
\langle 0|D_0^*|c\bar{q}(0^+)\rangle=\sqrt{m_{D_0^*}},\quad&\langle 0|D_1^{\prime \mu}|c\bar{q}(1^+)\rangle=\epsilon^\mu\sqrt{m_{D_1^{\prime}}},\\
\langle 0|D_{1}^{\mu}|c\bar{q}(1^+)\rangle=\epsilon^\mu\sqrt{m_{D_{1}}},\quad&\langle 0|D_{2}^{*\mu\nu}|c\bar{q}(2^+)\rangle=\zeta^{\mu\nu}\sqrt{m_{D_{2}^*}},\\
\end{array}\right.
\end{eqnarray}
respectively. Here, $\epsilon^\mu_{m}\,(m=0,\pm1)$ and $\zeta^{\mu\nu}_{m}(m =0,\pm1,\pm2)$ correspond to the  polarization vector and tensor, respectively. In the static limit, they have the form of $\epsilon_{0}^{\mu}= \left(0,0,0,-1\right)$, $\epsilon_{\pm}^{\mu}= \left(0,\,\pm1,\,i,\,0\right)/\sqrt{2}$, and $\zeta^{\mu\nu}_{m}=\sum_{m1,m2}\langle 1, m_1; 1, m_2|2, m\rangle\epsilon^{\mu}_{m_1}\epsilon^{\nu}_{m_2}$ \cite{Cheng:2010yd}.

In order to obtain the concrete effective potentials, one need to further construct the wave functions for the investigated systems. They include the color part, the flavor part, the spin-orbit part, and the spatial wave functions. For the systems composed by colorless hadrons, the color wave functions are simply $\bm{1}$. In Table~\ref{flavorwave}, we summarize the flavor wave functions $|I, I_3\rangle$ for the $S$-wave $\mathcal{A}\overline{\mathcal{A}}$ and $\mathcal{A}\overline{\mathcal{B}}$ systems, here, notations $\mathcal{A}$ and $\mathcal{B}$ stand for the different charmed mesons, and $J$, $J_1$, and $J_2$ correspond to the total angular momentum quantum numbers of the discussed charmoniumlike systems $\mathcal{A}\overline{\mathcal{B}}$, the charmed (charmed-strange) mesons $\mathcal{A}$, and the charmed (charmed-strange) mesons $\mathcal{B}$, respectively. In particular, we need to emphasize the $C$ parity for the discussed systems is determined by $C=cx_1x_2(-1)^{J-J_1-J_2}$ with $c=\pm1$, where the charge conjugate transformation conventions satisfy $\mathcal{A} \leftrightarrow x_1\overline{\mathcal{A}}$ and $\mathcal{B} \leftrightarrow x_2\overline{\mathcal{B}}$ \cite{Liu:2007bf,Liu:2008xz,Liu:2008fh,Liu:2008tn,Sun:2012sy,Wang:2020dya,Li:2015exa,Li:2013bca,Hu:2010fg,Shen:2010ky,Dong:2021juy,Liu:2013rxa,Chen:2015add}.

\renewcommand\tabcolsep{0.23cm}
\renewcommand{\arraystretch}{1.50}
\begin{table}[!htbp]\centering
\caption{Flavor wave functions $|I, I_3\rangle$ for these discussed $S$-wave $\mathcal{A}\overline{\mathcal{A}}$ and $\mathcal{A}\overline{\mathcal{B}}$ systems. Here, the notations $\mathcal{A}$ and $\mathcal{B}$ stand for different charmed mesons, and $I$ and $I_3$ represent their isospin and the third component of these discussed charmoniumlike systems, respectively.}\label{flavorwave}
\begin{tabular}{c|l|c}
\toprule[1.0pt]
\toprule[1.0pt]
Systems&$|I, I_3\rangle$&Flavor wave functions\\
\midrule[1.0pt]
\multirow{5}{*}{$\mathcal{A}\overline{\mathcal{A}}$}&$|1, 1\rangle$ &$\mathcal{A}^{+}\overline{\mathcal{A}}^{0}$\\
                           &$|1, 0\rangle$ &$\dfrac{1}{\sqrt{2}}\left(\mathcal{A}^{0}\overline{\mathcal{A}}^{0}-\mathcal{A}^{+}\mathcal{A}^{-}\right)$\\
                           &$|1, -1\rangle$&${\mathcal{A}}^{0}\mathcal{A}^{-}$\\
                           &$|0, 0\rangle$ &$\dfrac{1}{\sqrt{2}}\left(\mathcal{A}^{0}\overline{\mathcal{A}}^{0}+\mathcal{A}^{+}\mathcal{A}^{-}\right)$\\
\midrule[1.0pt]
\multirow{5}{*}{$\mathcal{A}\overline{\mathcal{B}}$}&$|1, 1\rangle$ &$\dfrac{1}{\sqrt{2}}\left(\mathcal{A}^{+}\overline{\mathcal{B}}^{0}+c\mathcal{B}^{+}\overline{\mathcal{A}}^{0}\right)$\\
          &$|1,0\rangle$&$\dfrac{1}{2}\left[\left(\mathcal{A}^{0}\overline{\mathcal{B}}^{0}
          -\mathcal{A}^{+}{\mathcal{B}}^{-}\right)+c\left(\mathcal{B}^{0}\overline{\mathcal{A}}^{0}-\mathcal{B^+}{\mathcal{A}}^{-}\right)\right]$\\
          &$|1, -1\rangle$&$\dfrac{1}{\sqrt{2}}\left({\mathcal{A}}^{0}\mathcal{B}^{-}+c{\mathcal{B}}^{0}\mathcal{A}^{-}\right)$\\
          &$|0,0\rangle$&$\dfrac{1}{2}\left[\left(\mathcal{A}^{0}\overline{\mathcal{B}}^{0}
          +\mathcal{A}^{+}{\mathcal{B}}^{-}\right)+c\left(\mathcal{B}^{0}\overline{\mathcal{A}}^{0}+\mathcal{B^+}{\mathcal{A}}^{-}\right)\right]$\\

\bottomrule[1.0pt]
\bottomrule[1.0pt]
\end{tabular}
\end{table}

The spin-orbit wave functions $|^{2S+1}L_J\rangle$ for the $|{\mathcal D}_{J_1}{\overline {\mathcal D}}_{J_2}\rangle$ systems can be constructed as
\begin{eqnarray}
|{\mathcal D}_{0}{\overline {\mathcal D}}_{1}\rangle &=&\sum_{m,m_L}C^{J,M}_{1m,Lm_L}\epsilon_{m}^\mu|Y_{L,m_L}\rangle,\\
|{\mathcal D}_{0}{\overline {\mathcal D}}_{2}\rangle &=&\sum_{m,m_L}C^{J,M}_{2m,Lm_L}\zeta_{m}^{\mu\nu}|Y_{L,m_L}\rangle,\\
|{\mathcal D}_{1}{\overline {\mathcal D}}_{1}\rangle &=&\sum_{m,m^{\prime},m_S,m_L}C^{S,m_S}_{1m,1m^{\prime}}C^{J,M}_{Sm_S,Lm_L}\epsilon_{m}^\mu\epsilon_{m^{\prime}}^\nu|Y_{L,m_L}\rangle,\\
|{\mathcal D}_{1}{\overline {\mathcal D}}_{2}\rangle &=& \sum_{m,m^{\prime},m_S,m_L}C^{S,m_S}_{1m,2m^{\prime}}C^{J,M}_{Sm_S,Lm_L}\epsilon_{m}^\lambda\zeta_{m^{\prime}}^{\mu\nu}|Y_{L,m_L}\rangle.
\end{eqnarray}
In the above expressions, the notations ${\mathcal D}_0$, ${\mathcal D}_1$, and ${\mathcal D}_2$ denote the charmed (charm-strange) mesons with the total angular momentum quantum numbers $J = 0, 1$, and $2$, respectively. $C^{e,f}_{ab,cd}$ is the Clebsch-Gordan coefficient, and $|Y_{L,m_L}\rangle$ is the spherical harmonics function. In Table \ref{spin-orbit wave functions}, we summary the relevant spin-orbit wave functions  $|{}^{2S+1}L_{J}\rangle$ and the discussed channels under considering the coupled channel effect.

\renewcommand\tabcolsep{0.10cm}
\renewcommand{\arraystretch}{1.50}
\begin{table}[!htpb]
\centering
\caption{The relevant quantum numbers $J^{P}$ and possible channels $|{}^{2S+1}L_{J}\rangle$ involved in this work. Here, $...$ means that the $S-$wave components for the corresponding channels do not exist.}\label{spin-orbit wave functions}
\begin{tabular}{c|lllll}\toprule[1pt]\toprule[1pt]
 $J^{P}$ &${\mathcal D}_0\overline{{\mathcal D}}_0$&${\mathcal D}_0\overline{{\mathcal D}}_1$&${\mathcal D}_0\overline{{\mathcal D}}_2$&${\mathcal D}_1\overline{{\mathcal D}}_1$&${\mathcal D}_1\overline{{\mathcal D}}_2$\\\midrule[1.0pt]
$0^{\pm}$&$|{}^1\mathbb{S}_{0}\rangle$&$...$&$...$&$|{}^1\mathbb{S}_{0}\rangle/|{}^5\mathbb{D}_{0}\rangle$&$...$\\
$1^{\pm}$&$...$ &$|{}^3\mathbb{S}_{1}\rangle/|{}^3\mathbb{D}_{1}\rangle$&$...$&$|{}^3\mathbb{S}_{1}\rangle/|{}^{3,5}\mathbb{D}_{1}\rangle$&$|{}^3\mathbb{S}_{1}\rangle/|{}^{3,5,7}\mathbb{D}_{1}\rangle$\\
$2^{\pm}$&$...$&$...$&$|{}^5\mathbb{S}_{2}\rangle/|{}^5\mathbb{D}_{2}\rangle$&$|{}^5\mathbb{S}_{2}\rangle/|{}^{1,3,5}\mathbb{D}_{2}\rangle$&$|{}^5\mathbb{S}_{2}\rangle/|{}^{3,5,7}\mathbb{D}_{2}\rangle$\\
$3^{\pm}$&$...$&$...$&$...$&$...$&$|{}^7\mathbb{S}_{3}\rangle/|{}^{3,5,7}\mathbb{D}_{3}\rangle$\\
\bottomrule[1pt]\bottomrule[1pt]
\end{tabular}
\end{table}

\begin{figure}[!htbp]
\centering
\begin{tabular}{cc}
\includegraphics[width=0.21\textwidth]{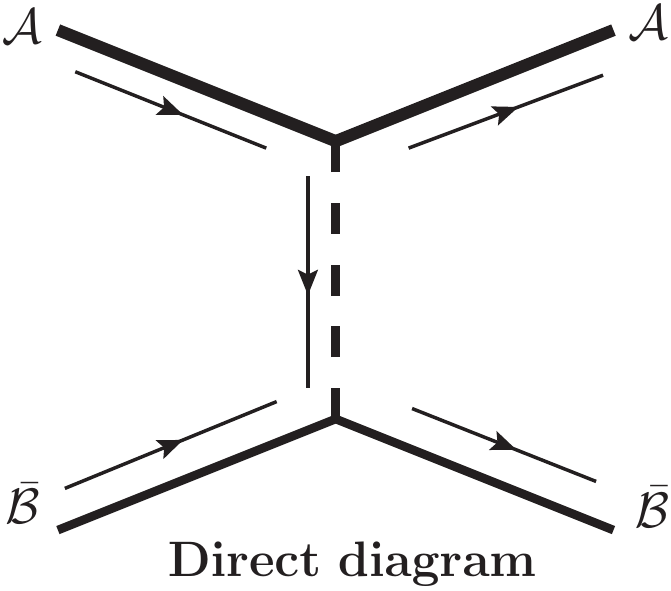}\quad\quad
\includegraphics[width=0.21\textwidth]{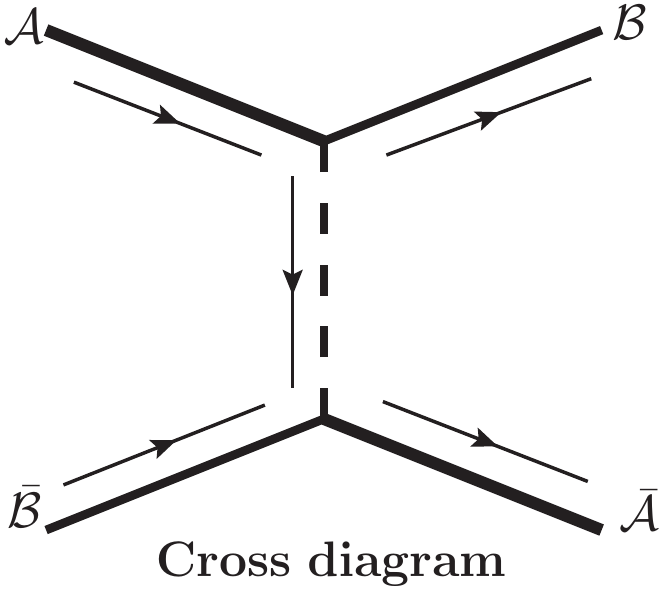}
\end{tabular}
\caption{The direct channel and cross channel Feynman diagrams for the $\mathcal{A}\overline{\mathcal{B}}\to \mathcal{A}\overline{\mathcal{B}}$ processes. Here, the notations $\mathcal{A}$ and $\mathcal{B}$ represent two different charmed (charmed-strange) mesons.}\label{fey}
\end{figure}

For the $\mathcal{A}\overline{\mathcal{B}}\to\mathcal{A}\overline{\mathcal{B}}$ processes, there exist the direct channel and cross channel Feynman diagrams~\cite{Liu:2007bf,Wang:2020dya}, where $\mathcal{A}$ and $\mathcal{B}$ stand for two different charmed (charm-strange) mesons. The total effective potentials can be written as \cite{Liu:2008fh,Liu:2008tn,Sun:2012sy,Wang:2020dya}
\begin{eqnarray}
\mathcal{V}_{Total}(\bm{r})=\mathcal{V}_{D}(\bm{r})+c\,\mathcal{V}_{C}(\bm{r}).
\end{eqnarray}
In Fig. \ref{fey}, we present the direct channel and cross channel Feynman diagrams. For the $\mathcal{A}\overline{\mathcal{A}}$ systems, there exist the direct channel contribution.

With the standard procedures of the OBE model \cite{Wang:2020dya,Wang:2019nwt,Wang:2019aoc,Wang:2020bjt,Wang:2021hql,Chen:2018pzd}, we can finally obtain the effective potentials in the coordinate space. In Appendix \ref{app01}, we collect the concrete expressions for all the OBE effective potentials. We estimate the coupling constants by fitting the reported experimental data and using several theoretical models \cite{Liu:2008xz,Casalbuoni:1996pg,Wang:2020dya,Falk:1992cx,Isola:2003fh,Cleven:2016qbn,Dong:2019ofp,He:2019csk,Wang:2019nwt,Wang:2019aoc,
Wang:2020lua,Riska:2000gd,Hu:2010fg,Bardeen:2003kt,Cheng:2004ru,He:2016pfa,Liu:2010hf,He:2013oma,He:2012zd}. In particular, we fix the phases between these coupling constants by the quark model \cite{Riska:2000gd}. In Table~\ref{parameters1}, we collect their values. In addition, we need to introduce the following parameters of the hadron masses, $m_{\sigma}=600.00~\rm{MeV}$, $m_{\pi}=137.27~\rm{MeV}$, $m_{\eta} =547.86~\rm{MeV}$, $m_{\rho}=775.49~\rm{MeV}$, $m_{\omega}=782.65~\rm{MeV}$, $m_{\phi}=1019.46~\rm{MeV}$, $m_D=1867.24~\rm{MeV}$, $m_{D^{\ast}}=2008.56~\rm{MeV}$, $m_{D_0^*}=2324.50~\rm{MeV}$, $m_{D_1^{\prime}}=2427.00~\rm{MeV}$, $m_{D_1}=2422.00~\rm{MeV}$, $m_{D^{\ast}_2}=2463.05~\rm{MeV}$, $m_{D_s}=1968.34~\rm{MeV}$, $m_{D_s^{\ast}}=2112.20~\rm{MeV}$, $m_{D_{s0}^*}=2317.80~\rm{MeV}$, $m_{D_{s1}^{\prime}}=2459.50~\rm{MeV}$, $m_{D_{s1}}=2535.11~\rm{MeV}$, and $m_{D^{\ast}_{s2}}=2569.10~\rm{MeV}$ \cite{Zyla:2020zbs}.
\renewcommand\tabcolsep{0.38cm}
\renewcommand{\arraystretch}{1.50}
\begin{table}[!htbp]
\caption{A summary of the coupling constants adopted in our calculations. Units of the coupling constants $h^{\prime}=(h_1+h_2)/\Lambda_{\chi}$, $\lambda$, $\lambda^{\prime}$, $\lambda^{\prime\prime}$, $\mu$, and $\mu_1$ are $\rm{GeV}^{-1}$, and the coupling constant $f_{\pi}$ is given in unit of GeV.}\label{parameters1}
\centering
\begin{tabular}{cccccc}
\toprule[1.0pt]
\toprule[1.0pt]
$g_{\sigma}$&$g^{\prime}_{\sigma}$&$g^{\prime\prime}_{\sigma}$&$|h_{\sigma}|$&$|h^{\prime}_{\sigma}|$&$g$                     \\
$-0.76$     &$0.76$              &$0.76$                     &$0.32$         &$0.35$                 &$0.59$             \\
$\tilde{k}$&  $k$        &$|h|$        &$|h^{\prime}|$&$f_{\pi}$         &$\beta$\\
$0.59$     &$0.59$     &$0.56$     &$0.55$       &$0.132$              &$-0.90$\\
$\beta^{\prime}$   &$\beta^{\prime\prime}$     &$\lambda$     &$\lambda^{\prime}$         &$\lambda^{\prime\prime}$ &  $|\zeta|$\\
$0.90$             &$0.90$                     &$-0.56$       &$0.56$                     &$0.56$                   &$0.727$ \\
$|\mu|$   &$|\zeta_1|$&$\mu_1$       &$g_V$\\
$0.364$   &$0.20$      &$0$           &$5.83$\\
\bottomrule[1.0pt]
\bottomrule[1.0pt]
\end{tabular}
\end{table}

\subsection{Numerical results and discussions}\label{sec32}

In this section, we attempt to find the loosely bound state solutions for the charmed (charm-strange) meson and anticharmed (anti-charm-strange) meson systems by solving the coupled channel Schr$\ddot{\rm{o}}$dinger equation. As the only one free parameter here, we vary the cutoff parameters in the range of $0.8$-$3.0~\rm{GeV}$. The loosely bound state with cutoff value around $1.0\rm~{GeV}$ can be the prime hadronic molecular candidate, since this value range is widely accepted as a reasonable input parameter based on the experience of studying the deuteron \cite{Tornqvist:1993ng,Tornqvist:1993vu,Wang:2019nwt}. As is well known, a reasonable loosely bound hadronic molecule should satisfy its binding energy is around several to several tens MeV, and its typical size should be larger than the size of all the component hadrons \cite{Chen:2016qju,Chen:2017xat}.

{Here, we need to emphasis that we mainly focus on the mass spectrum for the hidden-charm molecular tetraquark systems in this work, which is inspired by the abundant experimental data. In addition, we also give rough estimations of the branching ratios of the two-body hidden-charm decay behaviors for the $D^{*}\bar{D}^{*}$ molecular tetraquarks within the heavy quark symmetry analysis,
which is due to possible peculiar characteristic mass spectrum of the isoscalar $D^*\bar{D}^*$ molecular systems existing in the reported experimental data of the $B$ meson decays (see Fig. \ref{Jpsiomegaeta}).
For other obtained $\mathcal{D}\bar{\mathcal{D}}$ and $\mathcal{D}_s\bar{\mathcal{D}}_s$ molecules, we only simply list the two-body hidden-charm decay channels. Their decay behaviors will be further discussed in the future work.}
Our results will be categorized into three corresponding cases:
\begin{enumerate}
\item The isoscalar charmoniumlike molecular systems without hidden-strange quantum number,
\item The charmoniumlike molecular systems with hidden-strange quantum number,
\item The isovector hidden-charm molecular tetraquark systems.
\end{enumerate}

\subsubsection{Isoscalar charmoniumlike molecular systems without hidden-strange quantum number}

\paragraph{The isoscalar $D^*\bar{D}^*$ system.}
In Table \ref{bound1}, we present the corresponding bound state properties for the $S-$wave isoscalar $D^*\bar{D}^*$ system. When cutoff values vary from 0.8 to 3.0 GeV, we can obtain bound state solutions for the $S-$wave isoscalar $D^*\bar{D}^*$ states with $J^{PC}=0^{++}$, $1^{+-}$, and $2^{++}$. And we can obtain the relation of $\Lambda[0(0^{++})]<\Lambda[0(1^{+-})]<\Lambda[0(2^{++})]$. Suppose bound states with a smaller cutoff binds deeper when we set the same binding energy. We can find the isoscalar $D^*\bar{D}^*$ interaction with $I(J^{PC})=0(0^{++})$ is strongest attractive, followed by the states with $I(J^{PC})=0(1^{+-})$ and $0(2^{++})$. Thus, we can conclude these three states are possible isoscalar hidden-charm molecular tetraquark candidates, and their masses satisfy $M[0(0^{++})]<M[0(1^{+-})]<M[0(2^{++})]$. Here, our results are also consistent with the conclusions in Refs.~\cite{Liu:2008tn,Liu:2008mi,Liu:2009ei,Sun:2011uh,Sun:2012zzd,Zhao:2015mga,Liu:2017mrh,Dong:2021juy,Liu:2016kqx,Dai:2018nmw,Ding:2020dio,Yang:2017prf,Zhang:2006ix,Tornqvist:1993ng,DeRujula:1976zlg}.
\renewcommand\tabcolsep{0.10cm}
\renewcommand{\arraystretch}{1.50}
\begin{table}[!htbp]\centering
\caption{Bound state properties for the $S-$wave isoscalar $D^{\ast}\bar{D}^{\ast}$ system. Cutoff $\Lambda$, binding energy $E$, and root-mean-square (RMS) radius $r_{RMS}$ are in units of GeV, MeV, and fm, respectively. Here, we label the major probability for the corresponding channels in a bold manner.}\label{bound1}
\begin{tabular}{c|ccc|cccc}\toprule[1.0pt]\toprule[1.0pt]
\multicolumn{1}{c|}{Effect}&\multicolumn{3}{c|}{Single channel}&\multicolumn{4}{c}{$S-$$D$ wave mixing effect}\\\midrule[1.0pt]
$J^{PC}$&$\Lambda$ &$E$  &$r_{\rm RMS}$ &$\Lambda$ &$E$  &$r_{\rm RMS}$ &$P({}^1\mathbb{S}_{0}/{}^5\mathbb{D}_{0})$\\
\multirow{2}{*}{$0^{++}$}  &0.92&$-0.56$ &3.98         &0.91&$-0.61$ &3.89&\textbf{99.56}/0.44                   \\
                                            &0.99&$-11.81$ &1.09        &0.98&$-10.80$ &1.15&\textbf{99.42}/0.58                            \\\midrule[1.0pt]
$J^{PC}$&$\Lambda$ &$E$  &$r_{\rm RMS}$ &$\Lambda$ &$E$  &$r_{\rm RMS}$ &$P({}^3\mathbb{S}_{1}/{}^3\mathbb{D}_{1})$                              \\
\multirow{2}{*}{$1^{+-}$}  &1.07&$-0.38$ &4.54         &1.05&$-0.35$ &4.67&\textbf{99.49}/0.51                                \\
                                            &1.16&$-12.01$ &1.09        &1.15&$-12.35$ &1.10&\textbf{99.13}/0.87                                \\\midrule[1.0pt]
$J^{PC}$&$\Lambda$ &$E$  &$r_{\rm RMS}$ &$\Lambda$ &$E$  &$r_{\rm RMS}$ &$P({}^5\mathbb{S}_{2}/{}^1\mathbb{D}_{2}/{}^5\mathbb{D}_{2})$                            \\
\multirow{2}{*}{$2^{++}$}  &2.06&$-0.28$ &5.18         &1.82&$-0.33$ &5.05&\textbf{99.03}/0.07/0.90                               \\
                                            &3.00&$-12.35$ &1.26        &2.81&$-12.45$ &1.28&\textbf{98.00}/0.15/1.85                                   \\
\bottomrule[1.0pt]\bottomrule[1.0pt]
\end{tabular}
\end{table}

After that, we give rough estimations of the branching ratios of the two-body hidden-charm decay behaviors for these possible $S-$wave isoscalar $D^{\ast}\bar{D}^{\ast}$ molecular candidates by using the heavy quark symmetry. {As an approximate symmetry, the heavy quark symmetry is often applied to study the structures of the hadrons which contain the heavy quarks. In order to perform the heavy quark symmetry analysis for the two-body hidden-charm decay behaviors, we should first expand the spin wave functions of the heavy hadrons systems $|\ell_1 s_1 j_1, \ell_2 s_2 j_2, J M \big\rangle$ in terms of the heavy quark basis $|\ell_1 \ell_2 L, s_1 s_2 S, J M \big\rangle$, i.e.,
\begin{eqnarray}
&&\left|\ell_1 s_1 j_1, \ell_2 s_2 j_2, J M \right\rangle \nonumber\\
&&= \sum_{S,L}\hat{S} \hat{L} \hat{j_1} \hat{j_2}\left\{
\begin{array}{ccc}
\ell_1 & \ell_2 & L \\
s_1 & s_2 & S \\
j_1 & j_2 & J
\end{array}
\right\} |\ell_1 \ell_2 L, s_1 s_2 S, J M \big\rangle
\end{eqnarray}
with $\hat{A}=\sqrt{2A+1}$. Here, the 9-$j$ symbol is used to relate two bases $|\ell_1 s_1 j_1, \ell_2 s_2 j_2, J M \big\rangle$ and $|\ell_1 \ell_2 L, s_1 s_2 S, J M \big\rangle$ with the investigated system coupled in a different way \cite{Ozpineci:2013zas}.}

For these possible $S-$wave isoscalar $D^{\ast}\bar{D}^{\ast}$ molecular candidates, the two-body hidden-charm decay channels include the $\eta_c\eta$, $\eta_c\eta^{\prime}$, $\eta_c\omega$, $J/\psi\eta$, and $J/\psi\omega$ channels. We can expand their spin wave functions in the heavy quark spin symmetry basis, i.e.,
\begin{eqnarray}
\left|0^{++}\right\rangle &=& \frac{\sqrt{3}}{2}\left|0_{q\bar{q}}^{-+},0_{c\bar{c}}^{-+},0^{++}\right\rangle-\frac{1}{2}\left|1_{q\bar{q}}^{--},1_{c\bar{c}}^{--},0^{++}\right\rangle,\\
\left|1^{+-}\right\rangle &=& \frac{1}{\sqrt{2}}\left|1_{q\bar{q}}^{--},0_{c\bar{c}}^{-+},1^{+-}\right\rangle+\frac{1}{\sqrt{2}}\left|0_{q\bar{q}}^{-+},1_{c\bar{c}}^{--},1^{+-}\right\rangle,\\
\left|2^{++}\right\rangle &=& \left|1_{q\bar{q}}^{--},1_{c\bar{c}}^{--},2^{++}\right\rangle,
\end{eqnarray}
where $L_{q\bar{q}}^{P_q,C_q}$and $S_{c\bar{c}}^{P_Q,C_Q}$ stand for the spin parities for the light-flavor meson and charmonium state, respectively. In the heavy quark symmetry, we can estimate that
\begin{itemize}
  \item The isoscalar $D^*\bar{D}^*$ molecular state with $J^{PC}=0^{++}$ can decay into the $\eta_c\eta^{(\prime)}$ and $J/\psi\omega$ channels through the $S-$wave interaction, the relative decay ratio $\mathcal{B}_0=\Gamma_0[J/\psi\omega]/\Gamma_0[\eta_c\eta^{(\prime)}]$ is roughly $1:3$.

  \item For the isoscalar $D^*\bar{D}^*$ molecular state with $J^{PC}=1^{+-}$, it can decay into the $J/\psi\eta^{(\prime)}$ and $\eta_c\omega$ via the $S-$wave coupling. The relative decay branching ratio for the $J/\psi\eta^{(\prime)}$ and $\eta_c\omega$ channels is $\mathcal{B}_1=\Gamma_1[J/\psi\eta]/\Gamma_1[\eta_c\omega]=1:1$. Since the phase space for the $D^*\bar{D}^*\to J/\psi\eta$ is larger than that in the $\eta_c\omega$ final state around 100 MeV, the partial decay widths for these two hidden-charm decay processes satisfy $\Gamma_1[J/\psi\eta]>\Gamma_1[\eta_c\omega]$, it leads to the $J/\psi\eta$ channel is the prime decay channel to search for the isoscalar $D^*\bar{D}^*$ molecular state with $J^{PC}=1^{+-}$.

  \item For the isoscalar $D^*\bar{D}^*$ molecular state with $J^{PC}=2^{++}$, the $J/\psi\omega$  is the only one two-body hidden-charm decay mode by the $S-$wave interaction.
\end{itemize}

In addition, the isoscalar $D^*\bar{D}^*$ molecular state with $J^{PC}=0^{++}$ can strongly couple to the $D\bar{D}$ channel via the $S-$wave coupling, and the $D^*\bar{D}^*[0(2^{++})]$ can decay into the $D\bar{D}^*$ and $D\bar{D}$ channels through the $D$-wave interactions, which indicate that the two-body open-charm decay widths satisfy $\Gamma_{\text{Open}}[0(0^{++})]>\Gamma_{\text{Open}}[0(2^{++})]$. Thus, we can estimate that the strong decay width for the $D^{\ast}\bar{D}^{\ast}$ molecule with $I(J^{PC})=0(0^{++})$ is larger than that for the $D^{\ast}\bar{D}^{\ast}$ molecule with $I(J^{PC})=0(2^{++})$ \footnote{Here, we neglect the other decay modes with very small contribution, like the three-body decay modes, the two-body decay modes via the $D$-wave interaction.}. To summarize, if both the $D^{\ast}\bar{D}^{\ast}$ states with $I(J^{PC})=0(0^{++})$ and $0(2^{++})$ are the possible charmoniumlike molecular candidates, their mass and decay width should satisfy $M[0(0^{++})]<M[0(2^{++})]$ and $\Gamma[0(0^{++})]>\Gamma[0(2^{++})]$, respectively. These important characters on their mass spectrum and two-body strong decay behaviors provided here can help us to search and further identify the $D^{\ast}\bar{D}^{\ast}$ charmoniumlike molecules.

\begin{figure}[htbp]\centering
\includegraphics[width=7.5cm,keepaspectratio]{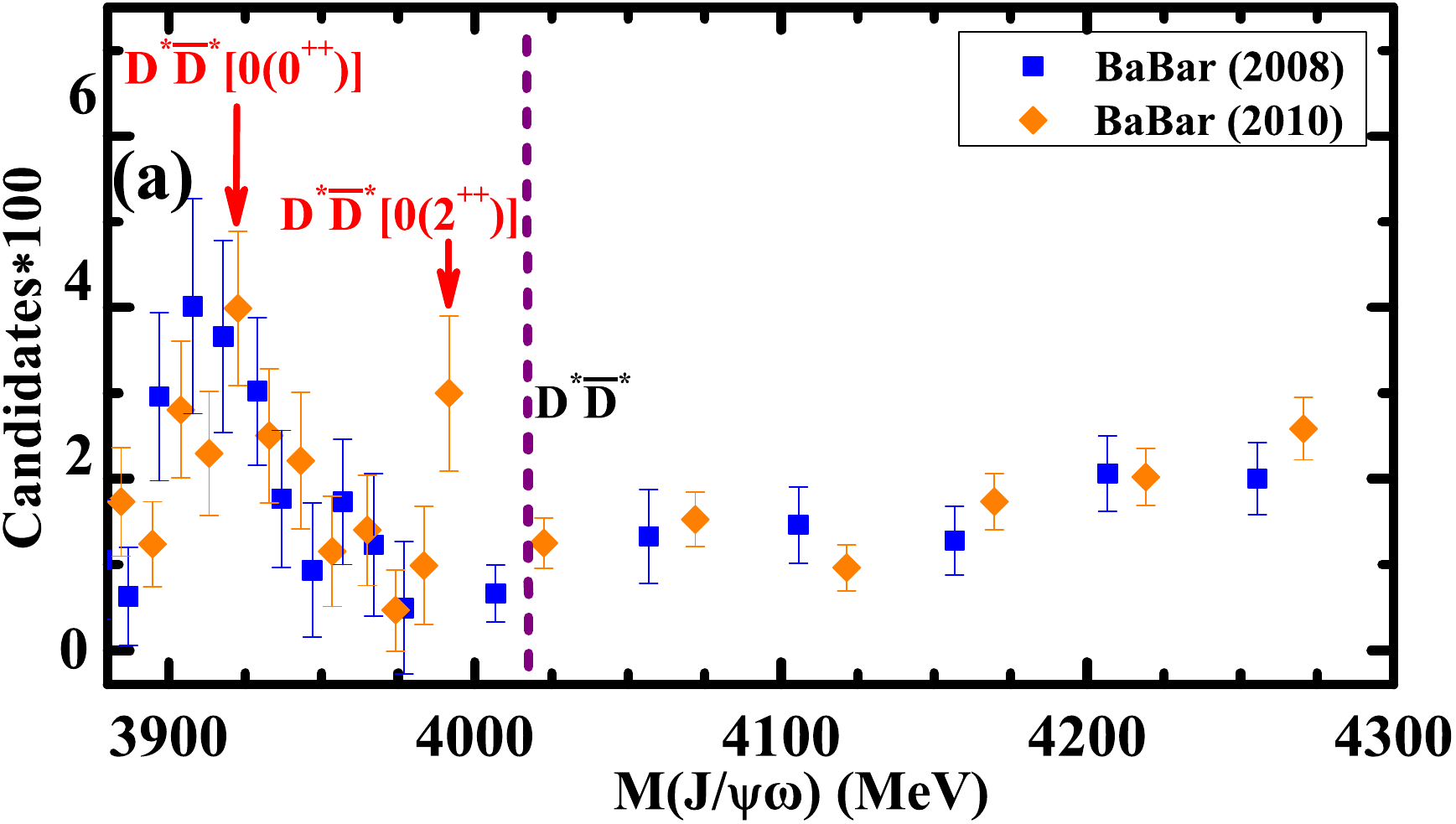}\quad
\includegraphics[width=7.5cm,keepaspectratio]{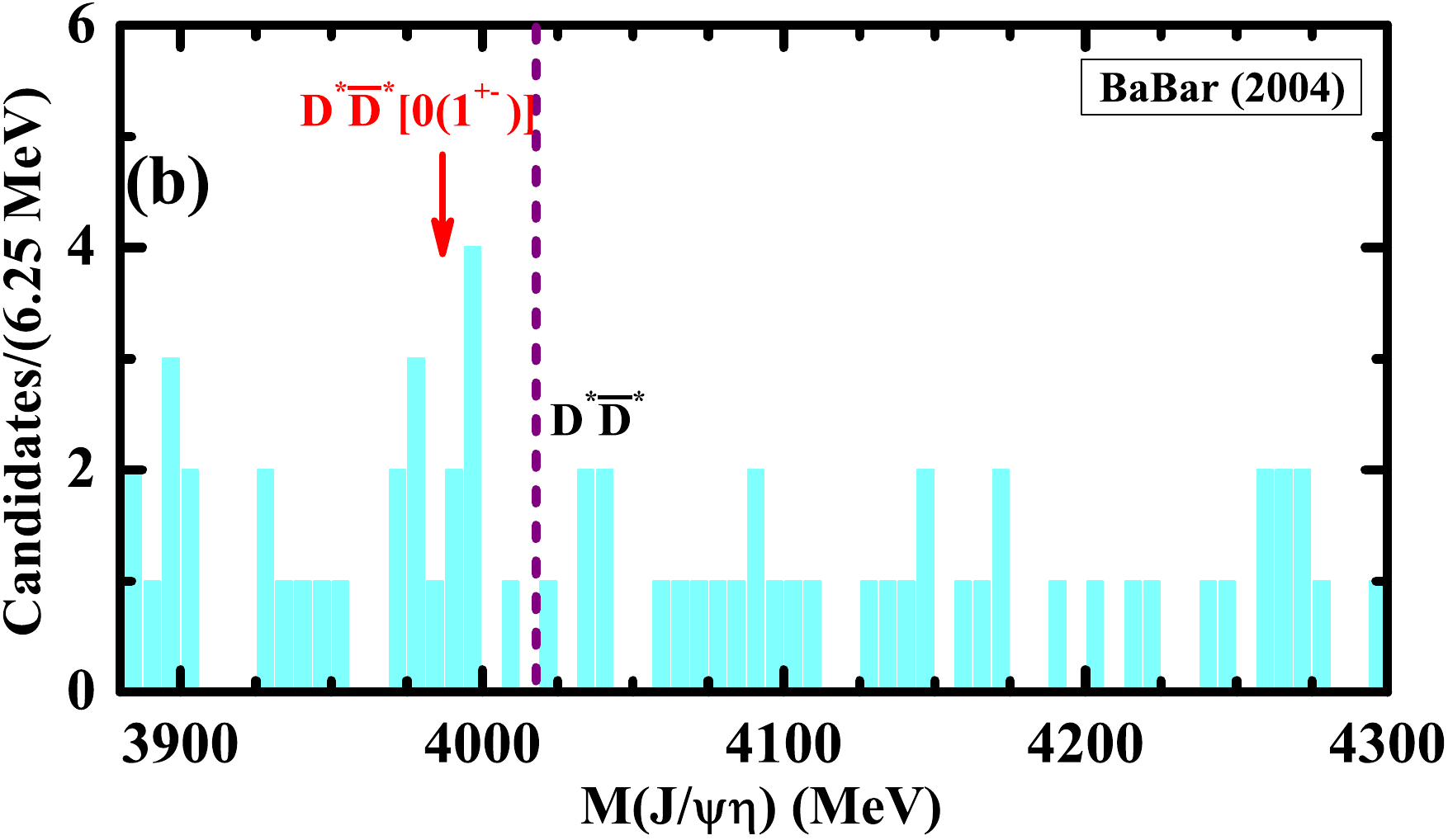}
\caption{The $J/\psi \omega$ and $J/\psi\eta$ invariant mass spectrum around the $D^*\bar{D}^*$ threshold in the $B \to  J/\psi \omega K$ \cite{Aubert:2007vj,delAmoSanchez:2010jr} and $B \to  J/\psi \eta K$ \cite{Aubert:2004fc}, respectively. }
\label{JpsiomegaetaDstarDstar}
\end{figure}
According to the above analysis, the $J/\psi\omega$ final state has the potential to observe the possible $D^{\ast}\bar{D}^{\ast}$ charmoniumlike molecules with $I(J^{PC})=0(0^{++})$ and $0(2^{++})$. Recalling the BABAR data presented in Fig. \ref{Jpsiomegaeta}, we can find possible evidence of the existence of two enhancement structures below the $D^*\bar{D}^*$ threshold by analyzing the $J/\psi\omega$ invariant mass spectrum of the $B\to J/\psi \omega K$ \cite{Aubert:2007vj,delAmoSanchez:2010jr}. In Fig. \ref{JpsiomegaetaDstarDstar} (a), we label the possible positions of the $D^{\ast}\bar{D}^{\ast}$ charmoniumlike molecules with $I(J^{PC})=0(0^{++})$ and $0(2^{++})$ in the $J/\psi \omega$ invariant mass spectrum in the $B \to  J/\psi \omega K$ \cite{Aubert:2007vj,delAmoSanchez:2010jr}. We look forward the future experiments with higher precision data can test our theoretical predictions.

Meanwhile, it is interesting to note that the $S-$wave $D^*\bar{D}^*$ state with $I(J^{PC})=0(1^{+-})$ is favored to be the possible charmoniumlike molecular candidate, and the $J/\psi\eta$ channel is the important two-body hidden-charm decay mode. Experimental, we may find an enhancement structure around 3.9 GeV in the $J/\psi\eta$ invariant mass spectrum of the $B\to J/\psi \eta K$ \cite{Aubert:2004fc}, which may correspond to the $S-$wave isoscalar $D^*\bar{D}^*$ molecular state with $J^{PC}=1^{+-}$ (see Fig. \ref{JpsiomegaetaDstarDstar} (b)). In addition, we notice that the BESIII Collaboration \cite{Ablikim:2012ht} and the Belle Collaboration \cite{Iwashita:2013wnn} respectively analyzed the $J/\psi\eta$ invariant mass spectrum in the $e^+ e^-$ annihilation process and the $B\to J/\psi \eta K$ process, unfortunately, no significant signal around 3.9 GeV was found. This could be because there are only several dozen events or less in the $J/\psi\eta$ invariant mass spectrum \cite{Aubert:2004fc,Ablikim:2012ht,Iwashita:2013wnn}, in comparison with the observations of the $J/\psi\omega$ invariant mass spectrum \cite{Abe:2004zs,Aubert:2007vj,delAmoSanchez:2010jr}, we expect more precision experimental data to further check the structure around 3.9 GeV in the $J/\psi\eta$ invariant mass spectrum.

In short, due to the lack of the sufficiently accurate experimental results, the present data sample from the $B$ meson decays was not large enough to analyze these possible charmoniumlike molecular tetraquark structures in the $J/\psi\omega$ and $J/\psi\eta$ invariant mass spectrum \cite{Abe:2004zs,Aubert:2007vj,delAmoSanchez:2010jr,Aubert:2004fc,Iwashita:2013wnn}, and the study of these possible charmoniumlike molecular tetraquark candidates will become an important research field at the precision frontier in future experiments. Thus, we strongly expect experimental colleagues to focus on the detailed structures around 3.9 GeV in the $J/\psi\omega$ and $J/\psi\eta$ invariant mass spectrum with more precise experimental data, like the LHCb, Belle II, and BESIII, it will provide strong evidence of existing charmoniumlike molecular tetraquark states if these possible enhancement structures can be confirmed in future experiments.

\paragraph{The others isoscalar $\mathcal{D}\bar{\mathcal{D}}$ systems.}
In this section, we mainly try to understand very broad structure around 4.3 GeV in the $J/\psi\omega$ invariant mass spectrum of the $B\to J/\psi\omega K$ process \cite{Abe:2004zs,Aubert:2007vj,delAmoSanchez:2010jr}. The mass thresholds of the $D^{\ast}\bar{D}_1$, $D^{\ast}\bar{D}^{\ast}_2$, $D\bar{D}^{\ast}_2$, and $D\bar{D}_1$ systems locate around this energy region. In general, the pion exchange interaction usually plays a crucial role in forming the hadronic molecular states \cite{Chen:2016qju}. In the following, we discuss these $S-$wave isoscalar charmoniumlike molecular tetraquark systems with three different groups, i.e., the $S-$wave isoscalar charmoniumlike molecular tetraquark systems with the pion exchange contribution occurring in the direct channel effective potentials, the $S-$wave isoscalar charmoniumlike molecular tetraquark systems with the pion exchange contribution occurring in the cross channel effective potentials, and the $S-$wave isoscalar charmoniumlike molecular tetraquark systems without the pion exchange process. For the sake of completeness, we also discuss the mass spectrum and the two-body hidden-charm decay channels for the $S-$wave isoscalar $D\bar{D}$ and $D\bar{D}^*$ systems.

\begin{description}
  \item[(i).] The $S-$wave isoscalar $D^{\ast}\bar{D}_1$ and $D^{\ast}\bar{D}^{\ast}_2$ systems.
\end{description}

By performing numerical calculations, we can obtain the loosely bound state solutions for the $S-$wave isoscalar $D^{\ast}\bar{D}_1$ and $D^{\ast}\bar{D}^{\ast}_2$ systems when the cutoff values are tuned from 0.8 to 3.0 GeV. In Table \ref{bound2}, we present the corresponding bound state solutions. Compared to the high spin states, we find that the low spin states can be easier to bind as charmoniumlike molecular candidates for the $S-$wave isoscalar $D^{\ast}\bar{D}_{1}$ and $D^{\ast}\bar{D}^{\ast}_{2}$ systems. Here, we also consider the coupled channel effect, and find the coupled channel effect plays a minor role in the above discussed systems. In fact, there are several papers on the predictions of the possible $S-$wave isoscalar $D^{\ast}\bar{D}_1$ and $D^{\ast}\bar{D}^{\ast}_2$ charmoniumlike molecular tetraquark states \cite{Close:2010wq,Li:2015exa,Close:2009ag,He:2017mbh,Li:2013bca,Zhu:2013sca}.
\renewcommand\tabcolsep{0.06cm}
\renewcommand{\arraystretch}{1.50}
\begin{table}[!htbp]
\caption{Bound state solutions for the $S-$wave isoscalar $D^{\ast}\bar{D}_1$ and $D^{\ast}\bar{D}^{\ast}_2$ systems. Conventions are the same as Table~\ref{bound1}.}\label{bound2}
\begin{tabular}{c|ccc|cccc }\toprule[1.0pt]\toprule[1.0pt]
\multicolumn{1}{c|}{Effect}&\multicolumn{3}{c|}{Single channel}&\multicolumn{4}{c}{$S$-$D$ wave mixing effect} \\\midrule[1.0pt]
\multicolumn{8}{c}{$D^{*}\bar D_1$}\\\midrule[1.0pt]
$J^{PC}$&$\Lambda$ &$E$  &$r_{\rm RMS}$ &$\Lambda$ &$E$  &$r_{\rm RMS}$ &$P({}^1\mathbb{S}_{0}/{}^5\mathbb{D}_{0})$   \\
\multirow{2}{*}{$0^{--}$}  &0.96&$-0.59$ &3.73         &0.95&$-0.52$ &3.92&\textbf{99.70}/0.30              \\
                                          &1.03&$-11.12$ &1.07        &1.03&$-12.40$ &1.03&\textbf{99.59}/0.41                                                       \\
\multirow{2}{*}{$0^{-+}$}  &0.92&$-0.56$ &3.91         &0.91&$-0.55$ &3.96&\textbf{99.53}/0.47                                                               \\
                                          &0.99&$-11.42$ &1.08        &0.99&$-12.82$ &1.05&\textbf{99.41}/0.59                                                           \\\midrule[1.0pt]
$J^{PC}$&$\Lambda$ &$E$  &$r_{\rm RMS}$ &$\Lambda$ &$E$  &$r_{\rm RMS}$ &$P({}^3\mathbb{S}_{1}/{}^3\mathbb{D}_{1}/{}^5\mathbb{D}_{1})$ \\
\multirow{2}{*}{$1^{--}$}  &1.10&$-0.48$ &4.11         &1.08&$-0.33$ &4.61&\textbf{99.64}/0.35/0.01                               \\
                                          &1.20&$-12.57$ &1.03        &1.19&$-12.74$ &1.03&\textbf{99.38}/0.61/0.01                                  \\
\multirow{2}{*}{$1^{-+}$}  &1.06&$-0.45$ &4.25         &1.04&$-0.34$ &4.68&\textbf{99.47}/0.52/0.01                                  \\
                                          &1.15&$-11.80$ &1.07        &1.14&$-11.75$ &1.09&\textbf{99.13}/0.86/0.01                                  \\\midrule[1.0pt]
$J^{PC}$&$\Lambda$ &$E$  &$r_{\rm RMS}$ &$\Lambda$ &$E$  &$r_{\rm RMS}$ &$P({}^5\mathbb{S}_{2}/{}^1\mathbb{D}_{2}/{}^3\mathbb{D}_{2}/{}^5\mathbb{D}_{2})$ \\
\multirow{2}{*}{$2^{--}$}  &2.56&$-0.32$ &4.89         &2.60&$-0.33$ &4.90&\textbf{98.90}/0.05/$o(0)$/1.05                                 \\
                                          &2.58&$-9.86$ &1.16        &2.74&$-8.65$ &1.33&\textbf{93.06}/4.18/$o(0)$/2.75                                    \\
\multirow{2}{*}{$2^{-+}$}  &1.73&$-0.81$ &3.68         &1.59&$-0.34$ &4.90&\textbf{99.30}/0.08/$o(0)$/0.62                                   \\
                                          &2.14&$-12.16$ &1.24        &2.07&$-12.74$ &1.24&\textbf{98.77}/0.27/$o(0)$/0.93                                 \\\midrule[1.0pt]

\multicolumn{8}{c}{$D^{*}\bar D_2^{*}$}\\\midrule[1.0pt]
$J^{PC}$&$\Lambda$ &$E$  &$r_{\rm RMS}$ &$\Lambda$ &$E$  &$r_{\rm RMS}$ &$P({}^3\mathbb{S}_{1}/{}^3\mathbb{D}_{1}/{}^5\mathbb{D}_{1}/{}^7\mathbb{D}_{1})$                               \\
\multirow{2}{*}{$1^{--}$}  &0.94&$-0.28$ &4.78         &0.94&$-0.56$ &3.90&\textbf{99.77}/0.07/$o(0)$/0.16                                                         \\
                                              &1.02&$-12.82$ &1.01        &1.02&$-13.86$ &0.99&\textbf{99.69}/0.05/$o(0)$/0.26                                                           \\
\multirow{2}{*}{$1^{-+}$}  &0.97&$-0.27$ &4.80         &0.97&$-0.51$ &4.00&\textbf{99.81}/0.09/$o(0)$/0.10                                                          \\
                                              &1.05&$-10.92$ &1.09        &1.05&$-11.81$ &1.06&\textbf{99.71}/0.19/$o(0)$/0.09                                                            \\\midrule[1.0pt]
$J^{PC}$&$\Lambda$ &$E$  &$r_{\rm RMS}$ &$\Lambda$ &$E$  &$r_{\rm RMS}$ &$P({}^5\mathbb{S}_{2}/{}^3\mathbb{D}_{2}/{}^5\mathbb{D}_{2}/{}^7\mathbb{D}_{2})$                              \\
\multirow{2}{*}{$2^{--}$}  &1.21&$-0.67$ &3.68         &1.19&$-0.28$ &4.84&\textbf{99.67}/$o(0)$/0.33/$o(0)$                                                              \\
                                              &1.36&$-12.76$ &1.04        &1.35&$-12.82$ &1.05&\textbf{99.38}/0.01/0.60/0.01                                                           \\
\multirow{2}{*}{$2^{-+}$}  &1.11&$-0.31$ &4.77         &1.10&$-0.28$ &4.84&\textbf{99.78}/$o(0)$/0.22/$o(0)$                                                                \\
                                              &1.21&$-11.55$ &1.08        &1.21&$-12.66$ &1.04&\textbf{99.59}/0.01/0.40/$o(0)$                                                             \\\midrule[1.0pt]
$J^{PC}$&$\Lambda$ &$E$  &$r_{\rm RMS}$ &$\Lambda$ &$E$  &$r_{\rm RMS}$ &$P({}^7\mathbb{S}_{3}/{}^3\mathbb{D}_{3}/{}^5\mathbb{D}_{3}/{}^7\mathbb{D}_{3})$                                \\
\multirow{2}{*}{$3^{--}$}  &1.90&$-0.32$ &4.94         &1.73&$-0.31$ &4.98&\textbf{99.21}/0.07/$o(0)$/0.72                                                             \\
                                              &2.87&$-12.26$ &1.23        &2.64&$-12.24$ &1.25&\textbf{98.36}/0.39/0.03/1.22                                                        \\
\multirow{2}{*}{$3^{-+}$}  &1.97&$-0.32$ &4.94         &1.77&$-0.33$ &4.91&\textbf{99.03}/0.04/$o(0)$/0.93                                                             \\
                                              &2.74&$-12.49$ &1.21        &2.51&$-12.41$ &1.24&\textbf{97.70}/0.19/0.03/2.07                                                            \\
\bottomrule[1.0pt]\bottomrule[1.0pt]
\end{tabular}
\end{table}

{In Fig.~\ref{bindingcutoff}, we present the cutoff parameter $\Lambda$ dependence of the binding energy $E$ for the $S$-wave $D^{\ast}\bar{D}_{1}$ state with $I(J^{PC})=0(0^{--})$, and there exist the loosely bound state solutions when the cutoff parameter is larger than 0.96 GeV, where the binding energy increases with the cutoff value monotonically. For simplicity, we only take $\Lambda$=0.96 and 1.03 GeV to present the loosely bound state solutions for the $S$-wave $D^{\ast}\bar{D}_{1}$ state with $I(J^{PC})=0(0^{--})$ in Table \ref{bound2}.
\begin{figure}[!htbp]
\centering
\includegraphics[width=0.40\textwidth]{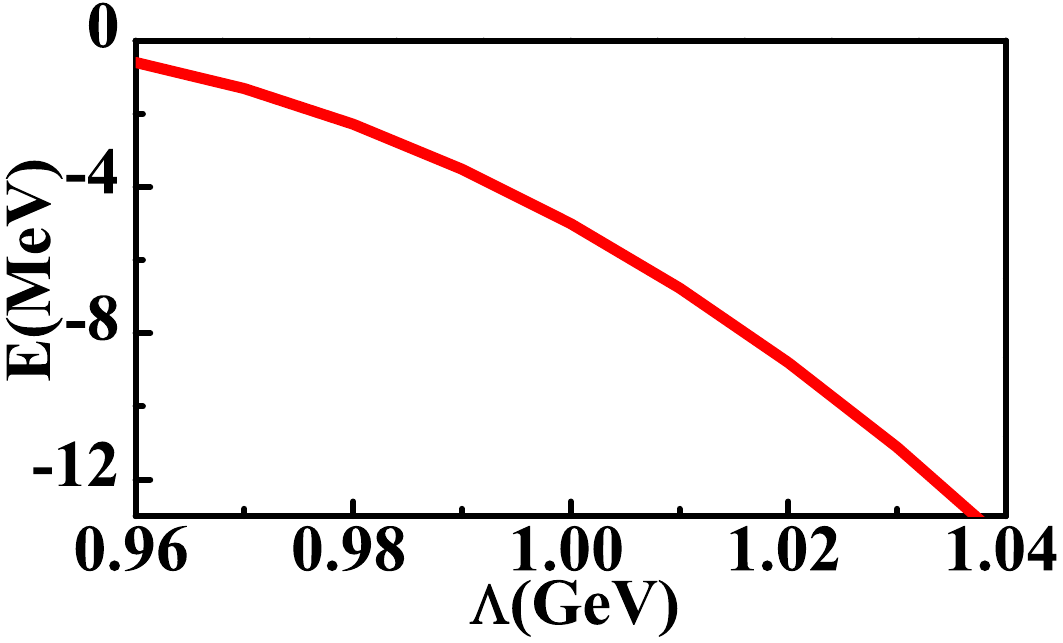}\\
\caption{(color online) The cutoff parameter $\Lambda$ dependence of the binding energy $E$ for the $S$-wave $D^{\ast}\bar{D}_{1}$ state with $I(J^{PC})=0(0^{--})$.}\label{bindingcutoff}
\end{figure}}

\begin{description}
  \item[(ii).] The $S-$wave isoscalar $D\bar{D}^{\ast}_2$ and $D\bar{D}^{\ast}$ systems.
\end{description}

For the $D\bar{D}^{\ast}_2$ and $D\bar{D}^{\ast}$ systems, the $\pi$ exchange occurs in the $D\bar{D}^{\ast}_2\to {D}^{\ast}_2\bar{D}$ and $D\bar{D}^{\ast}\to {D}^{\ast}\bar{D}$ processes, and the interaction Feynman diagram corresponds to the Cross diagram in Fig. \ref{fey}. {In Table \ref{bound3}, we collect the bound state solutions for the $S-$wave isoscalar $D\bar{D}^{\ast}_2$ and $D\bar{D}^{\ast}$ systems. It is obvious that these $S-$wave isoscalar $D\bar{D}^{\ast}_2$ and $D\bar{D}^{\ast}$ states can be possible charmoniumlike molecular candidates as their bound state solutions satisfy the typical characters for a loosely bound hadronic molecule\footnote{When the cutoff $\Lambda$ is taken around 1 GeV, the binding energy is around several to several tens MeV, and the size of bound state is larger than the size of its component.}. In fact, the $S-$wave isoscalar $D\bar{D}^*$ molecular states have been extensively studied in Refs. \cite{Wong:2003xk,Swanson:2003tb,Suzuki:2005ha,Liu:2008fh,Thomas:2008ja,Liu:2008tn,Lee:2009hy,Zhao:2014gqa,Li:2012cs,He:2014nya,Voloshin:2003nt,Close:2003sg,Tornqvist:2004qy,Sun:2011uh,Sun:2012zzd,Wang:2017dcq,Ding:2020dio,Sun:2017wgf,Yang:2017prf,Ding:2009vj,Zhang:2006ix,Liu:2017mrh,Tornqvist:1993ng,DeRujula:1976zlg}. However, our knowledge of the $S$-wave isoscalar $D\bar{D}^{\ast}_2$ molecular states is still not enough up to now \cite{Li:2015exa,Li:2013bca}.}
\renewcommand\tabcolsep{0.10cm}
\renewcommand{\arraystretch}{1.50}
\begin{table}[!htbp]
\caption{Bound state solutions for the $S-$wave isoscalar $D\bar{D}^{\ast}_2$ and $D\bar{D}^{\ast}$ systems. Conventions are the same as Table~\ref{bound1}.}\label{bound3}
\begin{tabular}{c|ccc|cccc}\toprule[1.0pt]\toprule[1.0pt]
Effect&\multicolumn{3}{c|}{Single channel}&\multicolumn{4}{c}{$S-$$D$ wave mixing effect}\\\midrule[1.0pt]
 $D\bar D_2^{*}[J^{PC}]$     &$\Lambda$ &$E$  &$r_{\rm RMS}$ &$\Lambda$ &$E$  &$r_{\rm RMS}$ &$P({}^5\mathbb{S}_{2}/{}^5\mathbb{D}_{2})$ \\
{$2^{--}$}  &1.46&$-0.25$ &5.08         &1.46&$-0.34$ &4.74&\textbf{99.99}/0.01        \\
                                          &1.85&$-12.22$ &1.11        &1.82&$-12.40$ &1.11&\textbf{99.91}/0.09          \\
{$2^{-+}$}  &1.30&$-0.36$ &4.69         &1.29&$-0.24$ &5.17&\textbf{99.99}/0.01          \\
                                          &1.47&$-12.03$ &1.11        &1.47&$-12.89$ &1.08&\textbf{99.93}/0.07            \\\midrule[1.0pt]

$D\bar D^{*}[J^{PC}]$&$\Lambda$ &$E$  &$r_{\rm RMS}$ &$\Lambda$ &$E$  &$r_{\rm RMS}$ &$P({}^3\mathbb{S}_{1}/{}^3\mathbb{D}_{1})$\\
\multirow{2}{*}{$1^{+-}$}  &1.62&$-0.36$ &4.70         &1.36&$-0.39$ &4.71&\textbf{97.38}/2.62         \\
                                        &1.77&$-12.51$ &1.07        &1.49&$-12.32$ &1.19&\textbf{92.15}/7.85             \\
\multirow{2}{*}{$1^{++}$}  &1.18&$-0.27$ &5.15         &1.08&$-0.27$ &5.22&\textbf{99.05}/0.95                                       \\
                                        &1.53&$-12.39$ &1.19        &1.30&$-12.09$ &1.23&\textbf{96.99}/3.01                                    \\
\bottomrule[1.0pt]\bottomrule[1.0pt]
\end{tabular}
\end{table}

\begin{description}
  \item[(iii).] The $S-$wave isoscalar $D\bar{D}_1$ and $D\bar{D}$ systems.
\end{description}

Despite the $\pi$ exchange does not contribute to the effective potentials for the $S-$wave $D\bar{D}_1$ system as the parity forbidden, the scalar and vector mesons exchange interactions may be strong enough to generate an bound state \cite{Chen:2017vai,Wang:2019nwt}. As shown in Table \ref{bound4}, our numerical results suggest the isoscalar $D\bar{D}_1$ states with $I(J^{PC})=0[(1^{--}), (1^{-+})]$ and the $S-$wave $D\bar{D}$ state with $I(J^{PC})=0(0^{++})$ can be possible charmoniumlike molecular candidates. In fact, the $S-$wave isoscalar $D\bar{D}_1$ molecular states were intensively discussed in Refs. \cite{Ding:2008gr,Dong:2019ofp,Close:2010wq,Li:2015exa,Close:2009ag,Li:2013bca,Wang:2020lua}, which may be related to the $Y(4260)$\footnote{In 2017, the BESIII gave more precise data of the $e^+e^-\to J/\psi\pi^+\pi^-$ \cite{Ablikim:2016qzw}, which shows that the $Y(4260)$ \cite{Aubert:2005rm} is split into two resonances $Y(4220)$ and $Y(4320)$.} \cite{Aubert:2005rm}. In Refs. \cite{Liu:2008tn,Liu:2008mi,Liu:2017mrh,Ding:2020dio,Yang:2017prf,Zhang:2006ix,DeRujula:1976zlg}, the $S-$wave $D\bar{D}$ bound state with $I(J^{PC})=0(0^{++})$ was estimated.
\renewcommand\tabcolsep{0.10cm}
\renewcommand{\arraystretch}{1.50}
\begin{table}[!htbp]
\caption{Bound state solutions for the $S-$wave isoscalar $D\bar{D}_1$ system. Conventions are the same as Table~\ref{bound1}.}\label{bound4}
\begin{tabular}{c|ccc|cccc}\toprule[1.0pt]\toprule[1.0pt]
Effect&\multicolumn{3}{c|}{Single channel}&\multicolumn{4}{c}{$S-$$D$ wave mixing effect} \\\midrule[1.0pt]
$D\bar D[J^{PC}]$&$\Lambda$ &$E$  &$r_{\rm RMS}$                 &    &        &    &                               \\
\multirow{2}{*}{$0^{++}$}  &1.46&$-0.29$ &5.08         &    &        &    &                               \\
                                    &1.76&$-12.55$ &1.15        &    &        &    &                               \\\midrule[1.0pt]
$D\bar D_1[J^{PC}]$&$\Lambda$ &$E$  &$r_{\rm RMS}$ &$\Lambda$ &$E$  &$r_{\rm RMS}$ &$P({}^3\mathbb{S}_{1}/{}^3\mathbb{D}_{1})$ \\
{$1^{--}$}  &1.39&$-0.36$ &4.72         &1.39&$-0.39$ &4.60&\textbf{99.98}/0.02                   \\
                                      &1.67&$-12.13$ &1.14        &1.67&$-12.42$ &1.13&\textbf{99.92}/0.08                    \\
{$1^{-+}$}  &1.38&$-0.29$ &4.92         &1.38&$-0.32$ &4.80&\textbf{99.98}/0.02                 \\
                                      &1.63&$-12.63$ &1.09        &1.63&$-12.89$ &1.08&\textbf{99.93}/0.07                 \\
\bottomrule[1.0pt]\bottomrule[1.0pt]
\end{tabular}
\end{table}

Let's give a short summary, we can predict a serial of possible charmoniumlike molecules composed by the $S-$wave isoscalar $D^{\ast}\bar{D}_1$, $D^{\ast}\bar{D}^{\ast}_2$, $D\bar{D}^{\ast}_2$, $D\bar{D}^{\ast}$, $D\bar{D}_1$, and $D\bar{D}$ systems. In Table \ref{sum1}, we summary their two-body hidden-charm decay information for all the possible $S-$wave isoscalar $D^{\ast}\bar{D}_1$, $D^{\ast}\bar{D}^{\ast}_2$, $D\bar{D}^{\ast}_2$, $D\bar{D}^{\ast}$, $D\bar{D}_1$, and $D\bar{D}$ charmoniumlike molecules. For example, the $J/\psi\omega$ channel is the two-body hidden-charm decay mode for the isoscalar $D^{*}\bar D_1[(0,1,2)^{-+}]$, $D^{*}\bar D_2^{*}[3^{-+}]$, and $D\bar{D}_1[1^{-+}]$ bound states, perhaps, it is possible to observe the experimental signal of these possible charmoniumlike molecules in the $J/\psi\omega$ final state. When we recall the experimental data of the $B\to J/\psi \omega K$ \cite{Abe:2004zs,Aubert:2007vj,delAmoSanchez:2010jr}, our predictions of these possible charmoniumlike molecules also reflect the complexity of the structures in the $J/\psi\omega$ invariant mass spectrum around 4.3 GeV. At present, it is a little difficult to definitely identify these possible charmoniumlike molecules, we hope further experiments can provide more precise measurement of the $B$ meson decay into a charmonium state plus a light-flavor meson. In addition, the $\chi_{cJ}(1P)\omega$ with $J=0,1,2$ are the two-body hidden-charm decay channels for the possible $S-$wave isoscalar $D^{\ast}\bar{D}_1$, $D^{\ast}\bar{D}^{\ast}_2$, $D\bar{D}^{\ast}_2$, and $D\bar{D}_1$ systems of negative $C$-parity in our calculations.
\renewcommand\tabcolsep{0.23cm}
\renewcommand{\arraystretch}{1.50}
\begin{table}[!htbp]\centering
\caption{A summary of the two-body hidden-charm decay channels for all the possible $S-$wave isoscalar $D^{\ast}\bar{D}_1$, $D^{\ast}\bar{D}^{\ast}_2$, $D\bar{D}^{\ast}_2$, $D\bar{D}^{\ast}$, $D\bar{D}_1$, and $D\bar{D}$ charmoniumlike molecules.}\label{sum1}
\begin{tabular}{c|c}\toprule[1.0pt]\toprule[1.0pt]
States   & Two-body hidden-charm decay channels  \\\midrule[1.0pt]
$D^{*}\bar D_1[0^{-+}]$        & ${J/\psi\omega}$,~$\chi_{c0}(1P)\eta$,~$\chi_{c0}(1P)\eta^{\prime}$    \\
$D^{*}\bar D_1[0^{--}]$               & ${\chi_{c1}(1P)\omega}$, $\eta_c\omega$, $J/\psi\eta$, $J/\psi\eta^{\prime}$  \\
$D^{*}\bar D_1[1^{-+}]$           & ${J/\psi\omega}$,~$\chi_{c1}(1P)\eta$, $\eta_c\eta$, $\eta_c\eta^{\prime}$,      \\
$D^{*}\bar D_1[1^{--}]$            & ${\chi_{c2}(1P)\omega}$,~$\chi_{c1}(1P)\omega$,~$\chi_{c0}(1P)\omega$, $\eta_c\omega$, $J/\psi\eta$, $J/\psi\eta^{\prime}$ \\
$D^{*}\bar D_1[2^{-+}]$           & ${J/\psi\omega}$,~$\chi_{c2}(1P)\eta$  \\
$D^{*}\bar D_1[2^{--}]$             & ${\chi_{c2}(1P)\omega}$,~$\chi_{c1}(1P)\omega$, $\eta_c\omega$, $J/\psi\eta$, $J/\psi\eta^{\prime}$   \\

$D^{*}\bar D_2^{*}[1^{-+}]$           & ${\chi_{c1}(1P)\eta}$, $\eta_c\eta$, $\eta_c\eta^{\prime}$, $J/\psi\omega$, $\chi_{c1}(1P)\eta^{\prime}$  \\
$D^{*}\bar D_2^{*}[1^{--}]$             & ${\chi_{c2}(1P)\omega}$, $\eta_c\omega$, $J/\psi\eta$, $J/\psi\eta^{\prime}$, $\chi_{c0}(1P)\omega$, $\chi_{c1}(1P)\omega$ \\
$D^{*}\bar D_2^{*}[2^{-+}]$      & ${\chi_{c2}(1P)\eta}$,~$J/\psi\omega$ \\
$D^{*}\bar D_2^{*}[2^{--}]$        & ${\chi_{c2}(1P)\omega}$, $\eta_c\omega$, $J/\psi\eta$, $J/\psi\eta^{\prime}$, $\chi_{c1}(1P)\omega$\\
$D^{*}\bar D_2^{*}[3^{-+}]$       & ${J/\psi\omega}$    \\
$D^{*}\bar D_2^{*}[3^{--}]$       & ${\chi_{c2}(1P)\omega}$ \\

$D\bar D_2^{*}[2^{-+}]$           & ${\chi_{c2}(1P)\eta}$,~$J/\psi\omega$    \\
$D\bar D_2^{*}[2^{--}]$              & ${\chi_{c1}(1P)\omega}$, $\eta_c\omega$, $J/\psi\eta$, $J/\psi\eta^{\prime}$  \\

$D\bar D^{*}[1^{++}]$           & $J/\psi\omega$    \\
$D\bar D^{*}[1^{+-}]$              & $\eta_c\omega$, $J/\psi\eta$  \\

$D\bar D_1[1^{-+}]$     & ${J/\psi\omega}$,~$\eta_{c}\eta^{\prime}$,~$\eta_{c}\eta$, $\chi_{c1}(1P)\eta$\\
$D\bar D_1[1^{--}]$             & ${\chi_{c0}(1P)\omega}$, $\eta_c\omega$, $J/\psi\eta$, $J/\psi\eta^{\prime}$\\

$D\bar D[0^{++}]$     & $\eta_{c}\eta$\\
\bottomrule[1.0pt]\bottomrule[1.0pt]
\end{tabular}
\end{table}
\subsubsection{Charmoniumlike molecular tetraquark systems with hidden-strange quantum number}

With the help of the $\rm{SU}(3)$ flavor symmetry, we can further study the interactions between the charm-strange meson and anti-charm-strange meson, in this section, we analyze the existence probability of the charmoniumlike molecules with hidden-strange quantum number and give their two-body hidden-charm decay channels.

\paragraph{The $D_s^*\bar{D}_s^*$ system.} For the $D_s^{*} \bar D_s^{*}$ system, the $\eta$ and $\phi$ exchanges contribute to the effective potentials in the OBE model. The relevant numerical results for the $S-$wave $D_s^{\ast}\bar D_s^{\ast}$ system are given in Table~\ref{bound5}, and the cutoff parameters are taken in the range from 1.0 to 3.0 GeV.
\renewcommand\tabcolsep{0.13cm}
\renewcommand{\arraystretch}{1.50}
\begin{table}[!htbp]
\caption{Bound state solutions for the $S-$wave $D_s^{\ast}\bar D_s^{\ast}$ system. Conventions are the same as Table~\ref{bound1}.}\label{bound5}
\begin{tabular}{c|ccc|cccc}\toprule[1.0pt]\toprule[1.0pt]
\multicolumn{1}{c|}{Effect}&\multicolumn{3}{c|}{Single channel}&\multicolumn{4}{c}{$S-$$D$ wave mixing effect}\\\midrule[1.0pt]
$J^{PC}$&$\Lambda$ &$E$  &$r_{\rm RMS}$ &$\Lambda$ &$E$  &$r_{\rm RMS}$ &$P({}^1\mathbb{S}_{0}/{}^5\mathbb{D}_{0})$ \\
\multirow{2}{*}{$0^{++}$}  &1.59&$-0.72$ &3.49         &1.58&$-0.23$ &4.98&\textbf{99.98}/0.02                                       \\
                                                &1.65&$-11.09$ &0.98        &1.65&$-11.34$ &0.97&\textbf{99.95}/0.05                                      \\\midrule[1.0pt]
$J^{PC}$&$\Lambda$ &$E$  &$r_{\rm RMS}$ &$\Lambda$ &$E$  &$r_{\rm RMS}$ &$P({}^3\mathbb{S}_{1}/{}^3\mathbb{D}_{1})$                           \\
\multirow{2}{*}{$1^{+-}$}  &1.89&$-0.52$ &4.00         &1.88&$-0.26$ &4.89&\textbf{99.98}/0.02                                       \\
                                                &2.00&$-12.37$ &0.95        &2.00&$-12.52$ &0.94&\textbf{99.96}/0.04                       \\
\bottomrule[1.0pt]\bottomrule[1.0pt]
\end{tabular}
\end{table}

For the $S-$wave $D_s^{*} \bar D_s^{*}$ states with $J^{PC}=0^{++}$ and $1^{+-}$, we can obtain the loosely bound state solutions for these states when the cutoff values are taken to be around 1.6 GeV and 1.9 GeV, respectively. Thus, the $S-$wave $D_s^{*} \bar D_s^{*}$ states with $J^{PC}=0^{++}$ and $1^{+-}$ can be the possible hidden-charm and hidden-strange molecular tetraquark candidates, especially the $D_s^{*} \bar D_s^{*}$ molecular state with $J^{PC}=0^{++}$ \cite{Ding:2009vd,Meng:2020cbk}. If taking a large cutoff $\Lambda>3.0$ GeV, there exist the loosely bound state solutions for the $D_s^{*} \bar D_s^{*}$ state with $J^{PC}=2^{++}$ in Ref. \cite{Liu:2009ei}. However, such cutoff parameter is far from the usual value around 1.0 GeV \cite{Tornqvist:1993ng,Tornqvist:1993vu,Wang:2019nwt}, which is consistent with our numerical results.

\paragraph{The $D_s\bar{D}_s$, $D_s\bar D_s^{\ast}$, $D_s\bar D_{s0}^{\ast}$, $D_s\bar D_{s1}^{\prime}$, $D_s^{\ast}\bar D_{s0}^{\ast}$, $D_s^{\ast}\bar D_{s1}^{\prime}$, ${D}_s^{(*)}\bar{D}_{s1}$, and $D_s^{(*)}\bar{D}_{s2}^*$ systems.} Besides the $S-$wave $D_s^{*} \bar D_s^{*}$ system, we also investigate the bound state properties of the $S-$wave $D_s\bar{D}_s$, $D_s\bar D_s^{\ast}$, $D_s\bar D_{s0}^{\ast}$, $D_s\bar D_{s1}^{\prime}$, $D_s^{\ast}\bar D_{s0}^{\ast}$, and $D_s^{\ast}\bar D_{s1}^{\prime}$ systems by tuning the cutoff values $\Lambda$ from 1.0 to 3.0 GeV, and the corresponding numerical results are listed in Table~\ref{bound6}. If we still adopt the general criterion of the loosely hadron-hadron molecule \cite{Wang:2020dya,Tornqvist:1993ng,Tornqvist:1993vu,Wang:2019nwt}, we can find
\begin{itemize}
  \item There may exist several possible charmoniumlike molecular tetraquark candidates with hidden-strange quantum number, such as the $D_s\bar D_{s}^{\ast}$ state with $J^{PC}=1^{+-}$ \cite{Liu:2017mrh,Meng:2020cbk}, the $D_s\bar D_{s0}^{\ast}$ state with $J^{PC}=0^{-\mp}$ \cite{Shen:2010ky,Liu:2010hf,He:2013oma,He:2016pfa}, the $D_s\bar{D}_{s1}^{\prime}$ state with $J^{PC}=1^{--}$, the $D_s^{\ast}\bar D_{s1}^{\prime}$ states with $J^{PC}=0^{-\mp}$ and $1^{-+}$. In particular, the coupled channel effect plays an important role in generating these loosely bound states, i.e., the $D_s\bar D_{s}^{\ast}$ state with $J^{PC}=1^{+-}$, the $D_s\bar D_{s0}^{\ast}$ state with $J^{PC}=0^{-\mp}$, and the $D_s\bar{D}_{s1}^{\prime}$ state with $J^{PC}=1^{--}$.

  \item If we increase the cutoff value around 2.0 GeV, we can obtain loosely bound state solutions for the $D_s \bar D_{s}^{\ast}$ state with $J^{PC}=1^{++}$, the $D_s^{\ast}\bar D_{s0}^{\ast}$ state with $J^{PC}=1^{--}$, and the $D_s^{\ast}\bar D_{s1}^{\prime}$ state with $J^{PC}=1^{--}$. They may be the possible charmoniumlike molecular candidates with hidden-strange quantum number.

  \item In addition, we do not obtain bound state solutions for the $D_s\bar{D}_s$ state with $J^{PC}=0^{++}$, the $D_s\bar D_{s1}^{\prime}$ state with $J^{PC}=1^{-+}$, and the $D_s^{\ast}\bar D_{s1}^{\prime}$ state with $J^{PC}=2^{-\mp}$ by tuning cutoff values from $1.0$ to $3.0$ GeV.
\end{itemize}
\renewcommand\tabcolsep{0.20cm}
\renewcommand{\arraystretch}{1.50}
\begin{table*}[!htbp]
\caption{Bound state solutions for the $S-$wave $D_s\bar D_s^{\ast}$, $D_s\bar D_{s0}^{\ast}$, $D_s\bar D_{s1}^{\prime}$, $D_s^{\ast}\bar D_{s0}^{\ast}$, and $D_s^{\ast}\bar D_{s1}^{\prime}$ systems. Conventions are the same as Table~\ref{bound1}.}\label{bound6}
\begin{tabular}{c|ccc|cccc|cccc}\toprule[1.0pt]\toprule[1.0pt]
\multicolumn{1}{c|}{Effect}&\multicolumn{3}{c|}{Single channel}&\multicolumn{4}{c|}{$S-$$D$ wave mixing effect}&\multicolumn{4}{c}{Coupled channel effect}\\\midrule[1.0pt]
States$[J^{PC}]$&$\Lambda$ &$E$  &$r_{\rm RMS}$ &$\Lambda$ &$E$  &$r_{\rm RMS}$ &$P({}^3\mathbb{S}_{1}/{}^3\mathbb{D}_{1})$&$\Lambda$ &$E$  &$r_{\rm RMS}$ &$P(D_s\bar D_s^{*}/D_s^{*}\bar D_s^{*})$\\
\multirow{2}{*}{$D_s\bar D_s^{*}[1^{+-}]$}  &2.65&$-0.26$ &4.96         &2.27&$-0.43$ &4.39&\textbf{99.25}/0.75         &1.63&$-0.13$ &5.41&\textbf{96.64}/3.36  \\
                                            &2.85&$-11.95$ &0.98        &2.43&$-11.85$ &1.03&\textbf{96.56}/3.44        &1.66&$-8.92$ &0.99&\textbf{78.85}/21.15     \\
\multirow{2}{*}{$D_s\bar D_s^{*}[1^{++}]$}  &$\times$&$\times$ &$\times$         &2.71&$-0.27$ &4.99&\textbf{99.46}/0.54         &    &        &    &                               \\
                                            &$\times$&$\times$ &$\times$         &3.00&$-5.69$ &1.49&\textbf{97.69}/2.31        &    &        &    &                               \\\midrule[1.0pt]

States$[J^{PC}]$&$\Lambda$ &$E$  &$r_{\rm RMS}$                 &    &        &    &                                 &$\Lambda$ &$E$  &$r_{\rm RMS}$ &$P(D_s\bar D_{s0}^*/D_s^{*}\bar  D_{s1}^{\prime})$\\
\multirow{2}{*}{$D_s\bar D_{s0}^*[0^{--}]$}  &$\times$&$\times$ &$\times$         &    &        &    &               &1.87&$-2.52$ &1.77&\textbf{85.11}/14.89 \\
                                        &$\times$&$\times$ &$\times$         &    &        &    &                    &1.88&$-8.21$ &0.94&\textbf{75.72}/24.48    \\
\multirow{2}{*}{$D_s\bar D_{s0}^*[0^{-+}]$}  &2.83&$-0.28$ &4.93         &    &        &    &                        &1.49&$-0.43$ &4.27&\textbf{96.97}/3.03  \\
                                             &3.00&$-0.92$ &3.32         &    &        &    &                        &1.52&$-10.05$ &1.00&\textbf{81.81}/18.19    \\\midrule[1.0pt]

State$[J^{PC}]$&$\Lambda$ &$E$  &$r_{\rm RMS}$ &$\Lambda$ &$E$  &$r_{\rm RMS}$ &$P({}^3\mathbb{S}_{1}/{}^3\mathbb{D}_{1})$&$\Lambda$ &$E$  &$r_{\rm RMS}$ &$P(D_s\bar D_{s1}^{\prime}/D_s^*\bar D_{s0}^*/D_s^{*}\bar  D_{s1}^{\prime})$\\
\multirow{2}{*}{$D_s\bar D_{s1}^{\prime}[1^{--}]$}  &2.17&$-0.33$ &4.65         &2.15&$-0.29$ &4.79&\textbf{99.98}/0.02              &1.87&$-0.31$ &4.69&\textbf{98.76}/1.24/$o(0)$\\
                                                    &2.55&$-12.60$ &1.00        &2.51&$-12.35$ &1.00&\textbf{99.82}/0.18             &1.99&$-12.64$ &0.92&\textbf{87.08}/12.92/$o(0)$\\\midrule[1.0pt]

State$[J^{PC}]$&$\Lambda$ &$E$  &$r_{\rm RMS}$ &$\Lambda$ &$E$  &$r_{\rm RMS}$ &$P({}^3\mathbb{S}_{1}/{}^3\mathbb{D}_{1})$&$\Lambda$ &$E$  &$r_{\rm RMS}$ &$P(D_s^*\bar D_{s0}^*/D_s^{*}\bar  D_{s1}^{\prime})$\\
\multirow{2}{*}{$D_s^*\bar D_{s0}^*[1^{--}]$}  &2.17&$-0.31$ &4.73         &2.17&$-0.31$ &4.72&\textbf{100.00}/$o(0)$         &2.17&$-0.31$ &4.72&\textbf{100.00}/$o(0)$  \\
                                               &2.53&$-12.42$ &1.00        &2.53&$-12.42$ &1.00&\textbf{100.00}/$o(0)$        &2.53&$-12.42$ &1.00&\textbf{100.00}/$o(0)$     \\\midrule[1.0pt]

States$[J^{PC}]$&$\Lambda$ &$E$  &$r_{\rm RMS}$ &$\Lambda$ &$E$  &$r_{\rm RMS}$ &$P({}^1\mathbb{S}_{0}/{}^5\mathbb{D}_{0})$&    &        &    &                               \\
\multirow{2}{*}{$D_s^{*}\bar  D_{s1}^{\prime}[0^{--}]$}  &1.37&$-0.41$ &4.33         &1.37&$-0.52$ &3.99&\textbf{99.96}/0.04         &    &        &    &                               \\
                                                         &1.45&$-12.34$ &0.98        &1.45&$-12.82$ &0.97&\textbf{99.90}/0.10        &    &        &    &                               \\
\multirow{2}{*}{$D_s^{*}\bar  D_{s1}^{\prime}[0^{-+}]$}  &1.87&$-0.33$ &4.24         &1.87&$-0.36$ &4.13&\textbf{99.99}/0.01         &    &        &    &                               \\
                                                         &1.91&$-10.27$ &0.85        &1.91&$-10.40$ &0.85&\textbf{99.99}/0.01        &    &        &    &                               \\ \midrule[1.0pt]
States$[J^{PC}]$&$\Lambda$ &$E$  &$r_{\rm RMS}$ &$\Lambda$ &$E$  &$r_{\rm RMS}$ &$P({}^3\mathbb{S}_{1}/{}^3\mathbb{D}_{1}/{}^5\mathbb{D}_{1})$&    &        &    &                               \\
\multirow{2}{*}{$D_s^{*}\bar  D_{s1}^{\prime}[1^{--}]$}  &2.29&$-0.59$ &3.53         &2.28&$-0.47$ &3.87&\textbf{99.97}/0.03/$o(0)$         &    &        &    &                               \\
                                                         &2.37&$-13.38$ &0.78        &2.36&$-13.47$ &0.79&\textbf{99.88}/0.12/$o(0)$        &    &        &    &                               \\
\multirow{2}{*}{$D_s^{*}\bar  D_{s1}^{\prime}[1^{-+}]$}  &1.55&$-0.22$ &5.10         &1.55&$-0.36$ &4.52&\textbf{99.95}/0.05/$o(0)$         &    &        &    &                               \\
                                                         &1.68&$-11.38$ &1.02        &1.68&$-12.17$ &1.00&\textbf{99.84}/0.16/$o(0)$        &    &        &    &                               \\
\bottomrule[1.0pt]\bottomrule[1.0pt]
\end{tabular}
\end{table*}

In our previous work \cite{Wang:2020dya}, we systematic study the interactions between a pair of charm-strange meson and anti-charm-strange meson in the $H$-doublet or $T$-doublet by using the OBE model and considering the $S-$$D$ wave mixing and the coupled channel effect, and we can predict several possible $H\bar{T}$-type charmoniumlike molecular states, i.e., the $D^{\ast}_s\bar{D}_{s1}$ molecular states with $J^{PC}=0^{-\pm}/1^{-\pm}$ and the $D^{\ast}_s\bar{D}^{\ast}_{s2}$ molecular states with $J^{PC}=1^{-\pm}/2^{-\pm}$.

In Table \ref{sum2}, we summary their bound properties and two-body hidden-charm decay channels for all the investigated possible $S-$wave charmoniumlike molecules with hidden-strange quantum number.
\renewcommand\tabcolsep{1.20cm}
\renewcommand{\arraystretch}{1.50}
\begin{table*}[!htbp]\centering
\caption{A summary of the bound state properties and two-body hidden-charm decay channels for all the investigated possible $S-$wave charmoniumlike molecules with hidden-strange quantum number. Here, notations $\surd$ and $\surd\!\!\!\setminus$ are marked the charmoniumlike molecular candidates with their bound state solutions with the cutoff $\Lambda$ around 1 to 2 GeV and around 2 to 3 GeV, respectively.}\label{sum2}
\begin{tabular}{c|c|c}\toprule[1.0pt]\toprule[1.0pt]
States & Bound state properties   & Two-body hidden-charm decay channels \\\midrule[1.0pt]
$D_s^*\bar{D}_s^*[0^{++}]$&   $\surd$ & ${\eta_c\eta^{\prime}}$,~$J/\psi\phi$, ${\eta_c\eta}$,~$\chi_{c1}(1P)\eta$\\
$D_s^*\bar{D}_s^*[1^{+-}]$&   $\surd$  & ${J/\psi\eta^{\prime}}$,~$\eta_c \phi$,~$J/\psi \eta$\\

$D_s\bar D_s^{*}[1^{++}]$&    $\surd\!\!\!\setminus$   & ${\chi_{c0}(1P)\eta}$, ${\chi_{c1}(1P)\eta}$ \\
$D_s\bar D_s^{*}[1^{+-}]$&   $\surd$    & ${\eta_c \phi}$,~$J/\psi\eta$,~$J/\psi\eta^{\prime}$\\

$D_s\bar D_{s0}^*[0^{-+}]$&   $\surd$    & ${J/\psi \phi}$,~$\chi_{c0}(1P)\eta$\\
$D_s\bar D_{s0}^*[0^{--}]$&   $\surd$    & ${\eta_c \phi}$,~$J/\psi\eta^{\prime}$,~$J/\psi\eta$\\

$D_s\bar D_{s1}^{\prime}[1^{--}]$&   $\surd$     & $J/\psi\eta^{\prime}$,~$J/\psi\eta$, ${\eta_c \phi}$\\

$D_s^{*}\bar D_{s0}^{*}[1^{--}]$&   $\surd\!\!\!\setminus$    & ${\eta_c \phi}$,~$J/\psi\eta^{\prime}$,~$J/\psi\eta$\\

$D_s^{*}\bar  D_{s1}^{\prime}[0^{-+}]$&  $\surd$ & ${J/\psi \phi}$,~$\chi_{c0}(1P)\eta^{\prime}$,~$\chi_{c0}(1P)\eta$\\
$D_s^{*}\bar  D_{s1}^{\prime}[0^{--}]$& $\surd$  & ${\chi_{c1}(1P)\phi}$,~$J/\psi\eta^{\prime}$,~$J/\psi\eta$, ${\eta_c \phi}$\\
$D_s^{*}\bar  D_{s1}^{\prime}[1^{-+}]$& $\surd$ & ${\chi_{c1}(1P)\eta^{\prime}}$,~$J/\psi \phi$,~$\chi_{c1}(1P)\eta$, ${\eta_c \eta}$, ${\eta_c \eta^{\prime}}$\\
$D_s^{*}\bar  D_{s1}^{\prime}[1^{--}]$&  $\surd\!\!\!\setminus$ & ${\chi_{c0}(1P)\phi}$,~$\chi_{c1}(1P)\phi$,~$J/\psi\eta^{\prime}$,~$J/\psi\eta$, ${\eta_c \phi}$\\

$D^{\ast}_s\bar{D}_{s1}[0^{-+}]$&      $\surd$      & ${\chi_{c0}(1P)\eta^{\prime}}$,~$\chi_{c0}(1P)\eta$,~$J/\psi \phi$\\
$D^{\ast}_s\bar{D}_{s1}[0^{--}]$&      $\surd$       & ${\chi_{c1}(1P)\phi}$,~$J/\psi\eta^{\prime}$,~$J/\psi\eta$, ${\eta_c \phi}$  \\
$D^{\ast}_s\bar{D}_{s1}[1^{-+}]$&      $\surd$        & ${J/\psi \phi}$, ${\chi_{c1}(1P)\eta^{\prime}}$,~$\chi_{c1}(1P)\eta$, ${\eta_c \eta}$, ${\eta_c \eta^{\prime}}$\\
$D^{\ast}_s\bar{D}_{s1}[1^{--}]$&      $\surd$        & ${\chi_{c0}(1P)\phi}$,~$\chi_{c2}(1P)\phi$,~$\chi_{c1}(1P)\phi$,~$J/\psi\eta^{\prime}$,~$J/\psi\eta$, ${\eta_c \phi}$\\

$D^{\ast}_s\bar{D}^{\ast}_{s2}[1^{-+}]$&  $\surd$   & ${\chi_{c1}(1P)\eta^{\prime}}$, $J/\psi \phi$,~$\chi_{c1}(1P)\eta$, ${\eta_c \eta}$, ${\eta_c \eta^{\prime}}$ \\
$D^{\ast}_s\bar{D}^{\ast}_{s2}[1^{--}]$&  $\surd$ & ${\chi_{c2}(1P)\phi}$,~$\chi_{c1}(1P)\phi$,~$\chi_{c0}(1P)\phi$,~$J/\psi\eta^{\prime}$,~$J/\psi\eta$, ${\eta_c \phi}$  \\
$D^{\ast}_s\bar{D}^{\ast}_{s2}[2^{-+}]$&  $\surd$ & ${\chi_{c2}(1P)\eta^{\prime}}$,~$\chi_{c2}(1P)\eta$, ${J/\psi \phi}$     \\
$D^{\ast}_s\bar{D}^{\ast}_{s2}[2^{--}]$&  $\surd\!\!\!\setminus$   & ${\chi_{c2}(1P)\phi}$,~$\chi_{c1}(1P)\phi$,~$J/\psi\eta^{\prime}$,~$J/\psi\eta$, ${\eta_c \phi}$   \\
\bottomrule[1pt]\bottomrule[1pt]
\end{tabular}
\end{table*}

The obtained results are theoretic reference to search for possible charmoniumlike molecular tetraquark with hidden-strange quantum number. So far, many charmoniumlike structures have been observed in the $J/\psi\phi$ invariant mass spectrum. In particular, the recent LHCb Collaboration updated the amplitude analysis of the $B^+\to J/\psi\phi K^+$ decay with the combined dataset collected in Run I plus Run II \cite{Aaij:2021ivw}, and they reported several enhancement structures, some of them locate close to the mass thresholds of our investigated charmoniumlike molecular tetraquark systems with hidden-strange quantum number, our study for these charmoniumlike molecules with hidden-strange quantum number may provide hints to understand the nature of the reported charmoniumlike states, i.e.,
\begin{enumerate}
  \item The masses of the $X(4140)$ with $J^{PC}=1^{++}$ and the $X(4150)$ with $J^{PC}=2^{-+}$ \cite{Aaij:2021ivw} are close to the $D_s^*\bar{D}_s^*$ threshold. It is obvious that their quantum configurations do not fit into the $S-$wave $D_s^*\bar{D}_s^*$ molecules. If their resonance parameters and quantum numbers can be further confirmed in the future, the assignment of these complicated structures as the $S-$wave $D_s^{*} \bar D_s^{*}$ molecular states will be facing great challenge.

  \item When we recall the other charmoniumlike structure $Y(4274)$ \cite{Aaltonen:2009tz,Chatrchyan:2013dma,Aaij:2016iza} (see Fig.~\ref{Jpsiphi} for more details), its mass are close to the $D_s\bar D_{s0}^{\ast}$ threshold, its spin-parity is favored as $J^{PC}=1^{++}$ by the LHCb Collaboration \cite{Aaij:2016iza,Aaij:2021ivw}, the similar situation happens, our results do not support the $Y(4274)$ with $J^{PC}=1^{++}$ as the $S-$wave $D_s\bar D_{s0}^{\ast}$ molecular state.

  \item In the mass region from 4.5 GeV to 4.7 GeV, the recent LHCb Collaboration reported the four enhancement structures $X(4500)$, $X(4630)$, $X(4685)$, and $X(4700)$ in the $J/\psi \phi$ mass spectrum from the $B \to  J/\psi \phi K$ decay \cite{Aaij:2021ivw}. Our results indicate there can exist six possible charmoniumlike molecular tetraquark candidates with hidden-strange quantum number with positive $C$-parity. Some of them can decay to the $J/\psi\phi$ final state. In our former work \cite{Yang:2021sue}, the $X(4630)$ can be assigned as the $D_s^{*} \bar D_{s1}$ charmoniumlike molecule with $J^{PC}=1^{-+}$. If these six predicted charomoniumlike molecules really exist, it is a big challenge to search and further identify separately by using the data from the $B \to  J/\psi \phi K$ process.

  \item In addition, our calculations can predict several possible charmoniumlike molecular candidates with negative $C$-parity, which deserve to attract the experiment's attentions. Because their two-body hidden-charm decay channels include the $\chi_{c0,1,2}(1P)\phi$, the $B\to\chi_{c0,1,2}(1P)\phi+K\to J/\psi\gamma\phi+K$ may be the prime decay process to search for these possible charmoniumlike molecules.
\end{enumerate}

In fact, the phenomena in high mass region are very complicated, beside these possible charmoniumlike molecular structures, there can exist several traditional charmonium states with their masses around these discussed thresholds. For example, the Lanzhou group indicated that the masses for the $\chi_{c0}(3{}^3P_0)$, $\chi_{c1}(3{}^3P_1)$, and $\chi_{c2}(3{}^3P_2)$ states are around 4177~MeV, 4197~MeV, and 4213~MeV \cite{Wang:2019mhs,Wang:2020prx}, which is close to the $D_s^{*} \bar D_s^{*}$ thresholds. Both the theoretical side and the experimental side will make many efforts to clarify this puzzling phenomenon in the future.

{By this study, we find a series of possible isoscalar $\mathcal{D}\bar{\mathcal{D}}$-type hidden-charm molecular tetraquarks. Some of the isoscalar charmoniumlike $XYZ$ states can be explained as the isoscalar hidden-charm molecules as their masses and quantum number configurations match the calculations of the isoscalar $\mathcal{D}\bar{\mathcal{D}}$-type loosely bound states. Even so, it is difficult to make a definitely decision if they are pure hidden-charm molecular tetraquarks or not. They are possible to be mixtures of the conventional charmonium states and the isoscalar hidden-charm molecular tetraquarks. However, the charged charmoniumlike states are genuine tetraquark states, if they are the real particles. In the following, we will search for possible isovector hidden-charm molecular tetraquarks, and check if the charged $Z$ isovector states have a close relation to these isovector hidden-charm molecular tetraquarks.}

\subsubsection{Isovector charmoniumlike molecular tetraquark systems}

In this section, let us discuss the final proposal whether there exist isovector charmoniumlike molecules from a charmed meson and an anticharmed meson interactions, where the charmed (anticharmed) mesons are in the $H/T$ doublets. In comparison with the isoscalar $\mathcal{D}\bar{\mathcal{D}}$ systems, the OBE effective potentials for the $S-$wave isovector $\mathcal{D}\bar{\mathcal{D}}$ systems are very different. Since the $\pi$ and $\rho$ exchanges couple to the isospin charge, their properties in the isovector $\mathcal{D}\bar{\mathcal{D}}$ systems are contrary to those in the isoscalar $\mathcal{D}\bar{\mathcal{D}}$ systems, and the corresponding effective potentials are three times weaker (see Eqs. (\ref{scalarisospin})-(\ref{vectorisopin}) for details). Therefore, the OBE effective potentials for the $S-$wave isovector $\mathcal{D}\bar{\mathcal{D}}$ systems may be not strong enough to bind isovector charmoniumlike molecular states.

When we input the corresponding OBE effective potentials, only the $S-$wave isovector $D^{\ast}\bar{D}_1$ state with $J^{PC}=2^{-+}$ can generate the loosely bound state with the cutoff parameter $\Lambda$ around 3.0 GeV. Even though we consider the $S-$$D$ wave mixing and the coupled channel effect, the cutoff $\Lambda$ is still much larger than 1.0 GeV \cite{Tornqvist:1993ng,Tornqvist:1993vu}. If we take the cutoff value from the deuteron as the reasonable input, the interactions from a pair of charmed meson and anticharmed meson are not strong enough to bind $S-$wave isovector charmoniumlike molecular tetraquark states.

\begin{figure}[!htbp]
\centering
\includegraphics[scale=0.25]{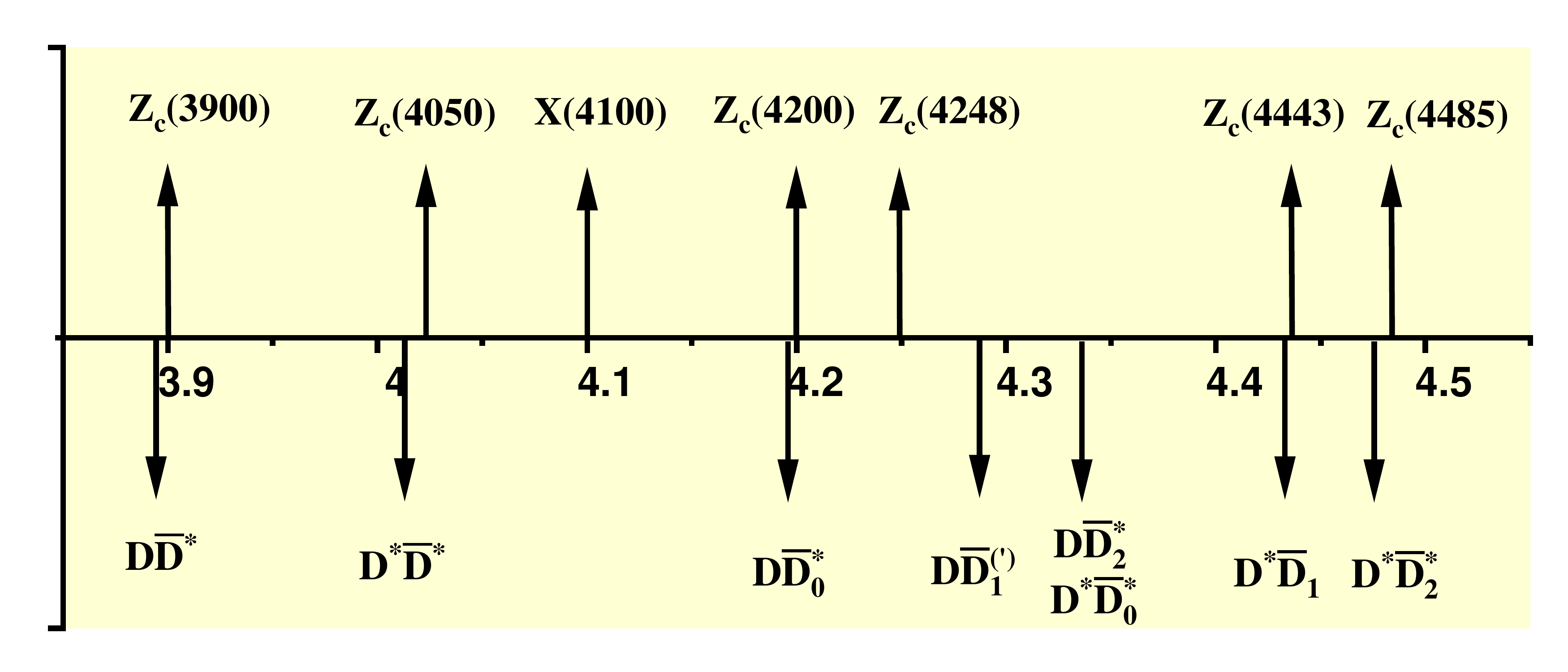}
\caption{The masses comparison between the charged charmoniumlike states \cite{Chen:2016qju,Liu:2013waa,Hosaka:2016pey,Liu:2019zoy,Brambilla:2019esw,Olsen:2017bmm,Guo:2017jvc} and the $\mathcal{D}\bar{\mathcal{D}}$ thresholds.}
\label{mass}
\end{figure}

Experimentally, there have accumulated abundant experimental observations of the charged charmoniumlike states from the $B$ meson decays as introduced in the Sec. \ref{sec23}. If they are real particles, they cannot be traditional mesons but the multiquark matters. As shown in Fig. \ref{mass}, their masses are all above the mass thresholds of the corresponding $\mathcal{D}\bar{\mathcal{D}}$ systems. Obviously, they cannot be assigned as the isovector charmoniumlike molecular states. In short, our results exclude the charged charmoniumlike states as the isovector meson-meson molecular states, which perfectly match the experimental observations. In fact, the other groups also support our conclusions \cite{Chen:2015add,Liu:2008tn,Sun:2012zzd,Zhao:2014gqa,Liu:2008mi,Ding:2008gr,He:2014nya,Liu:2007bf,Liu:2009wb,Liu:2019zoy,Uchino:2015uha,Liu:2017mrh,Meng:2021rdg,Wang:2020htx,Meng:2020ihj,Wang:2020dko}.

Before closing our discussion on the isovector charmoniumlike systems, we would like to emphasize that we only analyze whether the charged charmoniumlike states can be the isovector charmoniumlike molecular states composed of a pair of charmed meson and anticharmed meson. {Furthermore, the interaction of the off-diagonal couplings may be important to reproduce these charged charmoniumlike states by the threshold cusp \cite{HALQCD:2016ofq,Yamada:2021bzb,Dong:2020hxe,Ortega:2018cnm,He:2017lhy}, such as the $\pi J/\psi$-$DD$ and $\rho \eta_c$-$DD$ couplings, it is an interesting topic to consider the more complex coupled channels calculation to deal with these charged charmoniumlike states. In Ref. \cite{Chen:2013coa}, Lanzhou group once reproduced the $Z_c(3900)$ structure through the initial-single-pion-emission mechanism.} In fact, many theoretical groups propose other different explanations on the nature of the charged charmoniumlike states, like the compact tetraquark configurations, the meson-meson scattering states, and the kinematical effect (including the coupled channel cusp effect, the reflection mechanism, the interference effect, the initial single pion emission mechanism, the triangle singularities, the rescattering effect, and so on), people can refer articles \cite{Chen:2016qju,Liu:2019zoy,Guo:2017jvc,Wang:2020dmv,Wang:2020axi,Wang:2020kej,Chen:2010nv,Chen:2011kc,Chen:2015bft,Chen:2017uof,Chen:2011xk,Chen:2011pv,Chen:2013coa,Chen:2012yr,Liu:2015isc,Chen:2015eia,Huang:2019agb,Guo:2019twa} for more details.

\section{Summary}\label{sec4}
Exploration of the hadronic molecular states is an interesting and important research topics in the hadron physics. Since 2003, serials of charmoniumlike structures have been reported by different experiments in different beams, different reactions, and different energy regions, some of them are very close to a pair of charmed meson and anticharmed meson thresholds, which inspired theorists to explain them in the hadronic molecular scenario. Although the charmoniumlike hadronic molecular states are not forbidden by the quantum chromodynamics (QCD), one cannot give a precisely conclusion whether the charmoniumlike hadronic molecular states exist or not up to now.

The perfect match between experimental observations and the theoretical predictions on the three $P_c$ states provides a strong evidence of the existence of hidden-charm meson-baryon molecular pentaquark states. If replacing $ud$ quarks of $\Lambda_b$ by an antiquark $\bar{q}$, the production mechanisms of the $P_c$ states from $\Lambda_b$ baryon decays and the $XYZ$ states from $B$ meson decays are very similar. The $B\to XYZ+K$ decay should be the ideal process to produce charmoniumlike molecular states.

In such a situation, we firstly focus on the isoscalar $D^*\bar{D}^*$ systems with $J^{PC}=0^{++}$, $1^{+-}$, and $2^{++}$. And we find their interactions are attractive enough to generate bound charmoniumlike molecular states. After analyzing their two-body hidden-charm decay behaviors, we find that the $J/\psi\omega$ and $J/\psi\eta$ channels are the essential decay modes for the isoscalar $D^*\bar{D}^*$ molecular states with $J^{PC}=0^{++}/2^{++}$ and $1^{+-}$, respectively. If checking the data from the $J/\psi\omega$ invariant mass distribution in the $B\to J/\psi\omega K$ and the $J/\psi\eta$ invariant mass distribution in the $B\to J/\psi\eta K$, one can roughly find double enhancement structures in $J/\psi\omega$ invariant mass distribution and single structure in $J/\psi\eta$ invariant mass distribution, their masses are just below the $D^*\bar{D}^*$ threshold, which may correspond to the isoscalar $D^*\bar{D}^*$ charmoniumlike molecules with $J^P=0^{++}/2^{++}$ and $1^{+-}$, respectively. The behavior of the double enhancement structures around 3.9 GeV below the $D^*\bar{D}^*$ threshold from the $B\to J/\psi\omega K$ and the single structure in the same energy region from the $B\to J/\psi\eta K$ can provide crucial information to identify the charmoniumlike molecule. Thus, we suggest to systematically check the correlation of charmoniumlike molecular states and charmoniumlike structures existing in the $XYZ$ data of the $B\to XYZ+K$ decay.

After that, we promote our study to the interactions between a charmed (charm-strange) meson and an anticharmed (anti-charm-strange) meson, including the $D^{(*)}\bar{D}^{(*)}$, $\bar{D}^{(*)}\bar{D}_1$, $D^{(*)}\bar{D}_2^*$, $D_s^{(*)}\bar{D}_s^{(*)}$, ${D}_s^{(*)}\bar{D}_{s0}^*$, $D_s^{(*)}\bar{D}_{s1}^{\prime}$, ${D}_s^{(*)}\bar{D}_{s1}$, $D_s^{(*)}\bar{D}_{s2}^*$ systems. After input the OBE effective potentials, we can find a series of promising isoscalar charmoniumlike molecular tetraquark candidates as summarized in Tables \ref{sum1}. Our results exclude these discussed $S-$wave isovector charmoniumlike tetraquark states as the hadronic molecular candidates in the OBE model. Besides analyzing the mass spectrum, we also give their two-body hidden-charm decay channels of these possible molecular tetraquark candidates. In Table \ref{sum1}, we collect the two-body hidden-charm decay channels for these promising charmoniumlike molecular candidates. These obtained results can be very helpful to search and identify these discussed molecular tetraquark candidates.

In this work, our results indicate that the underlying phenomena behind the wide structure around 4.3 GeV in the $B\to J/\psi\omega K$ is very complicated. There can exist several possible isoscalar charmoniumlike molecular candidates, and the $J/\psi\omega$ channel is their two-body hidden-charm decay channel. In addition, we can predict many isoscalar charmoniumlike molecular states with negative $C$-parity in the same mass region, and their two-body hidden-charm decay modes are the $\chi_{cJ}(1P)\omega$ with $J=0,1,2$. The $B\to\chi_{cJ}(1P)\omega +K\to J/\psi\gamma\omega+K$ may be the good production process to search for these predicted isoscalar charmoniumlike molecular states with $C=-$.

Meanwhile, we also study the mass spectrum and give two-body hidden-charm decay channels for the possible charmoniumlike $\mathcal{D}_s\bar{\mathcal{D}_s}$ molecular states with hidden-strange quantum number summarized in Table \ref{sum2}, where $\mathcal{D}_s$ stands for the charm-strange meson. The $J/\psi\phi$ channel is the two-body hidden-charm decay channel for most of the possible $\mathcal{D}_s\bar{\mathcal{D}_s}$ molecular states. In order to further distinguish these structures, one need more precision experimental data, and more analysis on the other decay channel could be also useful.

Finding the charmoniumlike states has been going on for around 20 years. With the running of Belle II \cite{Kou:2018nap} and the accumulation of Run II and Run III data at LHCb \cite{Bediaga:2018lhg}, the study of the charmoniumlike states will step into a new area, here, we suggest to systematically check the correlation of charmoniumlike molecular states and charmoniumlike structures existing in the $XYZ$ data of the $B\to XYZ+K$ decay. The present work provides a good start point and new insight for identifying charmoniumlike molecule. We expect more theoretical and experimental efforts in exploring this important and intriguing research topic in the coming golden decade.

\appendix
\section{Relevant subpotentials}\label{app01}

Before presenting the effective potentials for these investigated tetraquark systems, we first define several operators $\mathcal{O}_k^{(\prime)}$ involved in this work, which include
\begin{eqnarray}\label{op}
\mathcal{O}_{1}&=&{\bm\epsilon^{\dagger}_4}\cdot{\bm\epsilon_2},~~~~~~\mathcal{O}_{2}={\bm\epsilon^{\dagger}_3}\cdot{\bm\epsilon_2},~~~~~~\mathcal{O}_{3}=T({\bm\epsilon^{\dagger}_3},{\bm\epsilon_2}),\nonumber\\
\mathcal{O}_{4}&=&\left({\bm\epsilon^{\dagger}_3}\cdot{\bm\epsilon_1}\right)\left({\bm\epsilon^{\dagger}_4}\cdot{\bm\epsilon_2}\right),~~~~~~\mathcal{O}_{5}=\left({\bm\epsilon^{\dagger}_3}\times{\bm\epsilon_1}\right)\cdot\left({\bm\epsilon^{\dagger}_4}\times{\bm\epsilon_2}\right),\nonumber\\
\mathcal{O}_{6}&=&T({\bm\epsilon^{\dagger}_3}\times{\bm\epsilon_1},{\bm\epsilon^{\dagger}_4}\times{\bm\epsilon_2}),\nonumber\\
\mathcal{O}_{7}&=&\mathcal{\sum}\left({\bm\epsilon^{\dagger}_{4m}}\cdot{\bm\epsilon_{2a}}\right)\left({\bm\epsilon^{\dagger}_{4n}}\cdot {\bm\epsilon_{2b}}\right),\nonumber\\
\mathcal{O}_{8}&=&\frac{2}{27}\mathcal{\sum}\left({\bm\epsilon^{\dagger}_{3m}}\cdot{\bm\epsilon_{2a}}\right)\left({\bm\epsilon^{\dagger}_{3n}}\cdot {\bm\epsilon_{2b}}\right),\nonumber\\
\mathcal{O}_{9}&=&\frac{1}{27}\mathcal{\sum}T({\bm\epsilon^{\dagger}_{3m}},{\bm\epsilon^{\dagger}_{3n}})T({\bm\epsilon_{2a}},{\bm\epsilon_{2b}})\nonumber\\
&&+\frac{2}{27}\mathcal{\sum}T({\bm\epsilon^{\dagger}_{3m}},{\bm\epsilon_{2a}})T({\bm\epsilon^{\dagger}_{3n}},{\bm\epsilon_{2b}}),\nonumber\\
\mathcal{O}_{10}&=&\frac{2}{27}\mathcal{\sum}\left({\bm\epsilon^{\dagger}_{3m}}\cdot{\bm\epsilon_{2a}}\right)T({\bm\epsilon^{\dagger}_{3n}},{\bm\epsilon_{2b}}),\nonumber\\
\mathcal{O}_{11}&=&-\frac{1}{3}\left({\bm\epsilon^{\dagger}_3}\cdot{\bm\epsilon_1}\right)\left({\bm\epsilon^{\dagger}_4}\cdot{\bm\epsilon_2}\right)
+\frac{1}{3}\left({\bm\epsilon^{\dagger}_3}\cdot{\bm\epsilon^{\dagger}_4}\right)\left({\bm\epsilon_1}\cdot{\bm\epsilon_2}\right),\nonumber\\
\mathcal{O}_{12}&=&\frac{2}{3}T({\bm\epsilon^{\dagger}_3},{\bm\epsilon_1})T({\bm\epsilon^{\dagger}_4},{\bm\epsilon_2})+\frac{1}{3}T({\bm\epsilon^{\dagger}_3},{\bm\epsilon^{\dagger}_4})T({\bm\epsilon_1},{\bm\epsilon_2}),\nonumber\\
\mathcal{O}_{13}&=&\frac{1}{6}\left({\bm\epsilon^{\dagger}_3}\cdot{\bm\epsilon^{\dagger}_4}\right)T({\bm\epsilon_1},{\bm\epsilon_2})+\frac{1}{6}\left({\bm\epsilon_1}\cdot{\bm\epsilon_2}\right)T({\bm\epsilon^{\dagger}_3},{\bm\epsilon^{\dagger}_4})\nonumber\\
&&-\frac{1}{3}\left({\bm\epsilon^{\dagger}_3}\cdot{\bm\epsilon_1}\right)T({\bm\epsilon^{\dagger}_4},{\bm\epsilon_2}),\nonumber\\
\mathcal{O}_{14}&=&\mathcal{\sum}\left({\bm\epsilon^{\dagger}_3}\cdot{\bm\epsilon_1}\right)\left({\bm\epsilon^{\dagger}_{4m}}\cdot{\bm\epsilon_{2a}}\right)\left({\bm\epsilon^{\dagger}_{4n}}\cdot {\bm\epsilon_{2b}}\right),\nonumber\\
\mathcal{O}_{14}^{\prime}&=&\mathcal{\sum}\left({\bm\epsilon^{\dagger}_4}\cdot{\bm\epsilon_2}\right)\left({\bm\epsilon^{\dagger}_{3m}}\cdot{\bm\epsilon_{1a}}\right)\left({\bm\epsilon^{\dagger}_{3n}}\cdot {\bm\epsilon_{1b}}\right),\nonumber\\
\mathcal{O}_{15}&=&\mathcal{\sum}\left({\bm\epsilon^{\dagger}_{4m}}\cdot{\bm\epsilon_{2a}}\right)\left[\left({\bm\epsilon^{\dagger}_{3}}\times{\bm\epsilon_{1}}\right)\cdot\left({\bm\epsilon^{\dagger}_{4n}}\times {\bm\epsilon_{2b}}\right)\right],\nonumber\\
\mathcal{O}_{15}^{\prime}&=&\mathcal{\sum}\left({\bm\epsilon^{\dagger}_{3m}}\cdot{\bm\epsilon_{1a}}\right)\left[\left({\bm\epsilon^{\dagger}_{4}}\times{\bm\epsilon_{2}}\right)\cdot\left({\bm\epsilon^{\dagger}_{3n}}\times {\bm\epsilon_{1b}}\right)\right],\nonumber\\
\mathcal{O}_{16}&=&\mathcal{\sum}\left({\bm\epsilon^{\dagger}_{4m}}\cdot{\bm\epsilon_{2a}}\right)T({\bm\epsilon^{\dagger}_{3}}\times{\bm\epsilon_{1}},{\bm\epsilon^{\dagger}_{4n}}\times {\bm\epsilon_{2b}}),\nonumber\\
\mathcal{O}_{16}^{\prime}&=&\mathcal{\sum}\left({\bm\epsilon^{\dagger}_{3m}}\cdot{\bm\epsilon_{1a}}\right)T({\bm\epsilon^{\dagger}_{4}}\times{\bm\epsilon_{2}},{\bm\epsilon^{\dagger}_{3n}}\times {\bm\epsilon_{1b}}),\nonumber\\
\mathcal{O}_{17}&=&\mathcal{\sum}\left({\bm\epsilon^{\dagger}_{3m}}\cdot{\bm\epsilon_1}\right)\left({\bm\epsilon^{\dagger}_4}\cdot{\bm\epsilon_{2a}}\right)\left({\bm\epsilon^{\dagger}_{3n}}\cdot {\bm\epsilon_{2b}}\right),\nonumber\\
\mathcal{O}_{17}^{\prime}&=&\mathcal{\sum}\left({\bm\epsilon^{\dagger}_{4m}}\cdot{\bm\epsilon_2}\right)\left({\bm\epsilon^{\dagger}_3}\cdot{\bm\epsilon_{1a}}\right)\left({\bm\epsilon^{\dagger}_{4n}}\cdot {\bm\epsilon_{1b}}\right),\nonumber\\
\mathcal{O}_{18}&=&\mathcal{\sum}\left({\bm\epsilon^{\dagger}_{3m}}\cdot{\bm\epsilon_1}\right)\left({\bm\epsilon^{\dagger}_4}\cdot{\bm\epsilon_{2a}}\right)T({\bm\epsilon^{\dagger}_{3n}}, {\bm\epsilon_{2b}}),\nonumber\\
\mathcal{O}_{18}^{\prime}&=&\mathcal{\sum}\left({\bm\epsilon^{\dagger}_{4m}}\cdot{\bm\epsilon_2}\right)\left({\bm\epsilon^{\dagger}_3}\cdot{\bm\epsilon_{1a}}\right)T({\bm\epsilon^{\dagger}_{4n}}, {\bm\epsilon_{1b}}),\nonumber\\
\mathcal{O}_{19}&=&\frac{1}{27}\mathcal{\sum}\left[\left({\bm\epsilon^{\dagger}_{3m}}\times{\bm\epsilon_{1}}\right)\cdot\left({\bm\epsilon^{\dagger}_{4}}\times{\bm\epsilon_{2a}}\right)\right]\left({\bm\epsilon^{\dagger}_{3n}}\cdot{\bm\epsilon_{2b}}\right)\nonumber\\
&&+\frac{1}{27}\mathcal{\sum}\left[\left({\bm\epsilon^{\dagger}_{3m}}\times{\bm\epsilon_{1}}\right)\cdot{\bm\epsilon_{2b}}\right]\left[{\bm\epsilon^{\dagger}_{3n}}\cdot\left({\bm\epsilon^{\dagger}_{4}}\times{\bm\epsilon_{2a}}\right)\right],\nonumber\\
\mathcal{O}_{19}^{\prime}&=&\frac{1}{27}\mathcal{\sum}\left[\left({\bm\epsilon^{\dagger}_{4m}}\times{\bm\epsilon_{2}}\right)\cdot\left({\bm\epsilon^{\dagger}_{3}}\times{\bm\epsilon_{1a}}\right)\right]\left({\bm\epsilon^{\dagger}_{4n}}\cdot{\bm\epsilon_{1b}}\right)\nonumber\\
&&+\frac{1}{27}\mathcal{\sum}\left[\left({\bm\epsilon^{\dagger}_{4m}}\times{\bm\epsilon_{2}}\right)\cdot{\bm\epsilon_{1b}}\right]\left[{\bm\epsilon^{\dagger}_{4n}}\cdot\left({\bm\epsilon^{\dagger}_{3}}\times{\bm\epsilon_{1a}}\right)\right],\nonumber\\
\mathcal{O}_{20}&=&\frac{1}{27}\mathcal{\sum}T({\bm\epsilon^{\dagger}_{3m}}\times{\bm\epsilon_{1}},{\bm\epsilon^{\dagger}_{3n}})T({\bm\epsilon^{\dagger}_{4}}\times{\bm\epsilon_{2a}},{\bm\epsilon_{2b}})\nonumber\\
&&+\frac{1}{27}\mathcal{\sum}T({\bm\epsilon^{\dagger}_{3m}}\times{\bm\epsilon_{1}},{\bm\epsilon^{\dagger}_{4}}\times{\bm\epsilon_{2a}})T({\bm\epsilon^{\dagger}_{3n}},{\bm\epsilon_{2b}})\nonumber\\
&&+\frac{1}{27}\mathcal{\sum}T({\bm\epsilon^{\dagger}_{3m}}\times{\bm\epsilon_{1}},{\bm\epsilon_{2b}})T({\bm\epsilon^{\dagger}_{3n}},{\bm\epsilon^{\dagger}_{4}}\times{\bm\epsilon_{2a}}),\nonumber\\
\mathcal{O}_{20}^{\prime}&=&\frac{1}{27}\mathcal{\sum}T({\bm\epsilon^{\dagger}_{4m}}\times{\bm\epsilon_{2}},{\bm\epsilon^{\dagger}_{4n}})T({\bm\epsilon^{\dagger}_{3}}\times{\bm\epsilon_{1a}},{\bm\epsilon_{1b}})\nonumber\\
&&+\frac{1}{27}\mathcal{\sum}T({\bm\epsilon^{\dagger}_{4m}}\times{\bm\epsilon_{2}},{\bm\epsilon^{\dagger}_{3}}\times{\bm\epsilon_{1a}})T({\bm\epsilon^{\dagger}_{4n}},{\bm\epsilon_{1b}})\nonumber\\
&&+\frac{1}{27}\mathcal{\sum}T({\bm\epsilon^{\dagger}_{4m}}\times{\bm\epsilon_{2}},{\bm\epsilon_{1b}})T({\bm\epsilon^{\dagger}_{4n}},{\bm\epsilon^{\dagger}_{3}}\times{\bm\epsilon_{1a}}),\nonumber\\
\mathcal{O}_{21}&=&\frac{1}{54}\mathcal{\sum}\left[\left({\bm\epsilon^{\dagger}_{3m}}\times{\bm\epsilon_{1}}\right)\cdot\left({\bm\epsilon^{\dagger}_{4}}\times{\bm\epsilon_{2a}}\right)\right]T({\bm\epsilon^{\dagger}_{3n}},{\bm\epsilon_{2b}})\nonumber\\
&&+\frac{1}{54}\mathcal{\sum}\left[\left({\bm\epsilon^{\dagger}_{3m}}\times{\bm\epsilon_{1}}\right)\cdot{\bm\epsilon_{2b}}\right]T({\bm\epsilon^{\dagger}_{3n}},{\bm\epsilon^{\dagger}_{4}}\times{\bm\epsilon_{2a}})\nonumber\\
&&+\frac{1}{54}\mathcal{\sum}\left({\bm\epsilon^{\dagger}_{3n}}\cdot{\bm\epsilon_{2b}}\right)T({\bm\epsilon^{\dagger}_{3m}}\times{\bm\epsilon_{1}},{\bm\epsilon^{\dagger}_{4}}\times{\bm\epsilon_{2a}})\nonumber\\
&&+\frac{1}{54}\mathcal{\sum}\left[{\bm\epsilon^{\dagger}_{3n}}\cdot\left({\bm\epsilon^{\dagger}_{4}}\times{\bm\epsilon_{2a}}\right)\right]T({\bm\epsilon^{\dagger}_{3m}}\times{\bm\epsilon_{1}},{\bm\epsilon_{2b}}),\nonumber\\
\mathcal{O}_{21}^{\prime}&=&\frac{1}{54}\mathcal{\sum}\left[\left({\bm\epsilon^{\dagger}_{4m}}\times{\bm\epsilon_{2}}\right)\cdot\left({\bm\epsilon^{\dagger}_{3}}\times{\bm\epsilon_{1a}}\right)\right]T({\bm\epsilon^{\dagger}_{4n}},{\bm\epsilon_{1b}})\nonumber\\
&&+\frac{1}{54}\mathcal{\sum}\left[\left({\bm\epsilon^{\dagger}_{4m}}\times{\bm\epsilon_{2}}\right)\cdot{\bm\epsilon_{1b}}\right]T({\bm\epsilon^{\dagger}_{4n}},{\bm\epsilon^{\dagger}_{3}}\times{\bm\epsilon_{1a}})\nonumber\\
&&+\frac{1}{54}\mathcal{\sum}\left({\bm\epsilon^{\dagger}_{4n}}\cdot{\bm\epsilon_{1b}}\right)T({\bm\epsilon^{\dagger}_{4m}}\times{\bm\epsilon_{2}},{\bm\epsilon^{\dagger}_{3}}\times{\bm\epsilon_{1a}})\nonumber\\
&&+\frac{1}{54}\mathcal{\sum}\left[{\bm\epsilon^{\dagger}_{4n}}\cdot\left({\bm\epsilon^{\dagger}_{3}}\times{\bm\epsilon_{1a}}\right)\right]T({\bm\epsilon^{\dagger}_{4m}}\times{\bm\epsilon_{2}},{\bm\epsilon_{1b}}).
\end{eqnarray}
Here, we define $\mathcal{S}=\sum_{m,n,a,b}C^{2,m+n}_{1m,1n}C^{2,a+b}_{1a,1b}$, and $T({\bm x},{\bm y})= 3\left(\hat{\bm r} \cdot {\bm x}\right)\left(\hat{\bm r} \cdot {\bm y}\right)-{\bm x} \cdot {\bm y}$ is the tensor force operator. For these operators $\mathcal{O}_k^{(\prime)}$, they should be sandwiched by the relevant spin-orbital wave functions $|{}^{2S+1}L_{J}\rangle$ for these investigated tetraquark systems, such as $\langle D\bar{D}^{\ast}({}^3\mathbb{S}_{1})|\mathcal{O}_1|D\bar{D}^{\ast}({}^3\mathbb{S}_{1})\rangle=1$. The obtained operator matrix elements $\mathcal{O}_k^{(\prime)}[J]$ are summarized in Tables~\ref{matrix1} and~\ref{matrix2}, which will be used in our calculations.

\renewcommand\tabcolsep{0.04cm}
\renewcommand{\arraystretch}{0.15}
\begin{table*}[!htbp]
\caption{The relevant operator matrix elements $\mathcal{O}_k^{(\prime)}[J]=\langle f|\mathcal{O}_k^{(\prime)}|i\rangle$.\label{matrix1}}
\begin{tabular}{l|l|l}\toprule[1.0pt]\toprule[1.0pt]
$\begin{array}{l}\mathcal{O}_1[1]=\rm {diag}(1,1)\\ \mathcal{O}_2[1]=\rm {diag}(1,1)\end{array}$ ~~ $\mathcal{O}_{3}[1]=\left(\begin{array}{cc} 0 & -\sqrt{2} \\ -\sqrt{2} & 1\end{array}\right)$ &$\begin{array}{l}\mathcal{O}_4[0]=\rm {diag}(1,1)\\ \mathcal{O}_5[0]=\rm {diag}(2,-1) \end{array}$~~$\mathcal{O}_{6}[0]=\left(\begin{array}{cc} 0 & \sqrt{2} \\ \sqrt{2} & 2\end{array}\right)$ & $\begin{array}{l} \mathcal{O}_4[1]=\rm {diag}(1,1,1) \\ \mathcal{O}_5[1]=\rm {diag}(1,1,-1) \end{array}$\\
$\mathcal{O}_{6}[1]=\left(\begin{array}{ccc} 0 & -\sqrt{2} &0 \\ -\sqrt{2} & 1 &0 \\ 0&0&1\end{array}\right)$&$\begin{array}{l} \mathcal{O}_4[2]=\rm {diag}(1,1,1,1) \\ \mathcal{O}_5[2]=\rm {diag}(-1,2,1,-1) \\ \mathcal{O}_7[2]=\rm {diag}(1,1)\\ \mathcal{O}_8[2]=\rm {diag}(\frac{2}{27},\frac{2}{27})\end{array}$ & $\mathcal{O}_{6}[2]=\left(\begin{array}{cccc} 0 & \frac{\sqrt{2}}{\sqrt{5}} & 0 &-\frac{\sqrt{14}}{\sqrt{5}} \\ \frac{\sqrt{2}}{\sqrt{5}} & 0 & 0 &-\frac{2}{\sqrt{7}} \\ 0 & 0 &-1 &0 \\ -\frac{\sqrt{14}}{\sqrt{5}}&-\frac{2}{\sqrt{7}}&0&-\frac{3}{7}\end{array}\right)$ \\
$\mathcal{O}_9[2]=\left(\begin{array}{cc} \frac{8}{135} & -\frac{4\sqrt{2}}{27\sqrt{35}} \\ -\frac{4\sqrt{2}}{27\sqrt{35}} & \frac{4}{27}\end{array}\right)$ & $\mathcal{O}_{10}[2]=\left(\begin{array}{cc} 0 & -\frac{\sqrt{7}}{27\sqrt{10}} \\ -\frac{\sqrt{7}}{27\sqrt{10}} & -\frac{1}{126}\end{array}\right)$&$\begin{array}{l} \mathcal{O}_{11}[0]=\rm {diag}(\frac{2}{3},-\frac{1}{3})\\ \mathcal{O}_{11}[1]=\rm {diag}(-\frac{1}{3},-\frac{1}{3},-\frac{1}{3})\\\mathcal{O}_{11}[2]=\rm {diag}(-\frac{1}{3},\frac{2}{3},-\frac{1}{3},-\frac{1}{3})\end{array}$ \\
$\mathcal{O}_{12}[0]=\left(\begin{array}{cc} \frac{4}{3} & -\frac{2\sqrt{2}}{3} \\ -\frac{2\sqrt{2}}{3} & 4\end{array}\right)$&$\mathcal{O}_{13}[0]=\left(\begin{array}{cc} 0 & \frac{\sqrt{2}}{15} \\ -\frac{8\sqrt{2}}{15} & -\frac{1}{15}\end{array}\right)$&$\mathcal{O}_{12}[1]=\left(\begin{array}{ccc} -\frac{2}{3} & -\frac{2\sqrt{2}}{3} &0 \\ -\frac{2\sqrt{2}}{3} & 0 &0 \\ 0&0&-\frac{4}{3}\end{array}\right)$\\
$\mathcal{O}_{13}[1]=\left(\begin{array}{ccc} 0 & -\frac{1}{30\sqrt{2}}&\frac{\sqrt{3}}{10\sqrt{2}} \\ -\frac{1}{30\sqrt{2}}&\frac{4}{105} &\frac{3\sqrt{3}}{70} \\ \frac{\sqrt{3}}{10\sqrt{2}}&\frac{3\sqrt{3}}{70}&-\frac{1}{105}\end{array}\right)$&$\mathcal{O}_{12}[2]=\left(\begin{array}{cccc} \frac{8}{15} &-\frac{2\sqrt{2}}{3\sqrt{5}} & 0 &-\frac{4\sqrt{2}}{3\sqrt{35}} \\-\frac{2\sqrt{2}}{3\sqrt{5}} &\frac{4}{3} & 0 &\frac{4}{3\sqrt{7}} \\ 0 &0 & -\frac{4}{3} &0 \\ -\frac{4\sqrt{2}}{3\sqrt{35}}&\frac{4}{3\sqrt{7}}&0&\frac{4}{3}\end{array}\right)$&$\mathcal{O}_{13}[2]=\left(\begin{array}{cccc} 0 &-\frac{1}{\sqrt{10}}& 0&\frac{\sqrt{2}}{3\sqrt{35}} \\ 0 &0& 0&-\frac{4}{21\sqrt{7}} \\0&0&-\frac{1}{14} &\frac{1}{14\sqrt{7}} \\ \frac{\sqrt{2}}{3\sqrt{35}}&\frac{23}{21\sqrt{7}}&\frac{1}{14\sqrt{7}}&\frac{13}{294}\end{array}\right)$\\
$\begin{array}{l} \mathcal{O}_{14}^{(\prime)}[1]=\rm {diag}(1,1,1,1) \\ \mathcal{O}_{15}^{(\prime)}[1]=\rm {diag}(\frac{3}{2},\frac{3}{2},\frac{1}{2},-1)\\\mathcal{O}_{17}^{(\prime)}[1]=\rm {diag}(\frac{1}{6},\frac{1}{6},\frac{1}{2},1)\\\mathcal{O}_{19}^{(\prime)}[1]=\rm {diag}(\frac{1}{18},\frac{1}{18},\frac{5}{54},-\frac{1}{27})\end{array}$& $\mathcal{O}_{16}[1]=\left(\begin{array}{cccc} 0&\frac{3}{5\sqrt{2}}&\frac{\sqrt{6}}{\sqrt{5}}&\frac{\sqrt{21}}{5\sqrt{2}} \\ \frac{3}{5\sqrt{2}}&-\frac{3}{10}&\frac{\sqrt{3}}{\sqrt{5}}&-\frac{\sqrt{3}}{5\sqrt{7}}\\\frac{\sqrt{6}}{\sqrt{5}}&\frac{\sqrt{3}}{\sqrt{5}}&\frac{1}{2}&\frac{2}{\sqrt{35}}\\\frac{\sqrt{21}}{5\sqrt{2}}
&-\frac{\sqrt{3}}{5\sqrt{7}}&\frac{2}{\sqrt{35}}&\frac{48}{35}\end{array}\right)$&
$\mathcal{O}_{16}^{\prime}[1]=\left(\begin{array}{cccc} 0&\frac{3}{5\sqrt{2}}&-\frac{\sqrt{6}}{\sqrt{5}}&\frac{\sqrt{21}}{5\sqrt{2}} \\ \frac{3}{5\sqrt{2}}&-\frac{3}{10}&-\frac{\sqrt{3}}{\sqrt{5}}&-\frac{\sqrt{3}}{5\sqrt{7}}\\-\frac{\sqrt{6}}{\sqrt{5}}&-\frac{\sqrt{3}}{\sqrt{5}}&\frac{1}{2}&-\frac{2}{\sqrt{35}}\\\frac{\sqrt{21}}{5\sqrt{2}}
&-\frac{\sqrt{3}}{5\sqrt{7}}&-\frac{2}{\sqrt{35}}&\frac{48}{35}\end{array}\right)$\\
 $\mathcal{O}_{18}[1]=\left(\begin{array}{cccc} 0&-\frac{23}{15\sqrt{2}}&-\frac{2\sqrt{2}}{\sqrt{15}}&-\frac{\sqrt{7}}{5\sqrt{6}} \\ -\frac{23}{15\sqrt{2}}&\frac{23}{30}&-\frac{2}{\sqrt{15}}&\frac{1}{5\sqrt{21}}\\\frac{2\sqrt{2}}{\sqrt{15}}&\frac{2}{\sqrt{15}}&\frac{1}{2}&-\frac{2}{\sqrt{35}}\\-\frac{\sqrt{7}}{5\sqrt{6}}
&\frac{1}{5\sqrt{21}}&\frac{2}{\sqrt{35}}&\frac{24}{35}\end{array}\right)$&$\mathcal{O}_{18}^{\prime}[1]=\left(\begin{array}{cccc} 0&-\frac{23}{15\sqrt{2}}&\frac{2\sqrt{2}}{\sqrt{15}}&-\frac{\sqrt{7}}{5\sqrt{6}} \\ -\frac{23}{15\sqrt{2}}&\frac{23}{30}&\frac{2}{\sqrt{15}}&\frac{1}{5\sqrt{21}}\\-\frac{2\sqrt{2}}{\sqrt{15}}&-\frac{2}{\sqrt{15}}&\frac{1}{2}&\frac{2}{\sqrt{35}}\\-\frac{\sqrt{7}}{5\sqrt{6}}
&\frac{1}{5\sqrt{21}}&-\frac{2}{\sqrt{35}}&\frac{24}{35}\end{array}\right)$& $\mathcal{O}_{20}[1]=\left(\begin{array}{cccc} \frac{2}{45}&-\frac{\sqrt{2}}{45}&0&\frac{\sqrt{2}}{15\sqrt{21}} \\ -\frac{\sqrt{2}}{45}&\frac{1}{15}&0&-\frac{8}{15\sqrt{21}}\\0&0&-\frac{1}{9}&-\frac{2\sqrt{5}}{27\sqrt{7}}\\\frac{\sqrt{2}}{15\sqrt{21}}
&-\frac{8}{15\sqrt{21}}&\frac{2\sqrt{5}}{27\sqrt{7}}&\frac{92}{945}\end{array}\right)$\\
$\mathcal{O}_{20}^{\prime}[1]=\left(\begin{array}{cccc} \frac{2}{45}&-\frac{\sqrt{2}}{45}&0&\frac{\sqrt{2}}{15\sqrt{21}} \\ -\frac{\sqrt{2}}{45}&\frac{1}{15}&0&-\frac{8}{15\sqrt{21}}\\0&0&-\frac{1}{9}&\frac{2\sqrt{5}}{27\sqrt{7}}\\\frac{\sqrt{2}}{15\sqrt{21}}
&-\frac{8}{15\sqrt{21}}&-\frac{2\sqrt{5}}{27\sqrt{7}}&\frac{92}{945}\end{array}\right)$& $\mathcal{O}_{21}[1]=\left(\begin{array}{cccc} 0&-\frac{7}{90\sqrt{2}}&0&\frac{\sqrt{7}}{30\sqrt{6}}\\ -\frac{7}{90\sqrt{2}}&\frac{7}{180}&0&-\frac{1}{30\sqrt{21}}\\0&0&\frac{1}{108}&\frac{\sqrt{5}}{27\sqrt{7}}\\\frac{\sqrt{7}}{30\sqrt{6}}&-\frac{1}{30\sqrt{21}}&-\frac{\sqrt{5}}{27\sqrt{7}}&\frac{4}{315}\end{array}\right)$&$\mathcal{O}_{21}^{\prime}[1]=\left(\begin{array}{cccc} 0&-\frac{7}{90\sqrt{2}}&0&\frac{\sqrt{7}}{30\sqrt{6}}\\ -\frac{7}{90\sqrt{2}}&\frac{7}{180}&0&-\frac{1}{30\sqrt{21}}\\0&0&\frac{1}{108}&-\frac{\sqrt{5}}{27\sqrt{7}}\\\frac{\sqrt{7}}{30\sqrt{6}}&-\frac{1}{30\sqrt{21}}&\frac{\sqrt{5}}{27\sqrt{7}}&\frac{4}{315}\end{array}\right)$\\
\bottomrule[1.0pt]\bottomrule[1.0pt]
\end{tabular}
\end{table*}

\renewcommand\tabcolsep{0.04cm}
\renewcommand{\arraystretch}{0.15}
\begin{table*}[!htbp]
\caption{The relevant operator matrix elements $\mathcal{O}_k^{(\prime)}[J]=\langle f|\mathcal{O}_k^{(\prime)}|i\rangle$.\label{matrix2}}
\begin{tabular}{l|l|l}\toprule[1.0pt]\toprule[1.0pt]
$\begin{array}{l} \mathcal{O}_{14}^{(\prime)}[2]=\rm {diag}(1,1,1,1) \\ \mathcal{O}_{15}^{(\prime)}[2]=\rm {diag}(\frac{1}{2},\frac{3}{2},\frac{1}{2},-1)\\ \mathcal{O}_{17}^{(\prime)}[2]=\rm {diag}(\frac{1}{2},\frac{1}{6},\frac{1}{2},1) \\\mathcal{O}_{19}^{(\prime)}[2]=\rm {diag}(\frac{5}{54},\frac{1}{18},\frac{5}{54},-\frac{1}{27})\end{array}$&$\mathcal{O}_{16}[2]=\left(\begin{array}{cccc} 0&-\frac{3\sqrt{2}}{5}&-\frac{\sqrt{7}}{\sqrt{10}}&\frac{\sqrt{7}}{5} \\ -\frac{3\sqrt{2}}{5}&\frac{3}{10}&\frac{3}{\sqrt{35}}&-\frac{3\sqrt{2}}{5\sqrt{7}}\\-\frac{\sqrt{7}}{\sqrt{10}}&\frac{3}{\sqrt{35}}&-\frac{3}{14}&\frac{4\sqrt{2}}{7\sqrt{5}}\\\frac{\sqrt{7}}{5}
&-\frac{3\sqrt{2}}{5\sqrt{7}}&\frac{4\sqrt{2}}{7\sqrt{5}}&\frac{12}{35}\end{array}\right)$&$\mathcal{O}_{16}^{\prime}[2]=\left(\begin{array}{cccc} 0&\frac{3\sqrt{2}}{5}&-\frac{\sqrt{7}}{\sqrt{10}}&-\frac{\sqrt{7}}{5} \\ \frac{3\sqrt{2}}{5}&\frac{3}{10}&-\frac{3}{\sqrt{35}}&-\frac{3\sqrt{2}}{5\sqrt{7}}\\-\frac{\sqrt{7}}{\sqrt{10}}&-\frac{3}{\sqrt{35}}&-\frac{3}{14}&-\frac{4\sqrt{2}}{7\sqrt{5}}\\-\frac{\sqrt{7}}{5}
&-\frac{3\sqrt{2}}{5\sqrt{7}}&-\frac{4\sqrt{2}}{7\sqrt{5}}&\frac{12}{35}\end{array}\right)$\\
$\mathcal{O}_{18}[2]=\left(\begin{array}{cccc} 0&-\frac{2\sqrt{2}}{5}&-\frac{\sqrt{7}}{\sqrt{10}}&-\frac{\sqrt{7}}{5} \\ \frac{2\sqrt{2}}{5}&-\frac{23}{30}&-\frac{2}{\sqrt{35}}&\frac{\sqrt{2}}{5\sqrt{7}}\\-\frac{\sqrt{7}}{\sqrt{10}}&\frac{2}{\sqrt{35}}&-\frac{3}{14}&-\frac{4\sqrt{2}}{7\sqrt{5}}\\\frac{\sqrt{7}}{5}
&\frac{\sqrt{2}}{5\sqrt{7}}&\frac{4\sqrt{2}}{7\sqrt{5}}&\frac{6}{35}\end{array}\right)$&$\mathcal{O}_{18}^{\prime}[2]=\left(\begin{array}{cccc} 0&\frac{2\sqrt{2}}{5}&-\frac{\sqrt{7}}{\sqrt{10}}&\frac{\sqrt{7}}{5} \\ -\frac{2\sqrt{2}}{5}&-\frac{23}{30}&\frac{2}{\sqrt{35}}&\frac{\sqrt{2}}{5\sqrt{7}}\\-\frac{\sqrt{7}}{\sqrt{10}}&-\frac{2}{\sqrt{35}}&-\frac{3}{14}&\frac{4\sqrt{2}}{7\sqrt{5}}\\-\frac{\sqrt{7}}{5}
&\frac{\sqrt{2}}{5\sqrt{7}}&-\frac{4\sqrt{2}}{7\sqrt{5}}&\frac{6}{35}\end{array}\right)$&$\mathcal{O}_{20}[2]=\left(\begin{array}{cccc} \frac{2}{27}&0&-\frac{\sqrt{2}}{27\sqrt{35}}&\frac{2}{27\sqrt{7}} \\ 0&\frac{1}{45}&0&\frac{\sqrt{2}}{15\sqrt{7}}\\-\frac{\sqrt{2}}{27\sqrt{35}}&0&\frac{29}{189}&\frac{5\sqrt{10}}{189}\\-\frac{2}{27\sqrt{7}}
&\frac{\sqrt{2}}{15\sqrt{7}}&-\frac{5\sqrt{10}}{189}&-\frac{28}{135}\end{array}\right)$\\
$\mathcal{O}_{20}^{\prime}[2]=\left(\begin{array}{cccc} \frac{2}{27}&0&-\frac{\sqrt{2}}{27\sqrt{35}}&-\frac{2}{27\sqrt{7}} \\ 0&\frac{1}{45}&0&\frac{\sqrt{2}}{15\sqrt{7}}\\-\frac{\sqrt{2}}{27\sqrt{35}}&0&\frac{29}{189}&-\frac{5\sqrt{10}}{189}\\\frac{2}{27\sqrt{7}}
&\frac{\sqrt{2}}{15\sqrt{7}}&\frac{5\sqrt{10}}{189}&-\frac{28}{135}\end{array}\right)$&$\mathcal{O}_{21}[2]=\left(\begin{array}{cccc} 0&0&-\frac{\sqrt{7}}{54\sqrt{10}}&\frac{\sqrt{7}}{54}\\ 0&-\frac{7}{180}&0&-\frac{1}{15\sqrt{14}}\\-\frac{\sqrt{7}}{54\sqrt{10}}&0&-\frac{1}{252}&\frac{2\sqrt{10}}{189}\\-\frac{\sqrt{7}}{54}&-\frac{1}{15\sqrt{14}}&-\frac{2\sqrt{10}}{189}&\frac{1}{135}\end{array}\right)$ &$\mathcal{O}_{21}^{\prime}[2]=\left(\begin{array}{cccc} 0&0&-\frac{\sqrt{7}}{54\sqrt{10}}&-\frac{\sqrt{7}}{54}\\ 0&-\frac{7}{180}&0&-\frac{1}{15\sqrt{14}}\\-\frac{\sqrt{7}}{54\sqrt{10}}&0&-\frac{1}{252}&-\frac{2\sqrt{10}}{189}\\\frac{\sqrt{7}}{54}&-\frac{1}{15\sqrt{14}}&\frac{2\sqrt{10}}{189}&\frac{1}{135}\end{array}\right)$\\
$\begin{array}{l} \mathcal{O}_{14}^{(\prime)}[3]=\rm {diag}(1,1,1,1) \\ \mathcal{O}_{15}^{(\prime)}[3]=\rm {diag}(-1,\frac{3}{2},\frac{1}{2},-1)\\\mathcal{O}_{17}^{(\prime)}[3]=\rm {diag}(1,\frac{1}{6},\frac{1}{2},1)\\\mathcal{O}_{19}^{(\prime)}[3]=\rm {diag}(-\frac{1}{27},\frac{1}{18},\frac{5}{54},-\frac{1}{27})\end{array}$&$\mathcal{O}_{16}[3]=\left(\begin{array}{cccc} 0&\frac{3}{5\sqrt{2}}&-\frac{1}{\sqrt{5}}&-\frac{4\sqrt{3}}{5} \\ \frac{3}{5\sqrt{2}}&-\frac{3}{35}&-\frac{6\sqrt{2}}{7\sqrt{5}}&-\frac{6\sqrt{6}}{35}\\-\frac{1}{\sqrt{5}}&-\frac{6\sqrt{2}}{7\sqrt{5}}&-\frac{4}{7}&\frac{\sqrt{3}}{7\sqrt{5}}\\-\frac{4\sqrt{3}}{5}
&-\frac{6\sqrt{6}}{35}&\frac{\sqrt{3}}{7\sqrt{5}}&-\frac{22}{35}\end{array}\right)$&$\mathcal{O}_{16}^{\prime}[3]=\left(\begin{array}{cccc} 0&\frac{3}{5\sqrt{2}}&\frac{1}{\sqrt{5}}&-\frac{4\sqrt{3}}{5} \\ \frac{3}{5\sqrt{2}}&-\frac{3}{35}&\frac{6\sqrt{2}}{7\sqrt{5}}&-\frac{6\sqrt{6}}{35}\\\frac{1}{\sqrt{5}}&\frac{6\sqrt{2}}{7\sqrt{5}}&-\frac{4}{7}&-\frac{\sqrt{3}}{7\sqrt{5}}\\-\frac{4\sqrt{3}}{5}
&-\frac{6\sqrt{6}}{35}&-\frac{\sqrt{3}}{7\sqrt{5}}&-\frac{22}{35}\end{array}\right)$\\
$\mathcal{O}_{18}[3]=\left(\begin{array}{cccc} 0&-\frac{1}{5\sqrt{2}}&-\frac{1}{\sqrt{5}}&-\frac{2\sqrt{3}}{5} \\ -\frac{1}{5\sqrt{2}}&\frac{23}{105}&\frac{4\sqrt{2}}{7\sqrt{5}}&\frac{2\sqrt{6}}{35}\\\frac{1}{\sqrt{5}}&-\frac{4\sqrt{2}}{7\sqrt{5}}&-\frac{4}{7}&-\frac{\sqrt{3}}{7\sqrt{5}}\\-\frac{2\sqrt{3}}{5}
&\frac{2\sqrt{6}}{35}&\frac{\sqrt{3}}{7\sqrt{5}}&-\frac{11}{35}\end{array}\right)$& $\mathcal{O}_{18}^{\prime}[3]=\left(\begin{array}{cccc} 0&-\frac{1}{5\sqrt{2}}&\frac{1}{\sqrt{5}}&-\frac{2\sqrt{3}}{5} \\ -\frac{1}{5\sqrt{2}}&\frac{23}{105}&-\frac{4\sqrt{2}}{7\sqrt{5}}&\frac{2\sqrt{6}}{35}\\-\frac{1}{\sqrt{5}}&\frac{4\sqrt{2}}{7\sqrt{5}}&-\frac{4}{7}&\frac{\sqrt{3}}{7\sqrt{5}}\\-\frac{2\sqrt{3}}{5}
&\frac{2\sqrt{6}}{35}&-\frac{\sqrt{3}}{7\sqrt{5}}&-\frac{11}{35}\end{array}\right)$&$\mathcal{O}_{20}[3]=\left(\begin{array}{cccc} -\frac{4}{135}&\frac{\sqrt{2}}{105}&\frac{2\sqrt{5}}{189}&-\frac{4}{315\sqrt{3}} \\ \frac{\sqrt{2}}{105}&\frac{16}{315}&0&-\frac{\sqrt{2}}{35\sqrt{3}}\\-\frac{2\sqrt{5}}{189}&0&\frac{1}{21}&-\frac{\sqrt{5}}{63\sqrt{3}}\\-\frac{4}{315\sqrt{3}}
&-\frac{\sqrt{2}}{35\sqrt{3}}&\frac{\sqrt{5}}{63\sqrt{3}}&\frac{82}{945}\end{array}\right)$\\
$\mathcal{O}_{20}^{\prime}[3]=\left(\begin{array}{cccc} -\frac{4}{135}&\frac{\sqrt{2}}{105}&-\frac{2\sqrt{5}}{189}&-\frac{4}{315\sqrt{3}} \\ \frac{\sqrt{2}}{105}&\frac{16}{315}&0&-\frac{\sqrt{2}}{35\sqrt{3}}\\\frac{2\sqrt{5}}{189}&0&\frac{1}{21}&\frac{\sqrt{5}}{63\sqrt{3}}\\-\frac{4}{315\sqrt{3}}
&-\frac{\sqrt{2}}{35\sqrt{3}}&-\frac{\sqrt{5}}{63\sqrt{3}}&\frac{82}{945}\end{array}\right)$&$\mathcal{O}_{21}[3]=\left(\begin{array}{cccc} 0&\frac{1}{30\sqrt{2}}&\frac{\sqrt{5}}{54}&-\frac{1}{45\sqrt{3}}\\ \frac{1}{30\sqrt{2}}&\frac{1}{90}&0&-\frac{\sqrt{2}}{35\sqrt{3}}\\-\frac{\sqrt{5}}{54}&0&-\frac{2}{189}&\frac{\sqrt{5}}{126\sqrt{3}}\\-\frac{1}{45\sqrt{3}}&-\frac{\sqrt{2}}{35\sqrt{3}}&-\frac{\sqrt{5}}{126\sqrt{3}}&-\frac{11}{1890}\end{array}\right)$
&$\mathcal{O}_{21}^{\prime}[3]=\left(\begin{array}{cccc} 0&\frac{1}{30\sqrt{2}}&-\frac{\sqrt{5}}{54}&-\frac{1}{45\sqrt{3}}\\ \frac{1}{30\sqrt{2}}&\frac{1}{90}&0&-\frac{\sqrt{2}}{35\sqrt{3}}\\\frac{\sqrt{5}}{54}&0&-\frac{2}{189}&-\frac{\sqrt{5}}{126\sqrt{3}}\\-\frac{1}{45\sqrt{3}}&-\frac{\sqrt{2}}{35\sqrt{3}}&\frac{\sqrt{5}}{126\sqrt{3}}&-\frac{11}{1890}\end{array}\right)$      \\
\bottomrule[1.0pt]\bottomrule[1.0pt]
\end{tabular}
\end{table*}

In addition, the function $Y(\Lambda_i,m_i,r)$ is defined as
\begin{eqnarray}
Y_i\equiv \left\{
\begin{aligned}
|q_i|&\leqslant m,\ \frac{e^{-m_i r}-e^{-\Lambda^2_i r}}{4\pi r}-\frac{\Lambda^2_i-m^2_i}{8\pi\Lambda_i}e^{-\Lambda_i r};\\
|q_i|&>m,\ \frac{\mathrm{cos} (m^{\prime}_i r)-e^{-\Lambda_i r}}{4\pi r}-\frac{\Lambda^2_i+m^{\prime2}_i}{8\pi\Lambda_i}e^{-\Lambda_i r};
\end{aligned}
\right.
\end{eqnarray}
where $m_i=\sqrt{m^2-q^2_i}$, $m^{\prime}_i=\sqrt{q^2_i-m^2}$, and $\Lambda_i=\sqrt{\Lambda^2-q^2_i}$.

\subsection{Hidden-charm molecular tetraquark systems without hidden-strange quantum number}
For convenience, we define two functions $\mathcal{H}(I)Y(\Lambda,m_P,r)$ and $\mathcal{G}(I)Y(\Lambda,m_V,r)$ for these investigated hidden-charm tetraquark systems, i.e.,
\begin{eqnarray}
&&\mathcal{H}(0)Y(\Lambda,m_P,r)=\frac{3}{2}Y(\Lambda,m_{\pi},r)+\frac{1}{6}Y(\Lambda,m_{\eta},r),\label{scalarisospin}\\
&&\mathcal{H}(1)Y(\Lambda,m_P,r)=-\frac{1}{2}Y(\Lambda,m_{\pi},r)+\frac{1}{6}Y(\Lambda,m_{\eta},r),\\
&&\mathcal{G}(0)Y(\Lambda,m_V,r)=\frac{3}{2}Y(\Lambda,m_{\rho},r)+\frac{1}{2}Y(\Lambda,m_{\omega},r),\\
&&\mathcal{G}(1)Y(\Lambda,m_V,r)=-\frac{1}{2}Y(\Lambda,m_{\rho},r)+\frac{1}{2}Y(\Lambda,m_{\omega},r).\label{vectorisopin}
\end{eqnarray}
Here,$\mathcal{H}(I)$ and $\mathcal{G}(I)$ are the isospin factors for these investigated hidden-charm tetraquark systems, and $I$ denote the isospin quantum numbers.

Through the above preparation, we can write the effective potentials in the coordinate space for all of the investigated hidden-charm tetraquark systems, which include
\begin{enumerate}
\item The $D\bar{D}$ system:
\begin{eqnarray}
\mathcal{V}&=&-g^2_{\sigma}Y_{\sigma}-\frac{1}{2}\beta^2g^2_V\mathcal{G}(I)Y_V.
\end{eqnarray}

\item The $D\bar{D}^{\ast}$ system:
\begin{eqnarray}
\mathcal{V}_D&=&-g^2_{\sigma}\mathcal{O}_{1}Y_{\sigma}-\frac{1}{2}\beta^2g^2_V\mathcal{O}_{1}\mathcal{G}(I)Y_V,\nonumber\\
\mathcal{V}_C&=&-\frac{g^2}{3f^2_{\pi}}\left(\mathcal{O}_{2}\mathcal{Z}+\mathcal{O}_{3}\mathcal{T}\right)\mathcal{H}(I)Y_{P0}\nonumber\\
&&+\frac{2}{3}\lambda^2g^2_V\left(2\mathcal{O}_{2}\mathcal{Z}-\mathcal{O}_{3}\mathcal{T}\right)\mathcal{G}(I)Y_{V0}.
\end{eqnarray}

\item The $D^*\bar{D}^*$ system:
\begin{eqnarray}
\mathcal{V}&=&-g^2_{\sigma}\mathcal{O}_4Y_{\sigma}+\frac{g^2}{3f^2_{\pi}}\left(\mathcal{O}_{5}\mathcal{Z}+\mathcal{O}_{6}\mathcal{T}\right)\mathcal{H}(I)Y_P\nonumber\\
&&-\frac{1}{2}\beta^2g^2_V\mathcal{O}_4\mathcal{G}(I)Y_V\nonumber\\
&&+\frac{2}{3}\lambda^2g^2_V\left(2\mathcal{O}_{5}\mathcal{Z}-\mathcal{O}_{6}\mathcal{T}\right)\mathcal{G}(I)Y_V.\nonumber\\
\end{eqnarray}

\item The $D\bar{D}_1$ system:
\begin{eqnarray}
\mathcal{V}_D&=&g_{\sigma}g^{\prime\prime}_{\sigma}\mathcal{O}_{1}Y_{\sigma}+\frac{1}{2}\beta\beta^{\prime\prime}g^2_V\mathcal{O}_{1}\mathcal{G}(I)Y_V,\nonumber\\
\mathcal{V}_C&=&\frac{2h^{\prime2}_{\sigma}}{9f^2_{\pi}}\left(\mathcal{O}_{2}\mathcal{Z}+\mathcal{O}_{3}\mathcal{T}\right)Y_{\sigma1}+\frac{\zeta^2_1g^2_V}{3}\mathcal{O}_{2}\mathcal{G}(I)Y_{V1}.\nonumber\\
\end{eqnarray}

\item The $D\bar{D}^{\ast}_2$ system:
\begin{eqnarray}
\mathcal{V}_D&=&g_{\sigma}g^{\prime\prime}_{\sigma}\mathcal{O}_{7}Y_{\sigma}+\frac{1}{2}\beta\beta^{\prime\prime}g^2_V\mathcal{O}_{7}\mathcal{G}(I)Y_V,\nonumber\\
\mathcal{V}_C&=&\frac{h^{\prime2}}{f^2_{\pi}}\left[\mathcal{O}_{8}\mathcal{Z}\mathcal{Z}+\mathcal{O}_{9}\mathcal{T}\mathcal{T}+\mathcal{O}_{10}\{\mathcal{T},\mathcal{Z}\}\right]\mathcal{H}(I)Y_{P2}.\nonumber\\
\end{eqnarray}

\item The $D^{\ast}\bar{D}_1$ system:
\begin{eqnarray}
\mathcal{V}_D&=&g_{\sigma}g^{\prime\prime}_{\sigma}\mathcal{O}_4Y_{\sigma}+\frac{5gk}{18f^2_{\pi}}\left(\mathcal{O}_{5}\mathcal{Z}+\mathcal{O}_{6}\mathcal{T}\right)\mathcal{H}(I)Y_P\nonumber\\
&&+\frac{1}{2}\beta\beta^{\prime\prime}g^2_V\mathcal{O}_4\mathcal{G}(I)Y_V\nonumber\\
&&-\frac{5}{9}\lambda\lambda^{\prime\prime}g^2_V\left(2\mathcal{O}_{5}\mathcal{Z}-\mathcal{O}_{6}\mathcal{T}\right)\mathcal{G}(I)Y_V,\nonumber\\
\mathcal{V}_C&=&\frac{h^{\prime2}_{\sigma}}{18\pi^2_{\pi}}\left(\mathcal{O}_{5}\mathcal{Z}+\mathcal{O}_{6}\mathcal{T}\right)Y_{\sigma3}+\frac{\zeta^2_1g^2_V}{12}\mathcal{O}_5\mathcal{G}(I)Y_{V3}\nonumber\\
&&+\frac{h^{\prime2}}{6f^2_{\pi}}\left[\mathcal{O}_{11}\mathcal{Z}\mathcal{Z}+\mathcal{O}_{12}\mathcal{T}\mathcal{T}+\mathcal{O}_{13}\{\mathcal{T},\mathcal{Z}\}\right]\mathcal{H}(I)Y_{P3}.\nonumber\\
\end{eqnarray}

\item The $D^{\ast}\bar{D}^{\ast}_{2}$ system:
\begin{eqnarray}
\mathcal{V}_D&=&g_{\sigma}g^{\prime\prime}_{\sigma}\frac{\mathcal{O}_{14}+\mathcal{O}_{14}^{\prime}}{2}Y_{\sigma}+\frac{1}{2}\beta\beta^{\prime\prime}g^2_V\frac{\mathcal{O}_{14}+\mathcal{O}_{14}^{\prime}}{2}\mathcal{G}(I)Y_V\nonumber\\
&&+\frac{gk}{3f^2_{\pi}}\left(\frac{\mathcal{O}_{15}+\mathcal{O}_{15}^{\prime}}{2}\mathcal{Z}+\frac{\mathcal{O}_{16}+\mathcal{O}_{16}^{\prime}}{2}\mathcal{T}\right)\mathcal{H}(I)Y_P\nonumber\\
&&-\frac{2}{3}\lambda\lambda^{\prime\prime}g^2_V\left(2\frac{\mathcal{O}_{15}+\mathcal{O}_{15}^{\prime}}{2}\mathcal{Z}-\frac{\mathcal{O}_{16}+\mathcal{O}_{16}^{\prime}}{2}\mathcal{T}\right)\mathcal{G}(I)Y_V,\nonumber\\
\mathcal{V}_C&=&\frac{h^{\prime2}_{\sigma}}{3f^2_{\pi}}\left(\frac{\mathcal{O}_{17}+\mathcal{O}_{17}^{\prime}}{2}\mathcal{Z}+\frac{\mathcal{O}_{18}+\mathcal{O}_{18}^{\prime}}{2}\mathcal{T}\right)Y_{\sigma4}\nonumber\\
&&+\frac{h^{\prime2}}{f^2_{\pi}}\left[\frac{\mathcal{O}_{19}+\mathcal{O}_{19}^{\prime}}{2}\mathcal{Z}\mathcal{Z}+\frac{\mathcal{O}_{20}+\mathcal{O}_{20}^{\prime}}{2}\mathcal{T}\mathcal{T}\right.\nonumber\\
&&\left.+\frac{\mathcal{O}_{21}+\mathcal{O}_{21}^{\prime}}{2}\{\mathcal{T},\mathcal{Z}\}\right]\mathcal{H}(I)Y_{P4}\nonumber\\
&&+\frac{\zeta^2_1g^2_V}{2}\frac{\mathcal{O}_{17}+\mathcal{O}_{17}^{\prime}}{2}\mathcal{G}(I)Y_{V4}.
\end{eqnarray}
\end{enumerate}
In the above expressions, the operators are defined as $\mathcal{Z}=\frac{1}{r^2}\frac{\partial}{\partial r}r^2\frac{\partial}{\partial r}$, $\mathcal{T}=r\frac{\partial}{\partial r}\frac{1}{r}\frac{\partial}{\partial r}$, and $\{\mathcal{T},\mathcal{Z}\}=\mathcal{T}\mathcal{Z}+\mathcal{Z}\mathcal{T}$. Additionally, the variables $q_i$ are written as $q_0=m_{D^{\ast}}-m_{D}$, $q_1=m_{D_1}-m_{D}$, $q_2=m_{D^{\ast}_2}-m_{D}$, $q_3=m_{D_1}-m_{D^{\ast}}$, and $q_4=m_{D^{\ast}_2}-m_{D^{\ast}}$.

\subsection{Hidden-charm tetraquark systems with hidden-strange quantum number}
For these discussed hidden-charm tetraquark systems with hidden-strange quantum number, we consider the effective potentials from the $\eta$ and $\phi$ exchanges \cite{Wang:2020dya}. In the following, we collect the expressions of the effective potentials for these discussed systems.
\begin{itemize}
 \item  The $D_s\bar{D}_s$ system:
\begin{eqnarray}
\mathcal{V}&=&-\frac{1}{2}\beta^2g^2_VY_{\phi}.
\end{eqnarray}

\item The $D_s\bar{D}_s^{\ast}$ system:
\begin{eqnarray}
\mathcal{V}_D&=&-\frac{1}{2}\beta^2g^2_V\mathcal{O}_{1}Y_{\phi},\nonumber\\
\mathcal{V}_C&=&-\frac{2g^2}{9f^2_{\pi}}\left(\mathcal{O}_{2}\mathcal{Z}+\mathcal{O}_{3}\mathcal{T}\right)Y_{\eta5}\nonumber\\
&&+\frac{2}{3}\lambda^2g^2_V\left(2\mathcal{O}_{2}\mathcal{Z}-\mathcal{O}_{3}\mathcal{T}\right)Y_{\phi5}.
\end{eqnarray}

\item The $D_s^{*} \bar D_s^{*}$ system:
\begin{eqnarray}
\mathcal{V}&=&\frac{2g^2}{9f_\pi^2}\left(\mathcal{O}_{5}\mathcal{Z}+\mathcal{O}_{6}\mathcal{T}\right)Y_{\eta}-\frac{1}{2}\beta^2g_V^2\mathcal{O}_{4}Y_\phi\nonumber\\
    &&+\frac{2}{3}\lambda^2g_V^2\left(2\mathcal{O}_{5}\mathcal{Z}-\mathcal{O}_{6}\mathcal{T}\right)Y_{\phi}.
\end{eqnarray}

\item The $D_s\bar D_{s0}^{\ast}$ system:
\begin{eqnarray}
\mathcal{V}_D&=&\frac{1}{2}\beta\beta^{\prime}g^2_VY_{\phi},\nonumber\\
\mathcal{V}_C&=&-\frac{2h^2q_{6}^2}{3f^2_{\pi}}Y_{\eta6}.
\end{eqnarray}

\item The $D_s\bar D_{s1}^{\prime}$ system:
\begin{eqnarray}
\mathcal{V}_D&=&\frac{1}{2}\beta\beta^{\prime}g^2_V\mathcal{O}_{1}Y_{\phi},\nonumber\\
\mathcal{V}_C&=&\frac{1}{2}\left(\zeta^2g^2_V-4\mu^2g^2_Vq_{7}^2\right)\mathcal{O}_{2}Y_{\phi7}\nonumber\\
&&-\frac{2}{3}\mu^2g^2_V\left(\mathcal{O}_{2}\mathcal{Z}+\mathcal{O}_{3}\mathcal{T}\right)Y_{\phi7}.
\end{eqnarray}

\item The $D_s^{\ast}\bar D_{s0}^{\ast}$ system:
\begin{eqnarray}
\mathcal{V}_D&=&\frac{1}{2}\beta\beta^{\prime}g^2_V\mathcal{O}_{1}Y_{\phi},\nonumber\\
\mathcal{V}_C&=&-\frac{1}{2}\left(\zeta^2g^2_V-4\mu^2g^2_Vq_{8}^2\right)\mathcal{O}_{2}Y_{\phi8}\nonumber\\
&&+\frac{2}{3}\mu^2g^2_V\left(\mathcal{O}_{2}\mathcal{Z}+\mathcal{O}_{3}\mathcal{T}\right)Y_{\phi8}.
\end{eqnarray}

\item The $D_s^{\ast}\bar D_{s1}^{\prime}$ system:
\begin{eqnarray}
\mathcal{V}_D&=&\frac{2g\tilde{k}}{9f^2_{\pi}}\left(\mathcal{O}_{5}\mathcal{Z}+\mathcal{O}_{6}\mathcal{T}\right)Y_{\eta}+\frac{1}{2}\beta\beta^{\prime}g^2_V\mathcal{O}_4Y_{\phi}\nonumber\\
&&-\frac{2}{3}\lambda\lambda^{\prime}g^2_V\left(2\mathcal{O}_{5}\mathcal{Z}-\mathcal{O}_{6}\mathcal{T}\right)Y_{\phi},\nonumber\\
\mathcal{V}_C&=&-\frac{2h^2q_{9}^2}{3f^2_{\pi}}\mathcal{O}_4Y_{\eta9}-\frac{1}{2}\left(\zeta^2g^2_V-4\mu^2g^2_Vq_{9}^2\right)\mathcal{O}_{6}Y_{\phi9}\nonumber\\
&&+\frac{2}{3}\mu^2g^2_V\left(\mathcal{O}_{5}\mathcal{Z}+\mathcal{O}_{6}\mathcal{T}\right)Y_{\phi9}.
\end{eqnarray}
\end{itemize}
Here, the variables $q_i$ are defined as $q_5=m_{D_s^{\ast}}-m_{D_s}$, $q_6=m_{D_{s0}^{\ast}}-m_{D_s}$, $q_7=m_{D_{s1}^{\prime}}-m_{D_s}$, $q_8=m_{D_{s0}^{\ast}}-m_{D_{s}^{\ast}}$, and $q_9=m_{D_{s1}^{\prime}}-m_{D_s^{\ast}}$. When performing the numerical calculations, the operators $\mathcal{O}_k$ will be replaced by the corresponding numerical matrixes, which are summarized in Table~\ref{matrix1}.

\acknowledgments
F. L. Wang would like to thank J. Z. Wang and M. X. Duan for very helpful discussions. This work is supported by the China National Funds for Distinguished Young Scientists under Grant No. 11825503, National Key Research and Development Program of China under Contract No. 2020YFA0406400, the 111 Project under Grant No. B20063, and the National Natural Science Foundation of China under Grant No. 12047501. R. C. is supported by the National Postdoctoral Program for Innovative Talent.

\end{document}